\documentclass[aps,prb,twocolumn,superscriptaddress,longbibliography]{revtex4-2}

\usepackage[utf8]{inputenc}
\usepackage{braket}
\usepackage{enumitem}
\usepackage{natbib}
\usepackage{float} 
\usepackage{etoolbox}
\usepackage{balance}
\usepackage[table]{xcolor}
\usepackage{colortbl}
\usepackage{array} 
\usepackage{xcolor}

\usepackage{tabularx}
\usepackage{tabularray}
\usepackage[normalem]{ulem}
\usepackage{bm}

\usepackage[utf8]{inputenc} 
\usepackage[T1]{fontenc}
\usepackage{amsmath,amssymb,amsfonts}
\usepackage{graphicx}
\usepackage{dcolumn}   
\usepackage{bm}        
\usepackage{fancyhdr}
\usepackage{hyperref}

\renewcommand{\headrulewidth}{0.4pt}   
\renewcommand{\footrulewidth}{0pt}
\makeatletter
\def\ps@titlepage{%
  \fancyhf{}%
  \renewcommand{\headrulewidth}{0.4pt}
  \renewcommand{\footrulewidth}{0pt}
  \fancyhead[C]{Accepted as regular article in PHYSICAL REVIEW B: \href{https://journals.aps.org/prb/accepted/10.1103/wq2b-b9fq}{https://journals.aps.org/prb/accepted/10.1103/wq2b-b9fq}} 
  \fancyfoot[C]{\thepage}          
}
\makeatother

\hypersetup{
  colorlinks = true,
  linkcolor  = blue,
  citecolor  = blue,
  urlcolor   = blue,
  breaklinks = true
}

\newcolumntype{C}{>{\centering\arraybackslash\hspace{0pt}}X}

\newcommand{\proj}[2]{\ket{#1}\!\bra{#2}}

\makeatletter
\renewcommand*\frontmatter@RRAPformat[1]{%
  \vspace{1.5em}
  \par\frontmatter@RRAP@format{#1}%
}
\makeatother

\begin{document}

\title{Skyrmionic qubits stabilized by Dzyaloshinskii-Moriya interaction as platforms for qubits and quantum gates}

\author{Doru Sticlet}
\affiliation{National Institute for R\&D of Isotopic and Molecular Technologies, 67-103 Donat, 400293 Cluj-Napoca, Romania}

\author{Romulus Tetean}
\affiliation{Faculty of Physics, Babes-Bolyai University,  400084 Cluj-Napoca, Romania}

\author{Coriolan Tiusan}
\email{Corresponding author: coriolan.tiusan@ubbcluj.ro}
\affiliation{Faculty of Physics, Babes-Bolyai University,  400084 Cluj-Napoca, Romania}
\affiliation{National Center of Scientific Research-CNRS, France}

\date{\vspace{1.5em}Phys.\ Rev.\ B — Accepted 14 November 2025}

\begin{abstract}

Quantum computation departs from the classical paradigm of deterministic, bit-based processing by exploiting inherently quantum phenomena such as superposition and entanglement.
We propose a framework for qubit realization based on skyrmionic states stabilized by the Dzyaloshinskii-Moriya interaction (DMI) in two-dimensional spin lattices.
The model incorporates competing exchange interactions, perpendicular magnetic anisotropy, and Zeeman coupling, solved via exact diagonalization under periodic (PBC) and open boundary conditions (OBC).
A quantum skyrmionic phase emerges for PBC within a parameter space defined by DMI, exchange, field, and anisotropy, while OBC favor classical-like, topologically protected skyrmions.
Quantum logic gates (Pauli X, Y, Z, Hadamard) are implemented on both skyrmion types.
Energy density and entanglement entropy analyses reveal that quantum skyrmions suffer from DMI-driven decoherence and reduced gate fidelity, whereas classical-like skyrmions maintain stability.
Exact simulations of qubit dynamics, including drive effects and Lindblad decoherence, demonstrate tunable anharmonic energy levels and coherent Bloch-sphere manipulation, making these skyrmionic states promising candidates for qubit implementation.
Overall, the Dzyaloshinskii-Moriya interaction plays a dual role-stabilizing skyrmionic qubits while simultaneously inducing decoherence during gate operations.

\end{abstract}

\maketitle

\begin{center}
{\small \textsc{doi}: \href{https://doi.org/10.1103/wq2b-b9fq}{10.1103/wq2b-b9fq}}
\end{center}

\thispagestyle{titlepage}   

\section{Introduction}\label{sec:intro}

Quantum computing represents a paradigm shift from classical computation, leveraging quantum mechanical phenomena to process information in fundamentally different ways~\cite{Steane1998,Nielsen2010}.
While classical computers manipulate bits that exist in definite states of 0 or 1, quantum computers utilize \textit{quantum bits} or \textit{qubits} that can exist in quantum superposition of both states simultaneously.

The theoretical foundations of quantum computing were established in the 1980s by physicists such as Richard Feynman~\cite{Feynman1982} and David Deutsch~\cite{Deutsch1985}, who recognized that quantum systems could potentially solve certain computational problems exponentially faster than classical computers. This quantum advantage stems from uniquely quantum properties including superposition, entanglement, and interference.

As the fundamental unit of quantum information the qubit state can be represented as
\begin{equation}
\ket{\psi} = \alpha\!\ket{0} + \beta\!\ket{1},
\end{equation}
where $\alpha$ and $\beta$ are complex probability amplitudes satisfying the normalization condition $|\alpha|^2 + |\beta|^2 = 1$. The states $\ket{0}$ and $\ket{1}$ represent the computational basis states, analogous to classical bits.

In the quantum world, qubits possess unique properties that sets them apart from classical bits—most notably, superposition, entanglement, and their behavior under measurement.

\textbf{Superposition} is one the most important feature. 
While a classical bit must be either 0 or 1, a qubit can exist in a coherent blend of both states at once. 
This means a quantum algorithm can simultaneously explore many possible solutions, giving quantum computing its remarkable potential~\cite{DeutschJozsa1992}.

\textbf{Entanglement} adds a layer of complexity and power. 
When two or more qubits become entangled, their states become deeply linked, so much so that the state of the whole system cannot be described simply by the states of individual qubits. 
For instance, two entangled qubits might be in the state
\[
\ket{\Phi^+} = \frac{1}{\sqrt{2}}(\ket{00} + \ket{11}),
\]
where knowing the state of one immediately determines the state of the other, no matter the distance between them.

\textbf{Measurement}, however, introduces a fundamental shift. 
Once a qubit is measured, it no longer exists in a superposition but collapses into one of its basis states, either \( \ket{0} \) or \( \ket{1} \), with probabilities determined by the amplitudes \( |\alpha|^2 \) and \( |\beta|^2 \). 
This collapse marks the transition from quantum ambiguity to classical certainty, and, with it, the loss of quantum coherence.

Quantum computing platforms have seen rapid development over the past decades, with a wide variety of physical systems proposed and explored as candidates for realizing qubits.  

Among these, \textbf{superconducting qubits} have emerged as one of the leading technologies. Pioneered and commercialized by major companies such as IBM, Google, and Rigetti~\cite{Arute2019,Krantz2019,Kjaergaard2020,Blais2021}, these systems are based on Josephson junctions~\cite{Josephson1962}, insulating barriers placed between superconducting materials. 
At millikelvin temperatures, these junctions support quantum states that correspond to distinct charge or flux configurations in a superconducting circuit. 
Several variants exist, including transmon qubits that mitigate charge noise, flux qubits utilizing persistent current loops, and phase qubits that rely on energy levels within current-biased junctions. Despite their maturity, superconducting qubits face several challenges, such as limited coherence times,  stringent cryogenic requirements, complex microwave control systems, and significant cross-talk in densely packed qubit arrays. The energy relaxation time, $T_1$, defined as the characteristic timescale over which a qubit undergoes spontaneous energy dissipation from the excited state to the ground state, typically falls within the range of $10$--$100~\mu\text{s}$. In contrast, the phase relaxation time, $T_2$, which characterizes the timescale over which quantum phase coherence between superposed states is lost due to dephasing mechanisms, is generally observed in the range of $1$--$50~\mu\text{s}$.

Another well-established approach is the use of \textbf{trapped ion qubits}, which encode quantum information in the internal electronic states of individual ions confined by electromagnetic fields. 
This platform, grounded in precise laser manipulation and radio-frequency trapping~\cite{Paul1990, Cirac1995, Wineland2013}, offers several compelling advantages. 
These include high-fidelity quantum gates exceeding 99\%, long coherence times that can extend to minutes, and the inherent uniformity of qubits derived from identical atomic species. 
However, scaling trapped ion systems remains challenging due to the complexity of the required laser infrastructure, slower gate speeds compared to superconducting systems, issues with anomalous heating, and increasing difficulty in individually addressing qubits as the system grows~\cite{Bruzewicz2019}.

In parallel, \textbf{photonic qubits} present a markedly different paradigm, where quantum information is carried by single photons~\cite{Knill2001,Kok2007}. 
Information is encoded in photon properties such as polarization, path, or time-bin. 
Operations are typically performed using linear optical elements and photon detectors. 
Photonic systems are particularly well-suited to quantum communication and networking due to their resistance to decoherence during propagation and their ability to function at room temperature. 
Despite these strengths, challenges such as the probabilistic nature of two-photon gates, photon loss, the need for high-quality single-photon sources, and scalability issues stemming from resource overheads remain significant hurdles.

\textbf{Silicon-based spin qubits} offer a pathway toward scalable quantum processors that may be integrated with existing semiconductor manufacturing technologies~\cite{Loss1998,Burkard2023}. 
These systems use the spin states of electrons or nuclei in silicon quantum dots to encode quantum information. 
The spin-up and spin-down states represent logical $\ket{0}$ and $\ket{1}$, and are manipulated through combinations of electric and magnetic fields applied via nanofabricated electrodes. 
Key advantages include compatibility with CMOS processes, small device footprints, potential for integration with classical electronics, and long coherence times—especially in isotopically purified silicon. On the other hand, this platform demands extremely precise control over voltage levels, faces device variability, and suffers from low readout fidelity and the complexity of implementing two-qubit gates.

A more theoretically oriented avenue is the pursuit of \textbf{topological qubits}, which aim to encode information in the non-local, topological properties of certain quantum systems~\cite{Kitaev2003,Nayak2008}. 
This concept relies on exotic quasi-particles known as anyons, particularly Majorana fermions, which exhibit non-Abelian braiding statistics~\cite{Kitaev2001,Beenakker2013,Aguado2017}. 
By braiding these particles around one another, one can perform quantum operations that are inherently fault-tolerant and naturally robust against local noise and decoherence. 
While this approach promises long coherence times and reduced error correction overhead, it remains largely unproven.   
Moreover, braiding operations alone do not provide a complete universal gate set, necessitating supplementary mechanisms for full quantum computation.
Despite recent claim of Microsoft that unveils Majorana 1, the world's first quantum processor powered by topological qubits~\cite{Microsoft2025},  there are still many challenging issues for the ultimate experimental realization of a topological qubit, and the materials and physical systems required to host Majorana fermions remain under active investigation. 

Finally, an emerging and highly exploratory direction in qubit research involves magnetic chiral textures, such as merons~\cite{Ezawa2022}, and more recently, \textbf{skyrmion-based qubits}~\cite{Psaroudaki2021, Psaroudaki2023, Xia2023, Petrovic2025}.
Skyrmions are nanoscale, topologically protected spin textures observed in certain magnetic materials~\cite{Fert2017, Zhang2023}. 
Their non-trivial topology confers robustness and resistance to perturbations~\cite{Ornelas2025}, and they can potentially be controlled via low-energy electric fields. 
Information can be encoded in a skyrmion's position, internal modes (such as helicity or the spin orientation of its core), or its mere presence, making skyrmions compelling candidates for quantum information processing. 
Additionally, their nanometric scale and demonstrated room-temperature stability in specific materials are compatible with the demands of scalable quantum hardware.
Recent theoretical developments have increasingly explored both {\em classical} and {\em quantum skyrmions}~\cite{Ochoa2019, Mazurenko2023}, addressing a broad range of phenomena: mechanisms for generating individual skyrmions and skyrmionic phases~\cite{Stepanov2019, Lohani2019, Siegl2022,Sotnikov2023}, skyrmion stability~\cite{Sotnikov2021, Gauyacq2019, Salvati2024}, formation of skyrmion lattices~\cite{Haller2022} or crystal phases~\cite{Mohylna2021}, as well as dynamics, manipulation, and phase transitions~\cite{Vijayan2023,Joshi2024}. 
Spin models have also been proposed for intrinsic antiferromagnetic skyrmions~\cite{Aldarawsheh2023}.
Notably, the concept of skyrmions and skyrmion-based qubits extends beyond magnetic systems. 
In nanophotonics, non-local optical skyrmions have recently been proposed as topologically robust, quantum-entangled states of light~\cite{Ornelas2024}, facilitated by advances in semiconductor cavity quantum electrodynamics~\cite{Ma2025}. 
Even with the growing interest and theoretical advances, skyrmion-based qubits remain at a very early stage of development. 
Key challenges include demonstrating quantum coherence, achieving precise control, implementing reliable readout mechanisms, optimizing materials for stable skyrmion formation, and developing a comprehensive theoretical framework to describe their quantum properties. 
While numerous physical platforms for qubit realization are actively being explored, each comes with its own set of advantages and limitations. 
The field continues to evolve rapidly, and no single approach has yet proven to be universally superior. 
Instead, specific platforms may be better suited to particular applications, and hybrid strategies could ultimately prove critical in achieving practical, large-scale quantum computing.

However, despite the diversity of physical implementations, all qubit platforms confront several overarching challenges that hinder progress toward practical, large-scale quantum computing. 
One of the most fundamental of these is \textbf{decoherence}, the process by which qubits lose their quantum coherence due to interactions with the surrounding environment. 
Decoherence sets a hard limit on the time available for quantum operations and remains a primary obstacle to sustaining reliable quantum computation. 
Even the most isolated systems are not immune, necessitating sophisticated techniques to preserve coherence over operationally relevant timescales. 
To address this, \textbf{quantum error correction} has emerged as a critical component of fault-tolerant quantum computing~\cite{Lidar2013, Cai2023}. 
Theoretical frameworks have been developed to encode logical qubits across many physical qubits, allowing quantum information to be preserved in the presence of errors. 
However, the overhead is substantial: current estimates suggest that thousands to millions of physical qubits may be required to reliably encode a single logical qubit. 
This imposes a significant burden on hardware development and control infrastructure, particularly for near-term devices.

Maintaining qubit performance over time also requires \textbf{precise control and continuous calibration}.  
Quantum systems are highly sensitive to fluctuations in their environment, and even small drifts in system parameters can degrade gate fidelity. 
As a result, advanced control electronics, feedback systems, and frequent re-calibration routines are essential to ensure consistent operation. 
This, in turn, places increasing demands on classical hardware and software infrastructure.

A central obstacle across all platforms is \textbf{scalability}. 
While small-scale quantum processors with tens or hundreds of qubits have been demonstrated, scaling up to the millions of qubits required for fault-tolerant quantum computing involves daunting engineering challenges. 
These include managing the growing complexity of control signals, ensuring adequate qubit connectivity, integrating components in a modular and manufacturable fashion, and maintaining low error rates throughout increasingly complex systems.

Despite these hurdles, the field of quantum computing continues to make remarkable strides. Recent demonstrations of \textbf{quantum advantage}, where quantum systems outperform classical computers on specific tasks, highlight the accelerating progress in both hardware and algorithms~\cite{Arute2019, Daley2022}. 
Nevertheless, realizing quantum computers capable of solving impactful, real-world problems remains an ambitious goal. 
Achieving this will require substantial advances in qubit fidelity, system architecture, and error correction capabilities~\cite{Preskill2018}.

Theoretical breakthroughs, such as Shor's algorithm for factoring large integers~\cite{Shor1994} and Grover's algorithm for unstructured search~\cite{Grover1996}, continue to underscore the transformative potential of quantum computing.  
Yet the path to practical realization is still unfolding. 
Each physical qubit technology offers unique advantages and limitations, and the ultimate architecture may depend on hybrid combinations or novel platforms not yet fully explored. 
The ongoing global investment in research and development across these diverse approaches reflects the enormous promise that quantum computing holds for fields as varied as cryptography, materials science, optimization, and fundamental physics.

This paper addresses the paradigm of skyrmionic qubits. 
Recent theoretical predictions indicate that the nanometer-sized skyrmions can encode quantum information in their helicity degree of freedom that can be manipulated using electric or magnetic fields, within a wide operating range providing a large anharmonicity~\cite{Psaroudaki2021, Psaroudaki2023}. 
A universal quantum computation scheme based on nanoscale helicity qubits in frustrated magnets has recently been proposed~\cite{Xia2023}. 
In parallel, pioneering strategies for manipulating skyrmion qubits include hybrid quantum systems, where qubits are strongly magnetically coupled to nanomechanical cantilevers, employing phonons as quantum interfaces~\cite{XFPan2024}. 
These type of skyrmions are stabilized by competing exchange interactions in the absence of the Dzyaloshinskii-Moriya interaction that would lock the helicity degree of freedom. Moreover, the helicity-unlocked chiral spin systems and their dynamics—for instance, in frustrated ferromagnetic films—are highly sensitive to external stimuli such as electric currents, which can induce helicity locking-unlocking transitions~\cite{Ezawa2017}.  
On the other hand, the conventional skyrmionic spintronic applications are mostly based on the DMI mechanisms and their control up to above the room temperature conditions is achieved  in skillfully tailored thin film architectures~\cite{Fert2017, Zhang2023}. 
To fulfill this criterium and profit on all existing know-how in skyrmionic materials and applications in information technologies~\cite{Zhang2023, Yu2023}, we investigated the classic and quantum skyrmionic states stabilized by DMI mechanisms~\cite{Fert2023, Camley2023}.  
Several other recent theoretical studies have pursued the direction of quantum skyrmions stabilized by DMI \cite{Siegl2022, Sotnikov2023, Sotnikov2021, Gauyacq2019, Salvati2024, Ma2025, Haller2022, Mohylna2021, Vijayan2023, Joshi2024}.

Our theoretical approach is based on an exact diagonalization framework developed for various two-dimensional spin lattice models with diverse geometries, symmetries, and sizes. However, in this study, we focus on lattices with triangular symmetry, under both periodic (PBC) and open boundary conditions (OBC). 

We demonstrate that, within a specific region of the quantum phase diagram---defined by the Dzyaloshinskii-Moriya interaction, direct exchange, anisotropy, and external magnetic field---a robust quantum skyrmionic phase can be stabilized under periodic boundary conditions. In contrast, similar analyses show that open boundary conditions predominantly give rise to classical skyrmions, akin to those described using micromagnetic models.  
We further demonstrate the possibility of manipulating both quantum and classical skyrmions to implement quantum logic gates, including Pauli $X$, $Y$, $Z$, and Hadamard gates. 
A detailed energy density analysis reveals that quantum skyrmions lack topological protection, whereas the classical-like skyrmions emerging under OBC are topologically protected. 
Time evolution analysis of the entanglement entropy during qubit manipulation indicates a general increase toward the von Neumann entropy limit. 
This arrow of time, directly governed by spin correlations driven by antisymmetric exchange interactions, is closely linked to the decoherence mechanisms responsible for the gradual decline in gate fidelity.  
Our results show that topological protection in classical skyrmionic qubits yields enhanced performance: slower growth of entanglement entropy and a reduced decay in gate fidelity over time, regardless of the gate type (Pauli or Hadamard). 
For skyrmionic states corresponding to spin lattices under OBC, our exact quantum calculations reveal an energy level diagram with clear anharmonicity, making them well-suited for qubit implementation. 
We simulate the manipulation of such qubits on the Bloch sphere using Pauli and Hadamard gates at multiple levels of approximation. 
First, we reduce the multi-spin system to an effective two-level system coupled to an external semi-classical electromagnetic field, simulating the effect of a coherent photon (e.g., in a microwave cavity). 
To make the model more realistic, we then introduce decoherence—both relaxation ($T_1$) and dephasing ($T_2$) via the Lindblad master equation~\cite{Lindblad1976,Gorini1976}, replacing the Schr\"odinger equation.  
Finally, we simulate the full Hamiltonian dynamics, including drive-induced effects, across the complete Hilbert space.  
Our analysis of the time evolution during gate operations reveals how entanglement entropy and gate fidelity are influenced by DMI-induced decoherence. 
A particularly challenging and intriguing finding emerges: the Dzyaloshinskii-Moriya interaction not only enables the formation of skyrmionic qubits but also acts as a major source of decoherence during their manipulation. 
Addressing this dual role calls for the design of advanced skyrmionic quantum materials in which DMI mechanisms are finely tuned to permit skyrmion formation, strengthen topological protection, and minimize the decoherence effects. 
In addition to these theoretical insights and phenomenological considerations, our work also explores some practical experimental aspects, including the physical measurement of a logical skyrmionic qubit and the use of antiferromagnetic skyrmions as two-qubit quantum gates for demanding quantum technology applications.

The structure of this paper is as follows. Section~\ref{sec:intro} provides a comprehensive introduction to the field of quantum computing, outlining the fundamental concepts, qubit paradigms, and current challenges. 
Section~\ref{sec:methods} details the methodology employed in our quantum approach, including the computational models and formalism, the definition of the relevant static and dynamic observables, and the strategies used for qubit manipulation (e.g., precessional dynamics and coupling to external periodic drives). 
This section also specifies the scales and units adopted in the calculations. Section~\ref{sec:results} presents and analyzes the results obtained for static and dynamic interacting two-dimensional spin lattices with both periodic and open boundary conditions. In this context, we demonstrate the manipulation of skyrmionic qubits and the construction of quantum gates via magnetic fields and photonic drives, considering different levels of description and Hilbert space dimensionalities. 
We further address the crucial issue of physically measuring a logical qubit state, which has significant implications for integration into quantum computing architectures. 
Finally, Section~\ref{sec:conclusions} provides a general discussion and concluding remarks, highlighting the principal findings of our work, the influence of the Dzyaloshinskii-Moriya interaction (DMI) on skyrmionic qubit fidelities, and outlining theoretical as well as experimental perspectives.

\section{Methods}\label{sec:methods}
\begin{figure}
\centering
\includegraphics[width=\columnwidth]{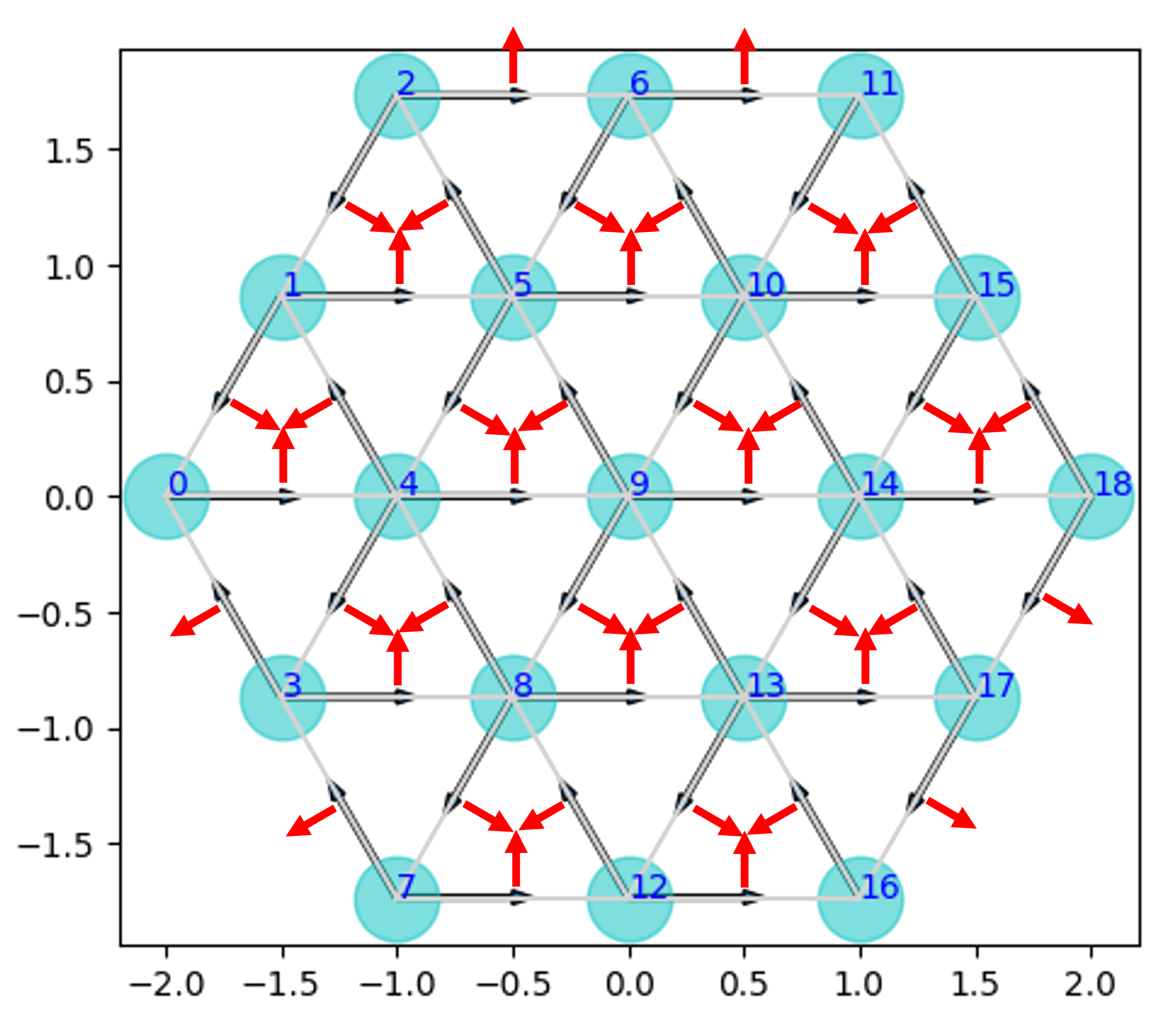}
\caption{Triangular spin grid with 19 spins. 
The arrows in the bondings define the bonding direction, assigning the same phase to the DMI vector $\vec D$ represented by the red arrows. 
Putting $\vec D \perp \vec r_{ij}$ will lead to N\'eel type skyrmions.}
\label{fig:1}
 \end{figure}

Our quantum approach and computational model are based on a triangular interacting spin lattice (see Figure~\ref{fig:1}) described by the following Hamiltonian:
 
\begin{equation}
\mathcal{H} = \sum_{\langle i,j\rangle} J_{ij} \vec{S}_i \cdot \vec{S}_j + \sum_{\langle i,j\rangle} \vec{D}_{ij} \cdot (\vec{S}_i \times \vec{S}_j) - \sum_i \vec{B} \cdot \vec{S}_i + K \sum_i S_z^2,
\label{eq:Hamiltonian}
\end{equation}
where the first term is the isotropic Heisenberg nearest-neighbor exchange interaction, the second term is the nearest-neighbor DMI interaction with a DMI vector $\vec D$  in plane and normal to an $i-j$ bond (see Figure~\ref{fig:1}), the third term is a Zeeman term,  e.g.,  $\bm{B}= (0, 0, B)$ when $\bm{B}\parallel$ OZ, and the last term is an anisotropy term, considering that the anisotropy easy axis is perpendicular to the 2D plane,  along $OZ$.  In our calculation we considered $J_{ij}=J$ and constant norm of the DMI vector $| \vec D_{ij}|=D$. 
The sums are over all lattice sites $i$ and all distinct nearest-neighbor pairs $\langle i,j \rangle$.

In our model, we deliberately neglected the contribution of the demagnetization energy (long-range dipolar interaction). This energy becomes important only when the number of spins (or the system volume) is large enough that long-range dipolar interactions can no longer be neglected compared to exchange interactions—in our case, DMI and direct exchange. At small scales (tens to hundreds of atoms/spins), we can reasonably assume that the exchange interactions are dominant, while the dipolar energy is several orders of magnitude smaller. For larger systems (nanoparticles, thin films, extended lattices)  the demagnetization energy need to be included, since in such cases the long-range dipolar interactions have a decisive influence on both the ground state and the excited states ~\cite{Nanomaterials2022}.

The first step is to solve numerically the Schrödinger equation $\mathcal{H} \ket{\psi}=E\ket{\psi} $ for the quantum many-body system corresponding to the 2D interacting spin lattice.  
The numerical approach is necessary because for a system composed by $n$ spins, the dimension of the Hilbert space is $2^n$,  and for the diagonalization one has to manipulate $2^n\times 2^n$ matrices. 
If for 1 and 2 spins the  $2\times 2$ and $4\times 4$ matrices can be still analytically easily diagonalized, for $n>3$ a numerical approach is preferred.  
In our calculations, we performed exact diagonalization (ED) calculations with the aid of the open-source {\sc Python} package {\sc QuSpin}~\cite{Quspin2017, Quspin2019}.
Note that ED is only feasible for systems with a few tens of spins, due to the exponential growth of the Hilbert space dimension with the size of the quantum system. Assuming the Hamiltonian is represented as a matrix of double-precision complex numbers, requiring approximately 4 TB of memory, we employed the Lanczos diagonalization algorithm, which constructs a sequence of orthonormal vectors (the Krylov subspace) and projects the large matrix onto a much smaller tridiagonal form.
Using this method, we diagonalized the Hamiltonian corresponding to a 19-spin lattice within 64 GB of RAM memory. 
We studied spins in a triangular lattice configuration (see Figure~\ref{fig:1}), where the inherent geometric frustration \cite{Moessner2006} combined with Dzyaloshinskii-Moriya interactions, not only aids in stabilizing non-collinear spin configurations such as helical and skyrmionic states but also serves as an additional source of quantum fluctuations.

Then,  once having the eigenvalues and the eigenvectors,  we measure various observables as the expectation values of various operators. 
Note that the expectation values can be calculated in the ground state and for all the excited levels considered for the output of the diagonalization.
\begin{itemize}
\item The {\bf onsite spin polarization:} 
\begin{equation}
\left< S^i_{\alpha}\right> =\frac{\hbar}{2}\bra{\psi}\sigma_\alpha^i\ket{\psi}
\end{equation} 
where  $\alpha=x,y,z$, and $\sigma_\alpha^i$ are the Pauli matrices acting on site $i$, that alongside with $\ket{\psi}$ are defined in the Hilbert space with $2^{n}$ dimensions, $\sigma_\alpha^{(i)} \in \mathbb{C}^{2^n \times 2^n}$:  
\begin{equation}
\sigma_\alpha^{i} =
\underbrace{I \otimes I \otimes \cdots \otimes I}_{(i-1)\ \text{factors}}
\ \otimes\ \sigma_\alpha\ \otimes\
\underbrace{I \otimes I \otimes \cdots \otimes I}_{(n-i)\ \text{factors}},
\end{equation}
with $\sigma^\alpha$ and $I$, being the $2\times2$ Pauli and identity matrices, respectively. 
This type of calculation gives us the instantaneous spin landscape corresponding to a given calculated eigenstate in the spin lattice, and can be further projected on the Bloch sphere $(S_x, S_y, S_z)$. 
\item The {\bf scalar spin chirality} is for a triangular spin lattice the equivalent to the classical topological number in micromagnetism, allowing one to classify various types of spin structures: e.g., helical, skyrmionic, fully saturated~\cite{Sotnikov2021}. 
It  is defined as the three-spin correlation function: 
\begin{equation}
Q=\sum_{i,j,k} \left\langle{\bm{\hat S}}_i \cdot ( {\bm{\hat S}}_j \times {\bm{\hat S}}_k) \right\rangle.
\end{equation}
Its physical signification is that of a solid angle $\Omega$ subtended by the spins, which acts as an emergent magnetic field for the electron coupled to spins, the effective (gauge) flux $\Phi$ penetrating the triangle being given by the product $\Omega \sim \vec S_i \cdot (\vec S_j \times \vec S_k)$. 
\item For classifying the classic skyrmions among the other spin textures, we  calculate the {\bf topological charge} for a discrete array of spins, equivalent to the classical skyrmion winding number as the accumulated rotation angle when going along a path from one side to the other of the spin grid (e.g., diagonal path, see Figure~\ref{fig:1}):
\begin{equation}
Q =\frac{1}{2\pi}\oint \nabla\theta\cdot dl =\frac{1}{2\pi} \left[  2\arccos \left( \frac{\left<\vec S_0\right> \cdot\left< \vec S_9\right>}{\left<S_0\right>\left<S_9\right>}\right)\right].
\end{equation}
\item The {\bf quantum fast Fourier transform (FFT) of spin-spin correlation function} 
\begin{equation}\label{eq:fft_S}
S_{\alpha\beta}(\bm{q}) = \sum_{\bm{r}\bm{r'}} e^{i\bm{q} \cdot (\bm{r'} - \bm{r})} \langle \hat{\bm{S}}_{\alpha\bm{r}} \hat{\bm{S}}_{\beta\bm{r'}} \rangle
\end{equation} 
 with $\alpha,\beta=x,y,z$. 
\item The {\bf elastic magnetic neutron scattering cross section} $d\sigma/d\Omega$ at momentum transfer vector $q$: 
\begin{equation} 
\frac{d\sigma}{d\Omega}(\bm{q}) \sim \sum_{\alpha\beta} \left( \delta_{\alpha\beta} - \hat{q}_\alpha \hat{q}_\beta \right) S_{\alpha\beta}(\bm{q}).
\end{equation}
This definition correspond to a neutron diffraction geometry  where the external applied magnetic field  $\bm{B} =B \hat{e}_z$ is parallel to the wave vector of the incoming neutron beam \cite{neutron}, the detector plane is spanned by the two components $q_x$ and $q_y$ of the scattering vector, and, in the limit of the small-angle approximation, one can assume $q_z = 0$. 
The elastic magnetic neutron scattering cross-section would be experimentally measured in a certain spin state: helical (H), skyrmionic (SK), or full polarized (FP) configuration.
\item The {\bf entanglement entropy density}: 
\begin{equation}
S_{\rm ent} = -\frac{1}{|A|} \text{tr}_A [\rho_A \log \rho_A]
\end{equation}
with  $\rho_A = \text{tr}_{A^c} \rho$ where $\rho$ is the density matrix of the entire system,  $\rho_A$ is the reduced density matrix of the subsystem $A$, and $A^c$ the complement of $A$ which is a subsystem of the spin lattice.
This entropy measures how entangled are subsystems $A$ and $A^c$.  
Concerning its time dependence, this would be a right measure for the decoherence arrow of time, i.e.,~the evolution of the decoherence in the system.
\end{itemize}

After determining the stationary eigenvalues, eigenstates, and associated observables, the dynamic evolution of the spin system was further investigated. This includes scenarios such as precessional experiments under the influence of an externally applied magnetic field with a specified orientation, or interactions between the spin system and a photonic drive. 
These dynamic simulations are particularly relevant for modeling skyrmionic qubit manipulation under various quantum gate operations, including Pauli $X$, $Y$, $Z$, and Hadamard gates. 

\subsection{Precessional manipulation}
The computational protocol for precessional `experiments' performed in the presence of an external magnetic field involves the following steps:

\begin{itemize}
\item At $t=0$, the Hamiltonian $\mathcal{H}_0$ of the system is described by equation~\eqref{eq:Hamiltonian} and solving by ED the Schr\o"dinger equation, we get the initial state $\ket{\psi(0)}$.
\item Then, we add the external field term
\begin{equation}
\mathcal{H}(t)=\mathcal{H}_0+\sum_i \bm{B}_{x,y,z}(t) \cdot \bm{S}_i
\end{equation}
that can be either: a static $\bm{B}_{x,y,z}(t>0)=(B_x, B_y, B_z)$ or periodic magnetic field $\bm{B}_{x,y,z}(t>0)=(B_x(t), B_y(t), B_z(t))$.
Its orientation will define the type of the quantum gate acting on the qubit. 
Under this new interaction, the quantum system initially into a stationary state will undergo a time evolution that we will calculate by numerically integrating the Schr\"odinger equation, and determining time-dependent eigenvalues $E(t)$ and eigenstates $\ket{\psi(t)}$.
\item Having the eigenvectors $\ket{\psi(t)}$, similarly to the stationary case, we calculate the time evolution of the observables: the on-site spin polarization $\left< S^i_{x,y,z}(t)\right>$, the quantum FFT of spin-spin correlations  $S_{\alpha,\beta}(\bm{q},t) $, and related neutron-diffraction cross-section $[d\sigma/d\Omega (t)]$ scalar spin chirality $Q(t)$, entanglement entropy $S_{\rm ent}(t)$, etc.
\end{itemize}

\subsection{Coupling with an external periodic drive}

For manipulation using an external periodic drive (e.g., photonic or microwave), we conducted the analysis through several steps and at various levels of approach complexity:

\begin{itemize}
\item We solve the eigenproblem corresponding to the static Hamiltonian $\mathcal{H}_0$ described by Eq.~\eqref{eq:Hamiltonian}.
\item Then, we chose the ground state $\{E_1, \ket{\psi_1}\}$ and the first excited state  $\{E_2, \ket{\psi_2}\}$ to build a 2 dimensional subspace, considering in a first-level approach that the dynamics of the system occurs mainly in the $\left\{ \ket{\psi_1}, \ket{\psi_2} \right\}$ subspace. 
This strategy leverages effective two-level physics in a large Hilbert space, and it is valid as long as one can safety stay within the subspace (no leakage to higher excited states), the drive is resonant with the energy gap $\Delta E=E_2-E_1$, and the other states are gapped far away. 
The chosen 2-level system represents our qubit space.
\item In this case, the full Hamiltonian $\mathcal{H}_0$ and the drive will be projected into a $2\times 2$ matrix,
$
\mathcal{H}_0 = (\begin{smallmatrix}
E_1 & 0 \\
0 & E_2
\end{smallmatrix}),
$
and the drive is implemented by choosing a drive operator, e.g., $D=\proj{\psi_1}{\psi_2}+\proj{\psi_2}{\psi_1}$, and then projecting it into the subspace spanned by $\left\{ \ket{\psi_1}, \ket{\psi_2} \right\}$.
\begin{equation}\label{eq:drive}
H_{\rm drive}(t)=A \cos(\omega t) H_{\rm gate},
\end{equation}
with $A$ and $\omega$ being the amplitude and frequency of the driving field. The $ H_{\rm gate}$ corresponds to the Pauli $X$ (bit flip, quantum NOT), $Y$ (bit flip $+$ phase), $Z$ (phase flip) and Hadamard (creating superposition) gate matrices:
$
\begin{aligned}
&\text{Pauli-}X : \quad H_{\rm gate} = \begin{pmatrix} 0 & 1 \\ 1 & 0 \end{pmatrix}, \\
&\text{Pauli-}Y : \quad H_{\rm gate} = \begin{pmatrix} 0 & -i \\ i & 0 \end{pmatrix}, \\
&\text{Pauli-}Z : \quad H_{\rm gate} = \begin{pmatrix} 1 & 0 \\ 0 & -1 \end{pmatrix}, \\
&\text{Hadamard :} \quad H_{\rm gate} = \frac{1}{\sqrt{2}} \begin{pmatrix} 1 & 1 \\ 1 & -1 \end{pmatrix}.
\end{aligned}
$
\item The time-dependent Hamiltonian will be 
\begin{equation}
\mathcal{H}(t)=\mathcal{H}_0+A \cos(\omega t)H_{\rm gate},
\end{equation}
for which we will solve and propagate the time evolution $\ket{\Psi(t)}$ that can be further used to measure the observables, e.g.
$\left< S_\alpha(t)\right>=\frac{\hbar}{2}\bra{\Psi(t)}\sigma_\alpha \ket{\Psi(t)}$, $\alpha=x,y,z$, $\sigma_\alpha$ being the Pauli matrices. 
More details about the derivation of the specific forms of the external magnetic fields required for the four distinct types of $H_{gate}$, along with the corresponding derivation is provided in Appendix A.

\item To simulate realistic time-dependent qubit behavior, we add decoherence including: {\em longitudinal relaxation} ($T_1$ decay), describing the decay from the excited state $\ket{1}$ to the ground state $\ket{0}$ and {\em transverse dephasing} ($T_2$ decay), describing the loss of phase coherence between $\ket{0}$ and $\ket{1}$, without population transfer.  
Instead of solving the Schr\"odinger equation for a pure state $\ket{\Psi(t)}$, we solve and evolve the density matrix $\rho(t)$, governed by the Lindblad equation,
\begin{equation}\label{eq:lindblad}
\frac{d\rho}{dt} = -\frac{i}{\hbar} [H(t), \rho] + \sum_{k} \left( L_k \rho L_k^\dagger - \frac{1}{2} \left\{ L_k^\dagger L_k, \rho \right\} \right),
\end{equation}
that includes both unitary evolution and dissipation from decoherence channels. 
$\rho$ is the density matrix, evolved under the Lindblad equation. The drive is included in the Hamiltonian as described before for the reduced 2-level system. 
The two Lindblad operators are $L_1=\sqrt{\gamma_1}  \ket{0}\bra{1}$ for relaxation with $\gamma_1=1/T_1$  being the relaxation rate and $L_2=\sqrt{\gamma_\Phi} \sigma_z$ with $\gamma_\Phi=1/T_2-1/2T_1$ being the pure dephasing rate, without energy loss.
\item With the eigenvectors issued from solving the Lindblad equation, we measure the observables, e.g., the spin polarization $\langle S_\alpha(t)\rangle$ that can be projected as a path on the Bloch sphere to directly visualize the gate dynamics.
\item A next level of approach gets beyond the 2 level projection and considers the full Hamiltonian photon drive. 
The corresponding Schrödinger equation reads
\begin{equation}\label{eq:td_schr}
i\hbar \frac{d}{dt} |\psi(t)\rangle = \left[ H_0 + A \cos(\omega t) D \right] |\psi(t)\rangle
\end{equation}
where $H_0$ is a large sparse or dense matrix  of the static Hamiltonian, $D$ is a rank-2 drive operator (e.g., for a Pauli-$X$ gate $D =\ket{\psi_1}\bra{\psi_2}+\ket{\psi_2}\bra{\psi_1}$), $\ket{\psi_{1,2}}$ being the eigenvectors of $H_0$ corresponding to the ground state and the first excited state, $A$ and $\omega$ , the amplitude and frequency of the driving field, respectively.
\item Computationally, the time evolution of the system coupled to the external drive is calculated using the following steps: (i) the action of the drive $D$ on an arbitrary vector $\ket{\phi}$ is calculated, e.g., for a Pauli-$X$ drive,  
\begin{equation}
D\ket{\phi}=\proj{\psi_1}{\psi_2}\phi\rangle + \proj{\psi_2}{\psi_1}\phi\rangle.
\end{equation}
(ii) The total time-dependent Hamiltonian action on the state is defined as
\begin{equation}
H(t) |\psi(t)\rangle = H_0 |\psi(t)\rangle + A\cos(\omega t) D |\psi(t)\rangle,
\end{equation}
with $\ket{\psi(t)}$,  being the state at time $t$. 
(iii) The ensuing Schr\"odinger equation is then solved to obtain the wave function at time $t$, $\ket{\psi(t)}$. 
With these, one calculates the observables  $\left< S_\alpha(t)\right>$, and represents them on the Bloch sphere. 
Note that for visualizing the qubit gate dynamics, one has to project the evolving state into {\em Logical Qubit Subspace} that is the two level subspace $\left\{ \ket{\psi_1}, \ket{\psi_2} \right\}$ where the qubit-like behavior exists. 
In this {\em Logical Qubit Subspace} the standard Pauli matrices are effectively defined:
\begin{equation}\label{eq:logical}
\begin{aligned}
& \sigma_x^{\rm logical}=\ket{\psi_1}\bra{\psi_2}+\ket{\psi_2}\bra{\psi_1},\\
&\sigma_y^{\rm logical}=i(\ket{\psi_2}\bra{\psi_1} -\ket{\psi_1}\bra{\psi_2}),\\
&\sigma_z^{\rm logical}=\ket{\psi_1}\bra{\psi_1} -\ket{\psi_2}\bra{\psi_2},\\
\end{aligned}
\end{equation}
so that the logical Bloch vector is:
\begin{equation}\label{eq:bloch_logical}
\bm{r}^{\rm logical}(t)=\bra{\psi(t)}\bm{\sigma}^{\rm logical}\ket{\psi(t)},
\end{equation}
with $\bm{\sigma}^{\rm logical}$, the vector of logical Pauli matrices.
\item Even if the Hamiltonian and the eigenvectors are embedded in the real full Hilbert space of the physical system of $n$ spins,  when mapping the dynamics on the qubit subspace defined by $\left\{ \ket{\psi_1}, \ket{\psi_2} \right\}$, we simulate the quantum evolution of our system considering a complete isolation of the 2-level subspace, assuming that there is no leakage into higher excited states, out of the qubit logical subspace.
\item The final and most computationally demanding approach addresses spin dynamics by considering the complete Hilbert space, which for our 19-spin lattice has a dimension of $2^{19}$. 
In this framework, the driven Hamiltonian is constructed using the vector $\ket{\Psi}$ in the full Hilbert space:
\begin{equation}\label{eq:full_drive}
H_{\rm drive} = A \cos(\omega t) G
\end{equation}
with $G$ being one of the  Pauli $X$, $Y$, $Z$ or Hadamard gate operators:
\begin{align}\label{eq:full_gate}
X&=\ket{\Psi_1}\bra{\Psi_2}+\ket{\Psi_2}\bra{\Psi_1},\notag\\
Y&=i(\ket{\Psi_2}\bra{\Psi_1} -\ket{\Psi_1}\bra{\Psi_2}), \notag\\
Z&=\ket{\Psi_1}\bra{\Psi_1} -\ket{\Psi_2}\bra{\Psi_2},\\
H&=\frac{1}{\sqrt{2}}(X_L+Z_L).\notag
\end{align}
Such drive operator couples the two states $ \ket{\Psi_1}$ and $\ket{\Psi_2}$. 
Under this driving, the system evolves toward the ground states of the driven Hamiltonian. 
In this case, the dynamics accounts for possible leakage into higher excited states, as the 2-level subspace is no longer completely isolated as in the previous case.
The necessary conditions to safely neglect leakage into the higher states would be: (1) the driving amplitude is small compared to the level spacing and (2) the drive frequency is off resonant with respect to the transitions from the two lower states to the rest of the spectrum.  
\end{itemize}

\subsection{Scales and units used in calculation}
In our quantum calculations based on the exact diagonalization (ED) method, implemented using the open-source {\sc Python} package {\sc QuSpin}, we employ dimensionless units. 
Specifically, all energy scales are normalized by setting the dominant interaction strength---the Dzyaloshinskii-Moriya interaction---to unity, i.e., $D=1$. 
Consequently, all relevant physical quantities such as the energy spectrum, magnetic field amplitude $B$, and exchange interaction $J$, are expressed in units of $D$.
Moreover, we adopt natural units by setting  $\hbar=1$, which implies that time is measured in units of the inverse energy scale. For instance, if $D=1\text{ meV}$, then time is expressed in femtoseconds as $t=\hbar/(1\text{ meV}) \simeq 0.66$ femtoseconds. 
This dimensionless framework provides significant advantages: the numerical results are universal and can be rescaled for any experimental system by simply adjusting the value of $D$ to match the specific material parameters. This flexibility is particularly useful in the study of tunable magnetic materials where both the DMI strength and exchange interaction can be engineered.

\section{Results and discussions}\label{sec:results}
We will discuss the results of our quantum calculations corresponding to the situations in which the triangular spin lattice has periodic aor open boundary conditions. 
\begin{figure*}[htbp]
  \centering
  \includegraphics[width=\textwidth]{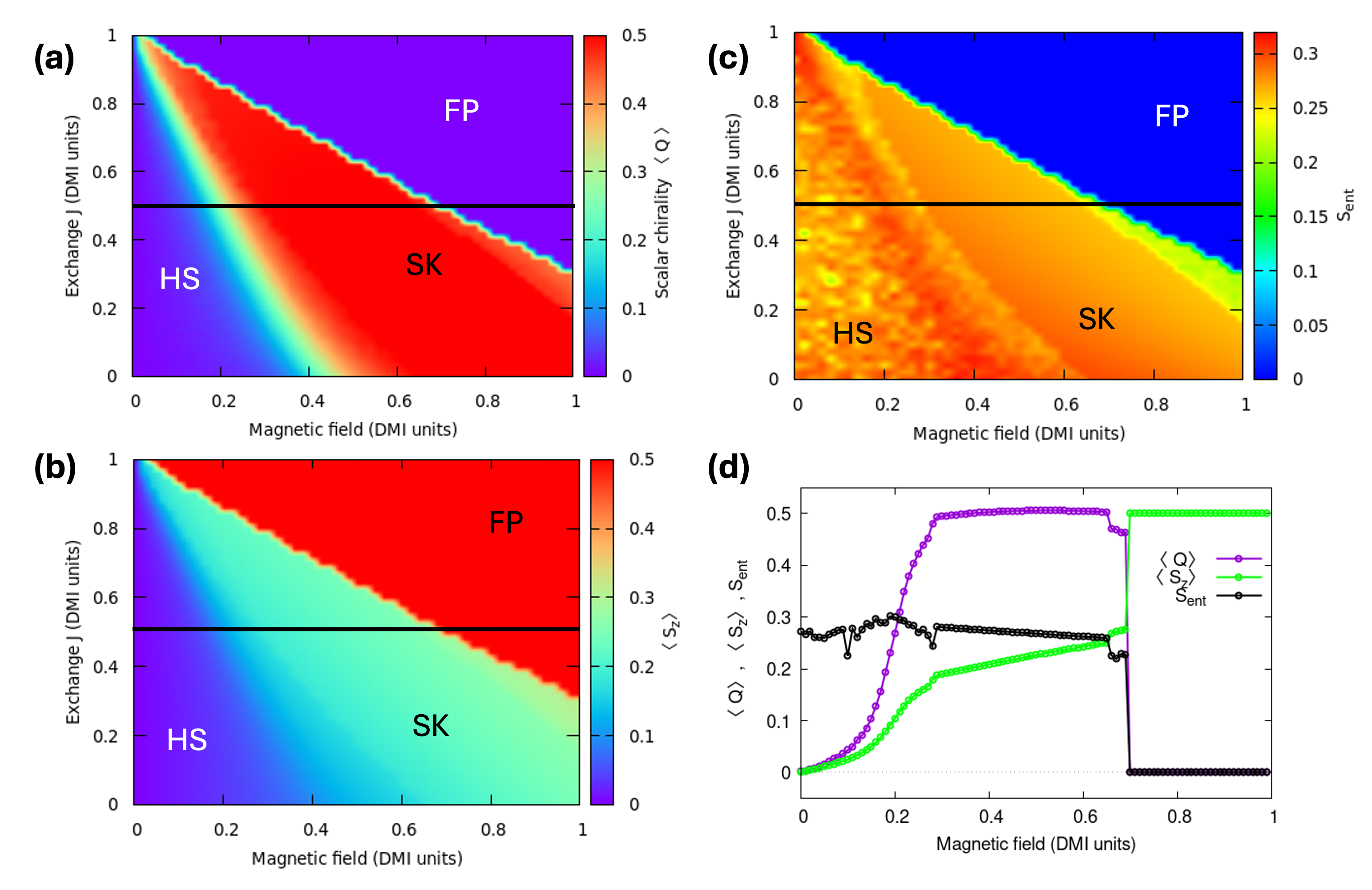}
  \caption{Quantum phase diagrams in the two variables space of magnetic field ($B$) and direct exchange ($J$), in units of DMI ($D=1$) illustrating the:  {\bf (a)} Scalar chirality, {\bf(b)} Average on-site polarization,  {\bf (c)} Entanglement entropy density,  {\bf (d)} Line analysis corresponding to $J=0.5$ and variable $B$ corresponding to the black line in each quantum phase diagrams. The quantum skyrmionic state (SK) corresponds to $\left<Q \right> \sim 0.5$, the helical states (HS) to $\left<Q \right> < 0.5$ and the fully polarized states to $\left<Q \right> =0$. The $\left<S_z \right> $ values are in units of $\hbar/2$.}
  \label{fig:2}
\end{figure*}
\begin{figure*}[htbp]
  \centering
  \includegraphics[width=0.8\textwidth]{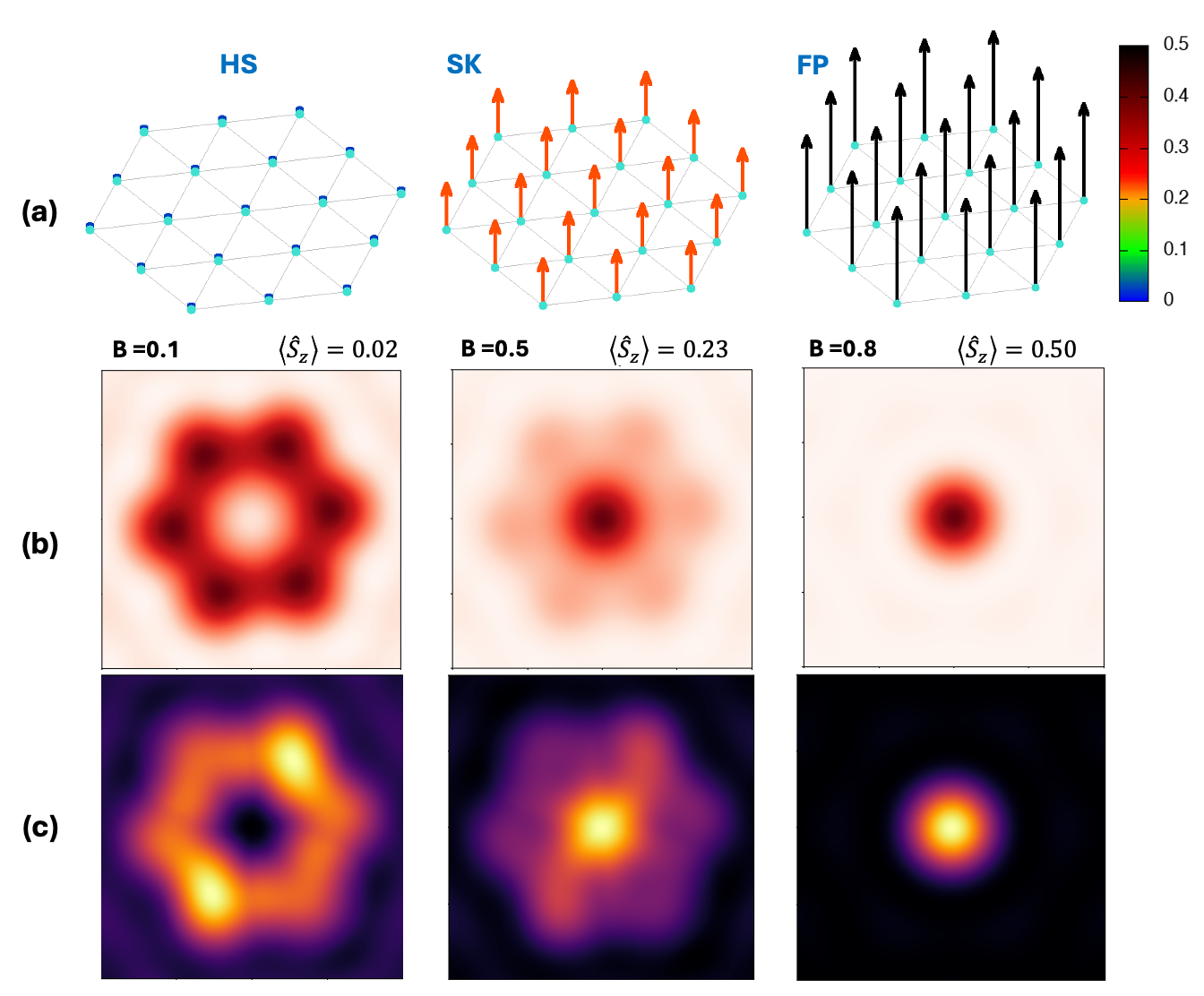}
  \caption{{\bf (a)} On-site average spin polarization corresponding to helical (HS), quantum skyrmionic (SK), and fully polarized (FP) spin states.  
  {\bf (b)} The FFT of the quantum spin-spin correlation functions and {\bf (c)} the elastic magnetic neutron scattering cross section $d\sigma/d\Omega$ at momentum transfer vector $\bm{q}$ corresponding  to H, SK and FP states.  
  Note that for (b) and (c) the images correspond to the 2D $1^{\rm st}$ Brillouin zone defined by $q_x/\pi=[-1,+1]$ and $q_y/\pi=[-1,+1]$.}
  \label{fig:3}
\end{figure*}

\subsection{Static quantum analysis of interacting 2D spin lattice with PBC}
Based on the formalism discussed in the previous section, we calculated different quantum phase diagrams (Figure~\ref{fig:2}), corresponding to the ground state, in which the states are classified with respect to the scalar chirality $Q$ (Fig.~\ref{fig:2}(a)), average spin polarization $\left< S_z\right>$ (Fig.~\ref{fig:2}(b)) and entanglement entropy density $S_{\rm ent}$ (Fig.~\ref{fig:2}(c)). 
Each diagram is built from $10^4$ individual quantum ED calculations. 
These results correspond to the interacting 2D spin lattice (Figure~\ref{fig:1}) with PBC described by the Hamiltonian from Eq.~\eqref{eq:Hamiltonian}. 
Noteworthy, the quantum phase diagrams presented in Figure~\ref{fig:2} correspond to an anisotropy value of $K=0$. 
Similar calculations performed for different values of anisotropy reveal that increasing $K$ results in a downward shift of all features in the phase diagram with respect to $J$. 
This behavior is consistent with the form of the Hamiltonian given in Eq.~\eqref{eq:Hamiltonian}, where the anisotropy term—chosen to represent anisotropy perpendicular to the lattice plane—acts as a correction to  $JS_i^zS_j^z$.
In the Figure~\ref{fig:2}(d), we illustrate a line projection corresponding to a direct exchange value $J/D =0.5$. 
The results corresponding to this line-projection are in perfect agreement with the similar one reported by Sotnikov et al.~\cite{Sotnikov2021}. 
The phase diagrams allow clearly identifying the three main quantum phases of the 19-spins triangular lattice.
\begin{enumerate}
\item {\bf The Helical states} correspond to a scalar chirality value $Q<0.5$, and average spin polarization $\left< S_z\right>$ gradually increasing with the external field, and a fluctuating entanglement entropy density. 
In the helical phase window, below a critical magnetic field $B$ linearly increasing with $J$, the system ground state energy is 6-fold degenerated in a high-symmetry state with multiple (6) spin configurations having same energy. 
A comparative study on square lattices, whose symmetry eliminates geometric frustration, clearly demonstrates that geometric frustration is the primary source of ground-state degeneracy in the helical phase, as well as the associated large fluctuations in entanglement entropy.
\item {\bf The Quantum skyrmionic states states}. 
In the intermediate magnetic field region, the ground state degeneracy is lifted, and the entanglement entropy $S_{\rm ent}$ become constant. 
This indicates that the magnetic field breaks some of the initial symmetries by introducing an energy bias for a spin configuration, the system's symmetry is reduced, and a specific unique ground state emerges. 
This state corresponds to the {\bf  quantum skyrmionic state} where the scalar chirality $Q \sim 0.5$.
\item Above a second critical field (high B), a narrow window of degeneracy reappears whose width competes with $J$ magnitude. 
Here, some new spin configurations become energetically competitive, the DMI creates alternative spin arrangements that become nearly degenerate, even under strong field.
\item {\bf The fully saturated spin state} corresponding to a scalar chirality $Q \sim 0.5$, an average spin value $\left< S_z\right> =0.5$ (in $\hbar/2$ units) and zero entanglement entropy density $S_{\rm ent}=0$, as expected for a perfectly ordered array of spins in the eigenstate corresponding to a strong magnetic field applied along $OZ$. Here, the degeneracy vanishes as all spins align with the magnetic field.
\end{enumerate}

Note that the PBC transform the spin system from a collection of individual spins to a  collective, synchronized quantum mechanical entity.
Figure~\ref{fig:3} reveals the following main results:

(1) Regardless of the state and the DMI strength, we have a constant value of the on-site spin polarization due to {\bf translation invariance of spin amplitudes} under PBC with DMI. 
This is a direct consequence of the fact that under PBC, the system Hamiltonian (see equation~\eqref{eq:Hamiltonian}) commutes with the translation operator $\bm{T_a}$, where $\bm{a}$ is a lattice vector: $\left[\mathcal{H}, \bm{T_a}\right]=0$. 
Therefore, the ground state can be chosen as an eigenstate of $\bm{T_a}$, leading to the translational invariance $\left< S^z_i \right>=\left <S^z_{i+\bm{a}}\right>$.

(2) Except for the fully saturated state, the $\left< S^z_i \right>$ is reduced ($\left< S^z_i \right> <0.5$). 
To explain this reduction, we consider an initial state that is fully polarized along the $z$ axis.
Looking at the Hamiltonian described by Eq.~\eqref{eq:Hamiltonian}, one can easily demonstrate that despite the fact that the DMI is not commuting with $S_z$, the PBC will ensure a global canceling of the DMI effect. 
However, even though PBC ensures global cancellation of DMI, the ground state shows quantum fluctuations. 
This can be analyzed within the perturbation theory. 
The corrected ground state due to the presence of DMI at first order in perturbation theory reads as
\begin{equation}
|GS\rangle = |GS\rangle_0 + \sum_{n} \frac{\langle n | H_{\rm DMI} | GS\rangle_0}{E_0 - E_n} |n\rangle,
\end{equation}
where $\ket{n}$ represents excited states with one or more flipped spins, and $E_n >E_0$, $E_0$, the ground state. 
With this new eigenfunction, the spin expectation value, including DMI effect, becomes
\begin{equation}
\langle S^z_i \rangle = \langle GS | S^z_i | GS \rangle \approx \frac{1}{2} \left( 1 - \sum_{n} \frac{|\langle n | H_{\rm DMI} | GS \rangle_0|^2}{(E_0 - E_n)^2} \right).
\end{equation}

In the case of weak DMI ($D \ll J$), the reduction in spin expectation value scales as $\langle S^z_i \rangle \approx \frac{1}{2} ( 1 -\alpha \frac{D^2}{J^2} )$, with $\alpha$ being a positive constant depending on lattice geometry and magnetic field strength.

For larger DMI values, when $D$ becomes comparable to $J$ (e.g.,  $J/D \sim 0.5$), the perturbative approach breaks down, and we need a more sophisticated treatment, leading to the following result: 
\begin{equation}
\langle S^z_i \rangle \approx \frac{1}{2} \left[ 1 - \alpha \frac{D^2}{J^2} - \beta \frac{D^4}{J^4} + O\left(\frac{D^6}{J^6}\right) \right].
\end{equation}
Depending on lattice geometry and field strength, the third and higher order terms can become important, and the reduction of $\langle S_z \rangle$ can reach $20-30$\%.

(3) Due to the constant on-site spin norm, related to the translational invariance imposed by the PBC, the non-colinear spin texture cannot be directly observed for the quantum skyrmions, as expected when compared to classical skyrmions (micromagnetic) in which the neighbor spins rotate. 
However, the skyrmionic type non-colinear spin textures emerge in the correlation functions, i.e., the ground state retains ferromagnetic amplitude uniformity while developing complex spin correlations, even for large DMI under PBC. 
Defining a connected (spin-spin) correlation function,
\begin{equation}
C_{ij}^{\alpha\beta} = \langle S_i^\alpha S_j^\beta \rangle - \langle S_i^\alpha \rangle \langle S_j^\beta \rangle,
\end{equation}
we can easily mathematically demonstrate that for large DMI, these correlations develop a spiral structure, in the Fourier space given by Eq.~\eqref{eq:fft_S}.

As illustrated in Figure~\ref{fig:3}(b), the correlations peak at a finite wave-vector $q \ne 0$, indicating spiral order in both helical (HS) and quantum skyrmionic state (SK), the real-space correlations showing oscillations, $$S_{\alpha\beta}(\bm{q}) \sim \cos(\bm{q}\cdot \bm{r}+ {\rm phase}).$$
The quantum skyrmionic state can be distinguished from the helical state by the presence of the main peak at $q=0$, with larger amplitude than the satellites at $q \ne 0$, with hexagonal symmetry.

Figure~\ref{fig:3}(c) illustrates the  elastic magnetic neutron scattering cross-section $d\sigma/d\Omega$ at momentum transfer vector $\bm{q}$, corresponding  to H, SK and FP states, i.e., what would be experimentally measured in a geometry where the externally applied magnetic field  $\bm{B} =B \hat{e}_z$  is parallel to the wave vector of the incoming neutron beam, the detector plane is spanned by the two components $q_x$ and $q_y$ of the scattering vector, and, in the limit of the small-angle approximation, $q_z = 0$. 
In the helical state, there is a superposition of spin spirals with wave vectors $q \ne 0$, resulting in six pronounced Bragg  peaks in $S_{zz}(q)$.
In the quantum skyrmionic state, there is both a Bragg peak at  $q = 0$ and the off-diagonal Bragg ``ring'' caused by the radial polarization winding of the skyrmion.
The fully saturated state is characterized by a single Bragg peak at $q=0$.

In conclusion, for this static analysis of the 2D interacting spin lattice with PBC, our calculations illustrate that translation symmetry and DMI create a quantum skyrmionic ground state with uniform ferromagnetic amplitude $\langle S_i^z\rangle$ (reduced by $D^p/J^p$ fluctuations, with $p$ scaling with the $D$ amplitude), coexisting with complex spin correlations. 
While single-site observables remain uniform, non-colinear DMI textures emerge exclusively in inter-site correlations, demonstrating simultaneous ferromagnetic coherence and intricate quantum entanglement patterns. 
As we will see in the next subsection, the DMI-induced quantum fluctuations critically impact coherent spin manipulation, presenting challenges and opportunities for skyrmion quantum gate implementation. 
The balance between quantum coherence and correlation structure exploitation becomes essential for skyrmion qubit design.

\subsection{Dynamic quantum analysis of interacting 2D spin lattice with PBC}
\begin{figure}[htbp]
  \centering
  \includegraphics[width=\columnwidth]{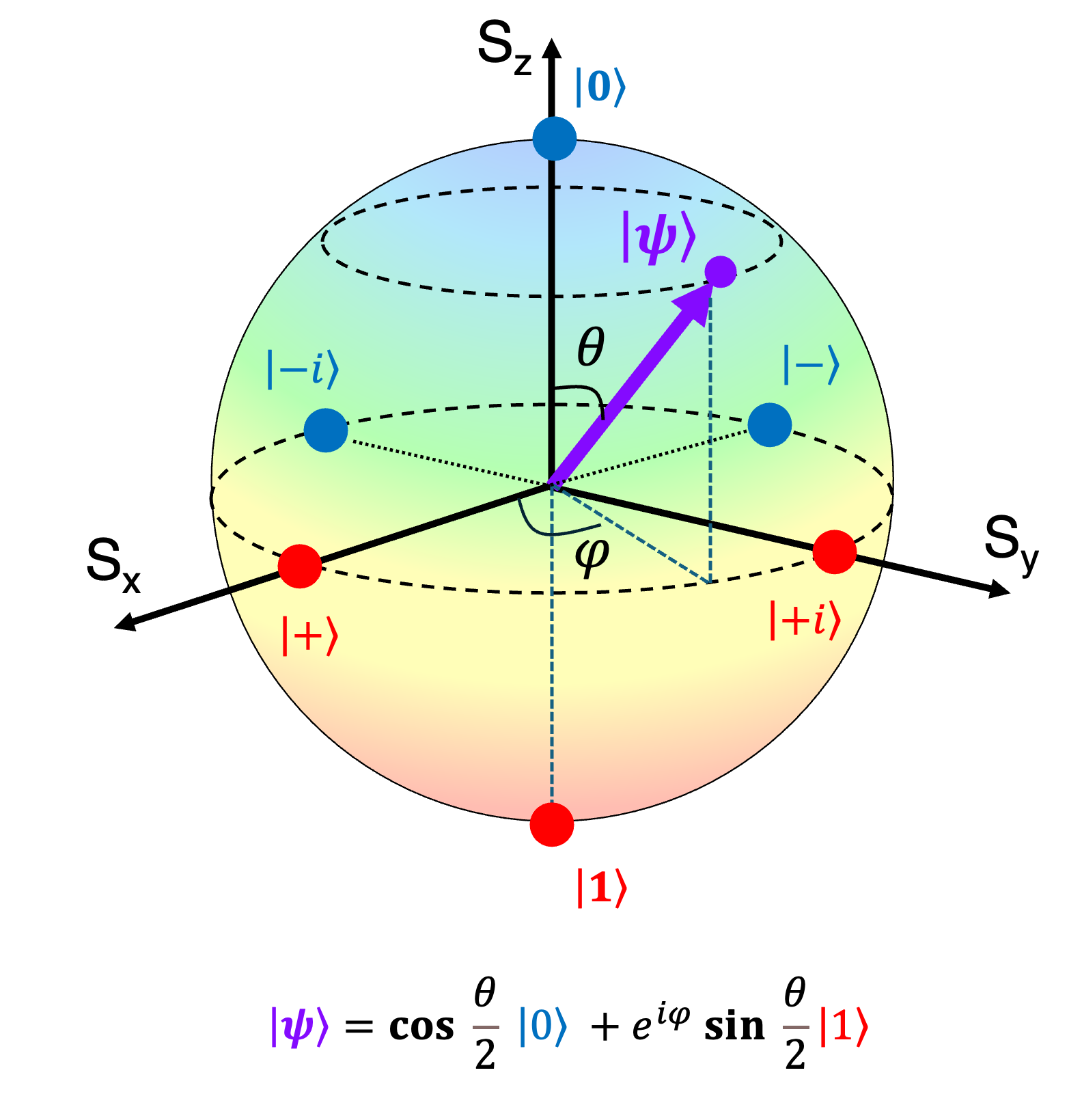}
  \caption{Bloch sphere representation for a 2-level state qubit, illustrating the main states $\ket{0}=$ ground state, $\ket{1}=$ excited state.}
  \label{fig:4}
\end{figure}

In this subsection, we investigate the precessional manipulation of a quantum skyrmionic state  in view of qubit applications, demonstrating the feasibility of several quantum gates: Pauli $X$, $Y$, $Z$, and Hadamard. 
The main concept of this type of manipulation is the spin precession around an effective field $\bm{B}_{\rm {eff}}$. 
The quantum information is stored in the value of $\langle S_z \rangle$, a quantum state $\ket{\Psi}$ being represented on the Bloch sphere (Figure~\ref{fig:4}).
\begin{figure*}[htbp]
  \centering
  \includegraphics[width=\textwidth]{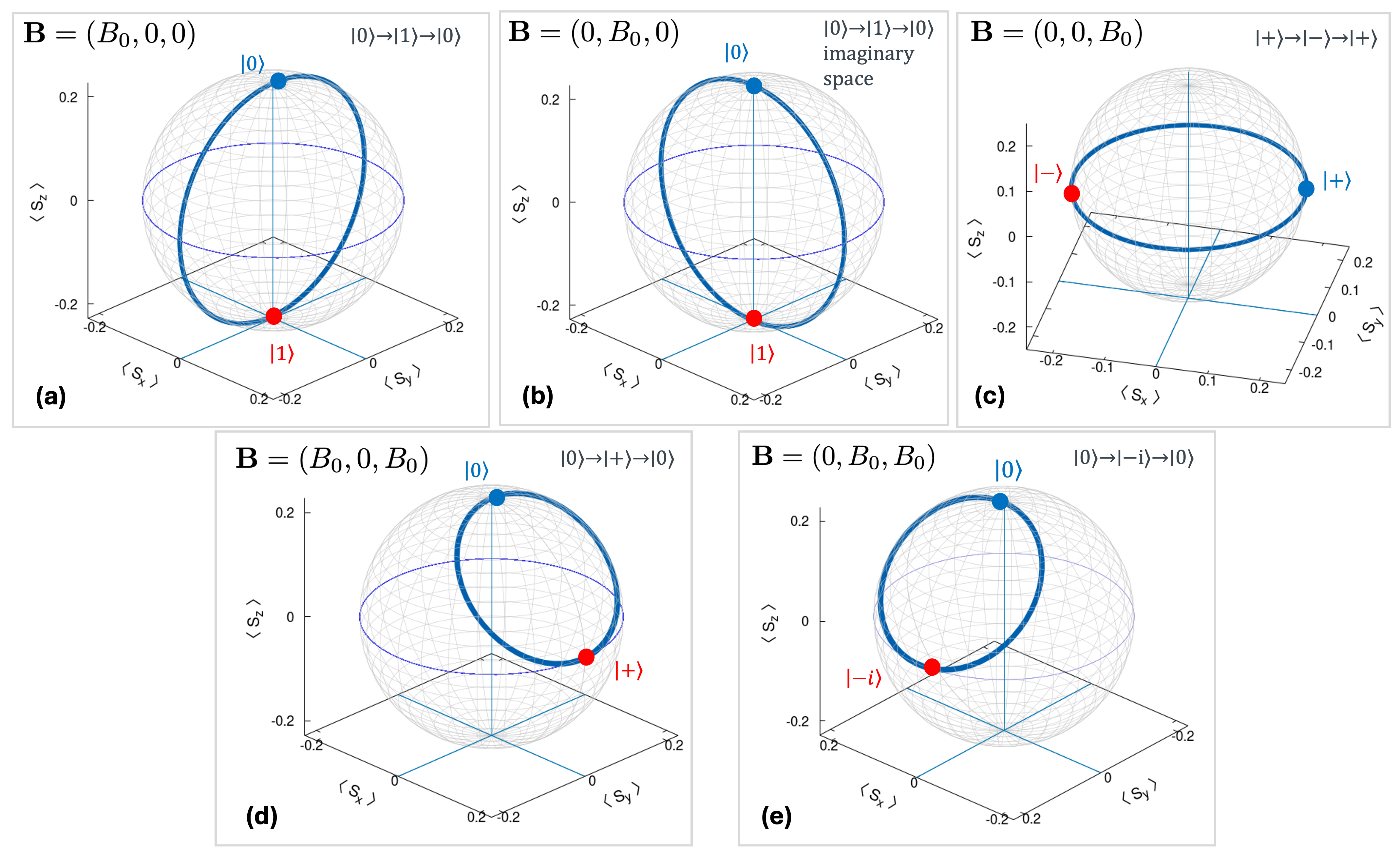}
  \caption{Simulated time dynamics trajectory on the Bloch sphere, corresponding to {\bf (a)} Pauli-$X$ {\bf(b)} Pauli-$Y$ {\bf (c)} Pauli-$Z$ (phase-shift) and {\bf (d, e)} Hadamard gates. 
  Each trajectory is constructed by applying the corresponding field drive term [see Eq.~\eqref{eq:drive}], and propagating the Schr\"odinger equation solution $\ket{\psi(t)}$ over 300 points grid of the time window corresponding to the Rabi period.}
  \label{fig:5}
\end{figure*}
We consider an initial quantum SK state, i.e., the spin lattice with spins primarily aligned along $z$ axis, due to the field $B_z$. 
To manipulate coherently all the spins, we need to apply a strong field that dominates the DMI interactions, the latter preferring to induce canted structures of adjacent spins. 
Because this is not allowed by the PBC constraints, the effect will be a strong reduction of  $\langle S_z \rangle$ by enhanced quantum fluctuation during the precessional manipulation with $\bm{B}_{\rm {eff}}$. 
To estimate the `critical' applied magnetic field for overcoming the DMI effect (decoupling), we recall the Eq.~\eqref{eq:Hamiltonian} with $K=0$.
The decoupling would correspond to a situation when the Zeeman term (e.g., with an applied field $(0,0, B_x)$ for a Pauli-$X$ gate) should dominate over both Heisenberg coupling $J$, $D$, and $B_z$.
Considering an arbitrary factor of 50, for a SK state defined by $J/D=0.5$, $B_z/D =0.5$ (see the phase diagram from Fig.~\ref{fig:2}), we find that $B_x\gg 50\sqrt{J^2+D^2+B_z^2}\approx 60$ to have decoupling. 
The spins will precess around an effective field $\bm{B}_{\rm {eff}}$  tilted from the $x$ axis by an angle $\theta =\arctan(B_z/B_x)$. 
This gives $\theta=0.47^\circ$ for $B_x=60$, and less than $0.3^\circ$ for  $B_x=100$, which is the field considered in our gate simulations. 
\begin{figure}[htbp]
  \centering
  \includegraphics[width=\columnwidth]{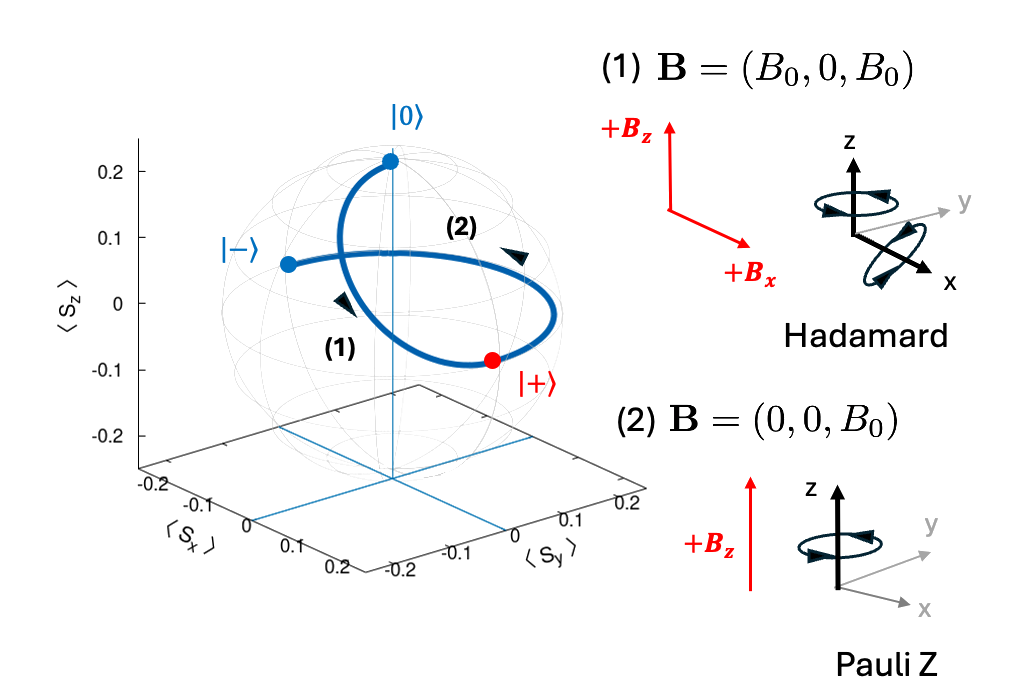}
  \caption{Simulated combined quantum-gate manipulation alternating Hadamard and Pauli-$Z$ gates. 
  The adjacent sketch illustrates the precession around the effective external field.}
  \label{fig:6}
\end{figure}
\begin{figure*}[htbp]
  \centering
  \includegraphics[width=\textwidth]{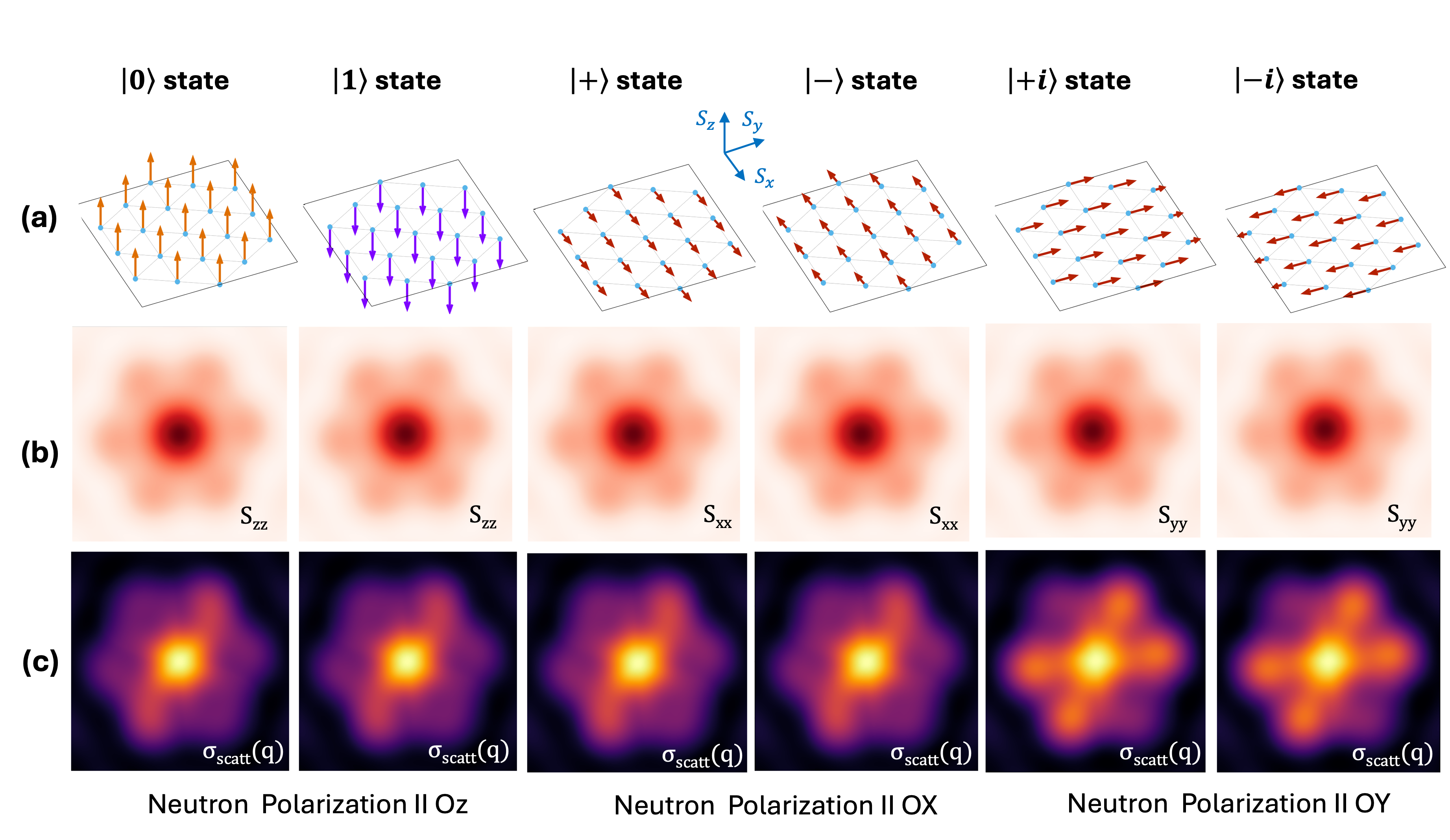}
  \caption{Robustness of the quantum SK state during the gate manipulation. 
  We illustrate the real space on-site spin polarization {\bf (a)} for the $\ket{0}$, $\ket{1}$, $\ket{+}$, $\ket{-}$, $\ket{+i}$, $\ket{-i}$ states, their corresponding FFT of the spin-spin correlation function $S_{zz}$ {\bf (b)}, and the cross-section {\bf (c)} of a neutron diffraction experiment with the beam polarization indicated in the figure insert. 
  The $S_{zz}$ and $\sigma_{\rm scatt}(q)$ images correspond to the 2D $1^{\rm st}$ Brillouin zone defined by $q_x/\pi\in[-1,+1]$ and $q_y/\pi\in[-1,+1]$.}
  \label{fig:7}
\end{figure*}
\begin{figure*}[htbp]
  \centering
  \includegraphics[width=0.8\textwidth]{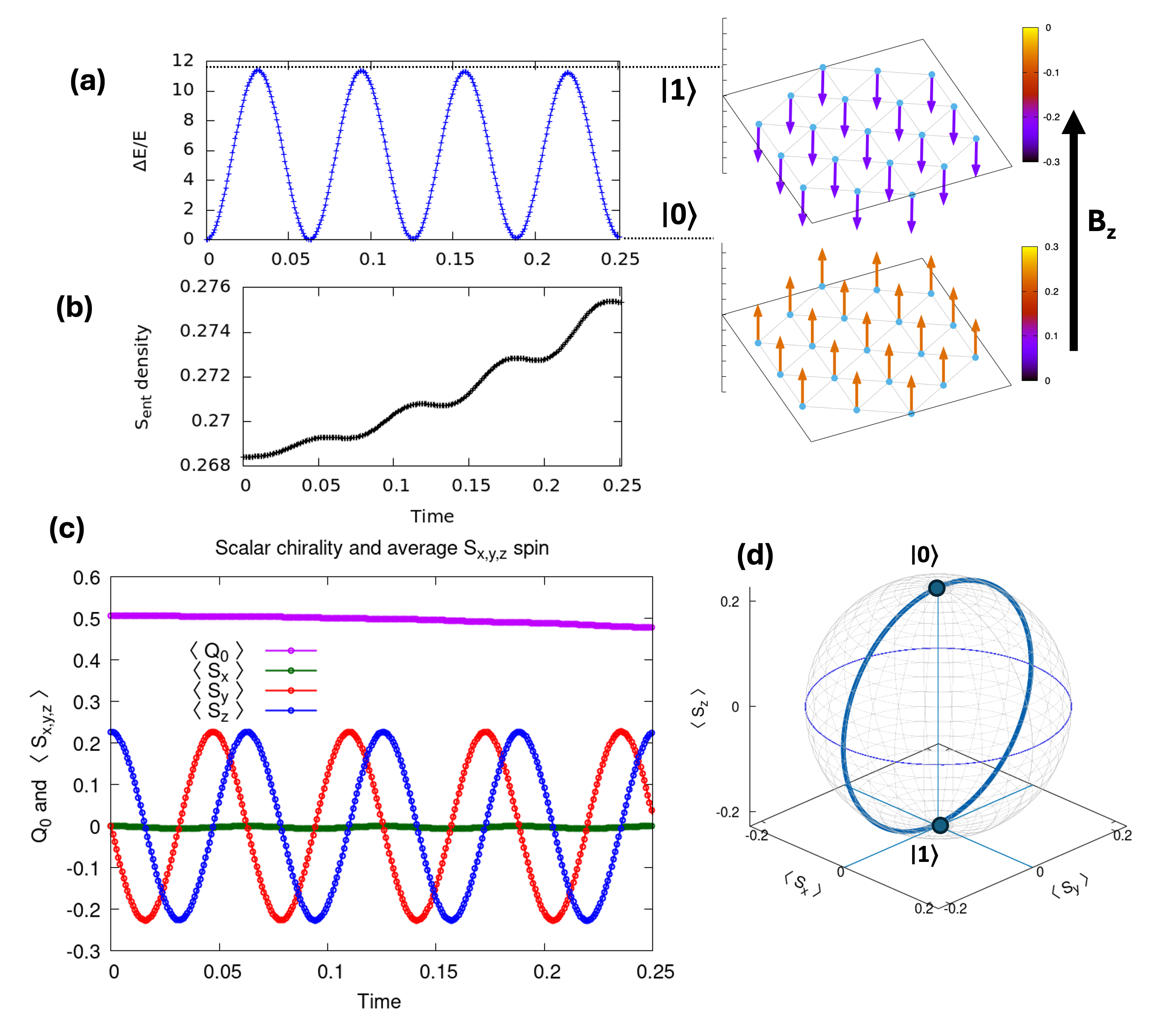}
  \caption{ {\bf (a)} Energy and {\bf (b)} entanglement entropy density variation during the precessional Rabi oscillations between the  $\ket{0}$ and $\ket{1}$  states corresponding to 4 periods timescale. {\bf (c)} Scalar chirality $Q_0$ and average value of $x$, $y$, $z$ spin polarization. {\bf (d)} Trajectory on the Bloch sphere corresponding to the Pauli-$X$ gate.}
  \label{fig:8}
\end{figure*}

In Figure~\ref{fig:5}, we illustrate the dynamics of the quantum SK state during the precessional manipulation,  using different orientations of a static applied magnetic field with chosen amplitude $B_0=100$.
The field orientation define the type of the gate. For the Hadamard-type gate, we apply simultaneously $B_x(B_y)$ and $B_z$. 
This type of gate imposes the uniform superposition of the system and allows mixing in a predictable way the initial state before starting a quantum computation. 
Each point from the trajectories on the Bloch sphere represents the average value $\langle S_z\rangle$ calculated with an instantaneous eigenvalue $\ket{ \psi(t)}$ of the Hamiltonian [see~Sec.~\ref{sec:methods}].

Such type of manipulations can be further combined, by a succession of pulse sequences, each pulse having a time width correspondingly chosen in agreement to the corresponding gate precession period. 
In Figure~\ref{fig:6},  we illustrate a combined manipulation of a quantum skyrmion alternating a Hadamard and a Pauli-$Z$ gate with properly chosen pulse lengths.

The results presented in the Figure~\ref{fig:7},  demonstrate at a first glance the robustness of the quantum skyrmionic state during the gate manipulation. 
This issue can be clearly observed  from the analysis of the real space on-site spin polarization $\langle S_z^i \rangle$ [Fig.~\ref{fig:7}(a)], each spin polarization being `measured' as $\langle S_\alpha^i \rangle=\frac{1}{2}\bra{\psi} \sigma_\alpha^i \ket{\psi}$, $\alpha=x,y,z$,  the   FFT of the spin-spin correlation function $S_{zz}$ [Fig.~\ref{fig:7}(b)], and the cross section of a neutron diffraction experiment with properly chosen beam polarization direction [Fig.~\ref{fig:7}(c)].  
However, this result mainly illustrates the conservation of the translational invariance versus the on-site spin norm, and of the skyrmionic correlations during the precessional manipulation of the quantum skyrmion qubit over the Bloch sphere. 
A more detailed examination, focusing on the time evolution of energy variation and entanglement entropy density, reveals that decoherence effects emerge during gate manipulation.

First, we analyze the manipulation of the quantum SK qubit in a Pauli-$X$ gate, within a relatively short time corresponding to 4 periods of precessional rotation between the states $\ket{0}$ and $\ket{1}$, see Figure~\ref{fig:8}. 
One clearly distinguishes the Rabi oscillations between these two levels and an oscillating increase of the entanglement entropy density. 
The oscillatory behavior is typical when coupling a drive to a 2-level system.
The periodic increase and decrease of the entanglement entropy follows the Rabi oscillations due to the coherent exchange of information between the system and the drive. 
Similar oscillatory behavior will be illustrated later when coupling a 2-level system with a photonic drive. 
Note that the entanglement entropy density has been calculated considering that the total system of spins is partitioned in two symmetric parts.
Figure~\ref{fig:8}(c) further confirms the robustness of the quantum SK state during the gate manipulation, over 4 periods/cycles, as the scalar chirality stays around a value of $0.5$ that is the one expected for an SK state [as illustrated in Fig.~\ref{fig:2}].
 \begin{figure}[htbp]
  \centering
  \includegraphics[width=\columnwidth]{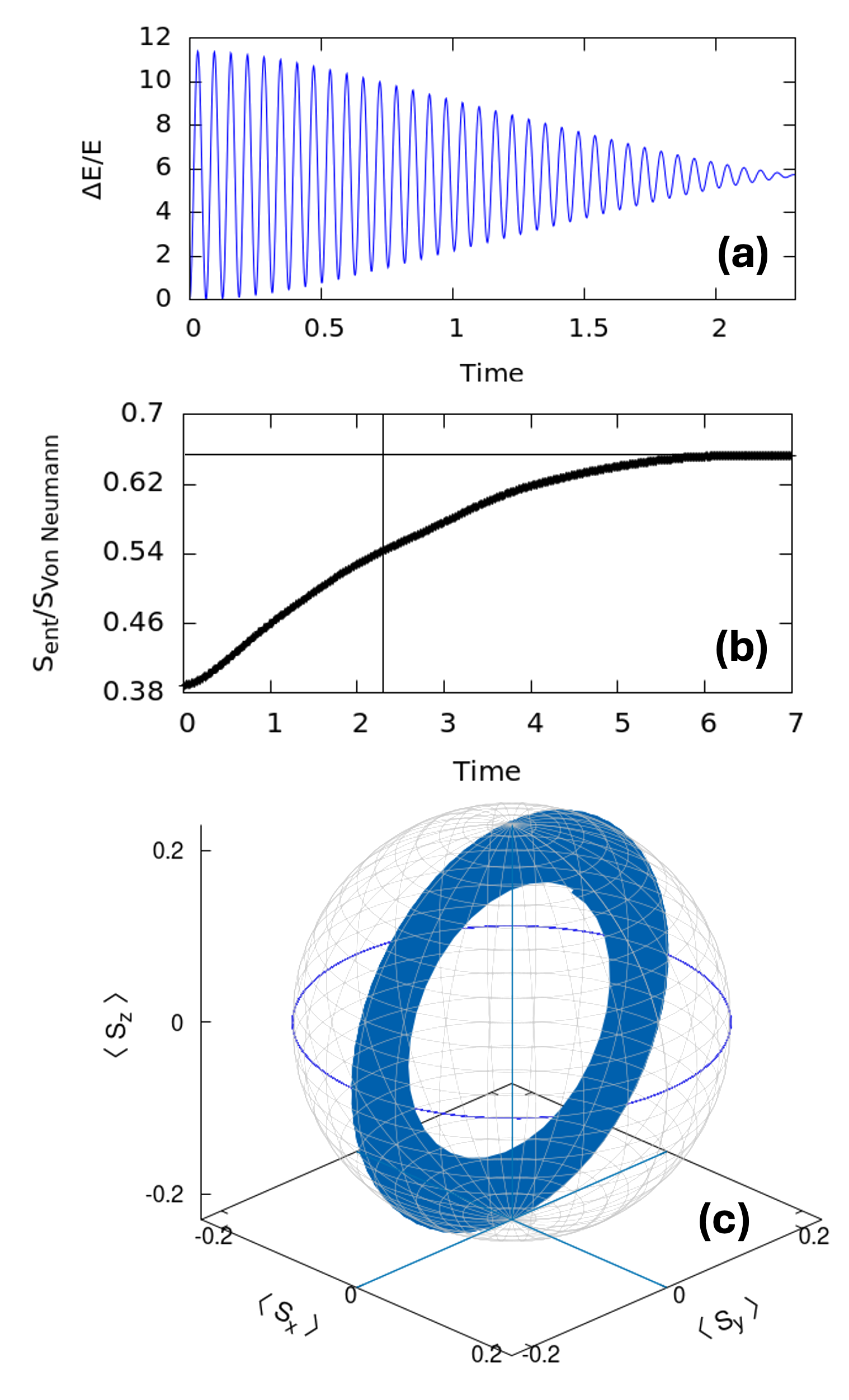}
  \caption{ {\bf (a)} Energy relaxation and {\bf (b)} entanglement entropy density variation during the precessional Rabi oscillations between the $\ket{0}$ and $\ket{1}$ states corresponding to a Pauli-$X$ gate manipulation of a quantum SK with PBC. {\bf (c)} Trajectory on the Bloch sphere and $\left< S_z \right>$ decay during the gate manipulation within the timescale of the energy relaxation depicted in (a).}
  \label{fig:9}
\end{figure}

The importance of the decoherence increase during the gate manipulation is better emphasized when extending the calculations to a larger time window, as illustrated in Figure~\ref{fig:9}. 
In this figure,  one can clearly distinguish the energy relaxation mechanism [Figure~\ref{fig:9}(a)] and the increase of the entanglement entropy density, as signature of the decoherence arrow of time, $S_{\rm ent}/S_{\mathrm{{von Neumann}}}$ being calculated at even larger timescale [Figure~\ref{fig:9}(b)] to observe its asymptotic behavior.  
In Figure~\ref{fig:9}(c), within the same timescale as the energy relaxation depicted in Figure~\ref{fig:9}(a), the spin dynamics of the quantum skyrmionic qubit shows that the average value of $\left< S_z \right>$ decays by about 30\% during the gate manipulation. 
This directly affects the gate fidelity. 
The origin of the energy and $\left< S_z \right>$ decay is directly related to the decoherence mechanisms induced by the DMI. 
During the precessional manipulation, the DMI interactions would prefer to induce canted structures of adjacent spins, which is not allowed by the PBC. 
The direct effect will be the enhancement of quantum fluctuations with direct effect on spin polarization and population of $\ket{0}$ and $\ket{1}$ states, as illustrated by the Figure~\ref{fig:9}(a). 
This is similar to a longitudinal energy relaxation mechanism described by a $T_1$ relaxation time. 
Note that absolutely similar analysis leads to exactly similar behavior for all the Pauli $X$, $Y$, $Z$, and Hadamard gates.
\begin{figure*}[htbp]
  \centering
  \includegraphics[width=0.8\textwidth]{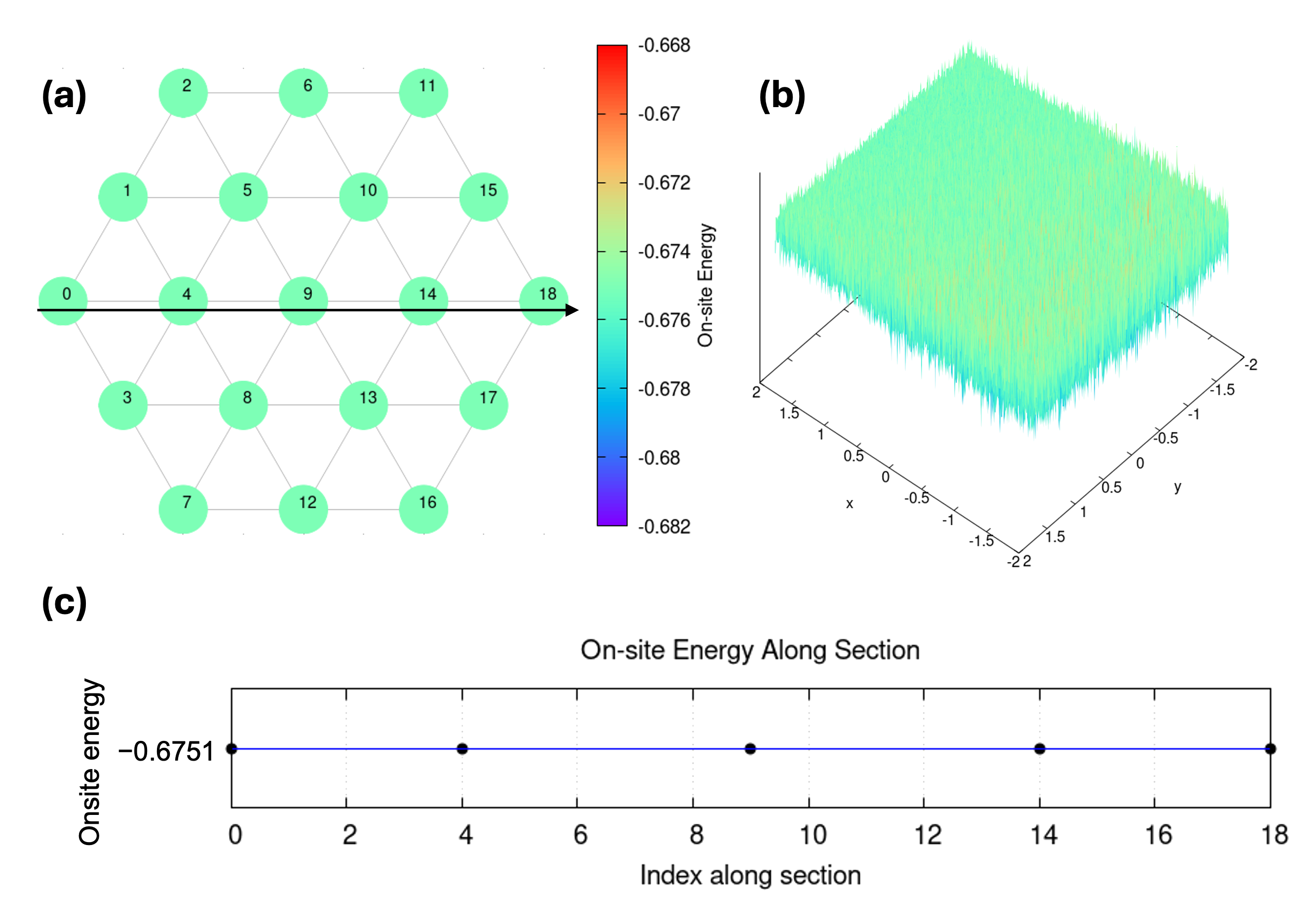}
  \caption{ {\bf (a)}  On-site energy density in DMI units (projected ground state eigenvalue) on each site of the 2D spin lattice.  {\bf (b)} Extrapolated energy density surface. {\bf (c)} On-site energy along selected spin path: $1 \leftrightarrow 4  \leftrightarrow 9  \leftrightarrow 14  \leftrightarrow 18$.}
  \label{fig:10}
\end{figure*}

One of the last questions addressed with respect to the quantum skyrmions stabilized by DMI in triangular lattices with PBC is their topological protection. 
This topological protection would promote higher coherence and stability during the qubit manipulation.  
To answer to this question, we calculated the on-site energy density by quantum-mechanically measuring the individual spin ground state eigenvalue. 
The result, shown in Figure~\ref{fig:10}, demonstrates that {\em quantum skyrmions lack topological protection}, a fact that is, to some extent, intuitively expected. 
Indeed, due to the translational invariance, all spins are equivalent, the energy/site will be equal, there will be no energy quantum well for spins, and, therefore, there is no protection. 
This issue is completely different for the classical-like skyrmions, as we will see in the next subsections. 
Thus, the PBC quantum skyrmion is not topologically protected but represents merely what is commonly referred to as a ``quantum analog'' of the classical skyrmion~\cite{Sotnikov2021}.
Nevertheless, the PBC quantum skyrmion exhibits a well-defined scalar spin chirality within a specific parameter range and can be identified in the spin-spin correlation functions, observable in neutron scattering experiments.
The emergence of such correlations is an indicator that a genuinely topologically protected skyrmion state could arise under open boundary conditions. 
 
\subsection{Static quantum analysis of interacting 2D spin lattice with OBC}
\begin{figure*}[htbp]
  \centering
  \includegraphics[width=\textwidth]{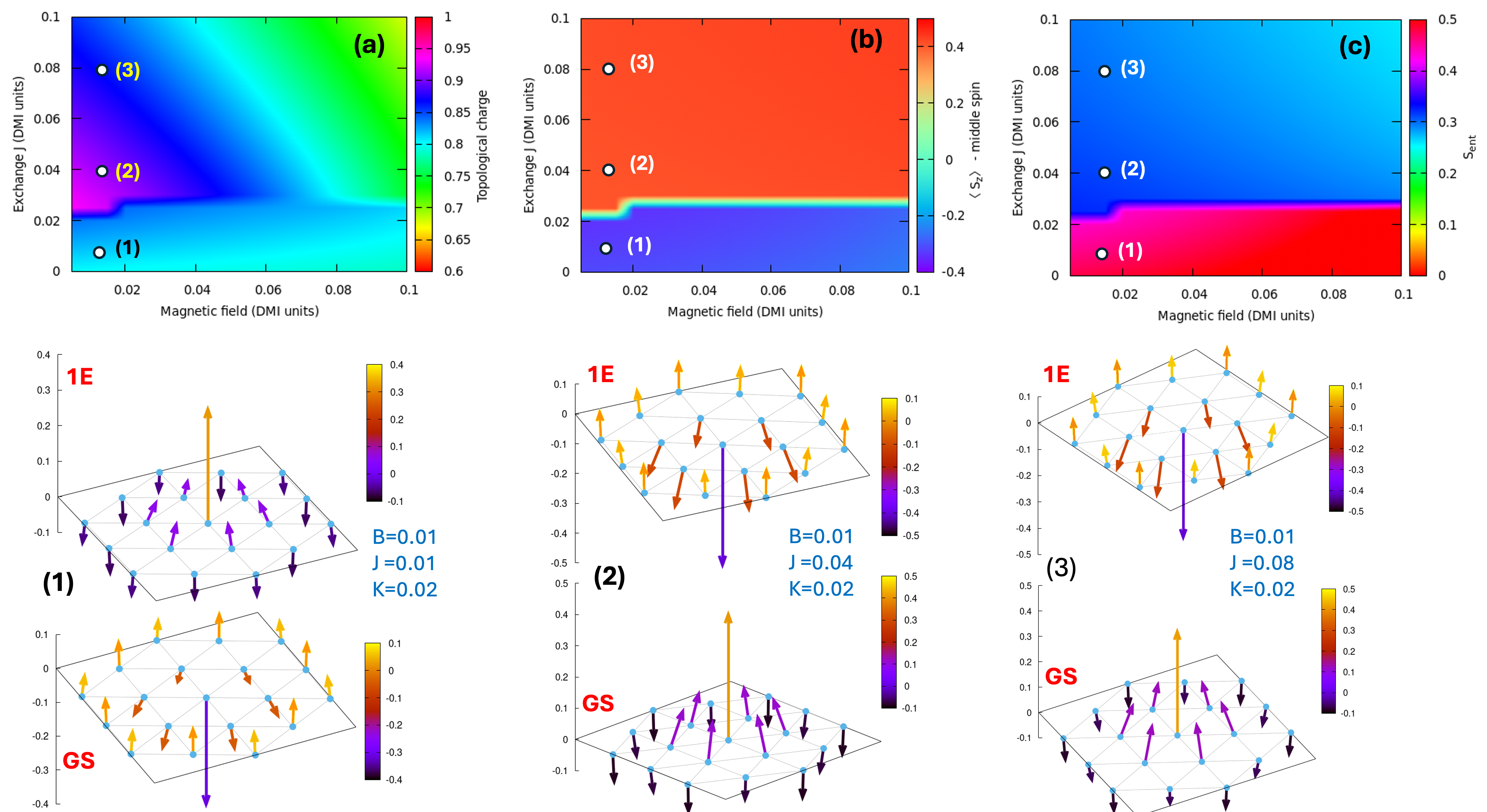}
  \caption{Quantum phase diagrams as a function of magnetic field and exchange interactions. {\bf (a)} Topological charge,   {\bf (b)} average $\langle S_z \rangle$ for the central ($9^{\rm th}$) spin in the 2D lattice, and {\bf (c)} entanglement entropy density. 
  Bottom panels illustrate the ground state {\bf GS} and first excited state {\bf 1E} of classical skyrmion state corresponding to the points (1), (2) and (3) indicated on the phase diagrams for which $B$ and $K$ are fixed and $J$ is varied.}
  \label{fig:11}
\end{figure*}

When solving the Schr\"odinger equation for the 2D interacting spin lattice with DMI, considering open boundary conditions, in a well-defined window in the $J$-$D$-$B$-$K$ phase diagram,  ``classical-like skyrmions'' with topological charge or winding number $\approx 1$ can be stabilized, as illustrated by the Figure~\ref{fig:11}.

Before proceeding with our analysis, let us clarify the resemblance between the “classical-like” skyrmion obtained from quantum calculations with open boundary conditions (OBC) and a classical skyrmion found in micromagnetic simulations. First, the two approaches operate on vastly different length scales: in exact-diagonalization (ED) quantum simulations, the spin system is extremely small (19 spins in our case), whereas micromagnetic simulations typically involve much larger systems, from several nanometers up to micrometers. Moreover, the nature of the spin variables is fundamentally different. Quantum calculations treat individual quantum spins—local quantum observables that may arise from superposed states, can take values smaller than 0.5 (in units of $\hbar$), and may vary from site to site under OBC. By contrast, micromagnetic simulations use classical macrospins, each representing a magnetic volume of the order of the exchange length, modeled as unit-norm vectors identical in magnitude across the simulation grid.
Despite these conceptual and scale differences, the spin textures of the “classical-like” skyrmions obtained in ED closely resemble those of purely classical skyrmions in micromagnetics. This correspondence is further confirmed by their topological charge: in both quantum OBC systems and classical (micromagnetic or atomistic) simulations, the skyrmion number $Q$ takes comparable values, typically close to $+1$ for a classical-like skyrmion. We refer to the skyrmions obtained in quantum ED with OBC as “classical-like” skyrmions to distinguish them from “quantum skyrmions” stabilized in systems with periodic boundary conditions (PBC), which exhibit a fundamentally different spin structure characterized by uniform on-site $\langle S_z \rangle$ polarization due to translational invariance.

In the absence of PBC constraints, the presence of the Dzyaloshinskii-Moriya interaction (DMI) allows individual spins to cant and form chiral skyrmionic structures in order to minimize the total interaction energy. 
The sign of the DMI parameter $D$ determines the chirality or helicity of these structures. 
This contrasts with systems lacking DMI, where skyrmions are stabilized by competing exchange interactions. 
In such cases, helicity becomes a quantum property, with the two helicity states forming a quantum superposition~\cite{Psaroudaki2021,Xia2023}.

As in the case with periodic boundary conditions, we again observe that introducing and increasing $K$ leads to a downward shift of all features in the phase diagram with respect to $J$. 
The behavior is consistent with the form of the Hamiltonian in Eq.~\eqref{eq:Hamiltonian}, where the anisotropy term—representing anisotropy perpendicular to the lattice plane appears as a correction to  $JS^z_iS^z_j$.

\begin{figure*}[htbp]
  \centering
  \includegraphics[width=\textwidth]{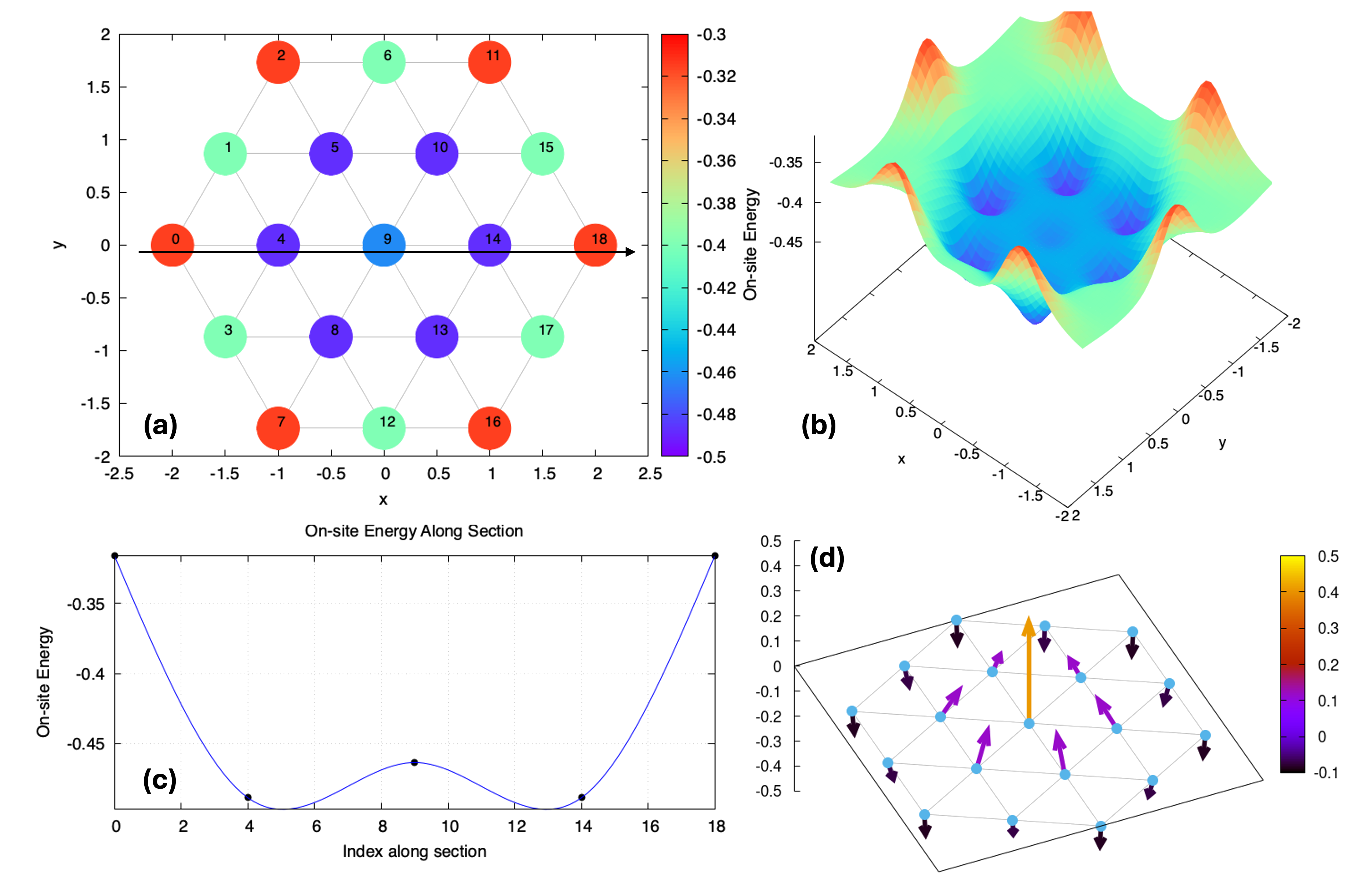}
  \caption{{ \bf (a)}  On-site energy density (projected ground state eigenvalue) on each site of the 2D spin lattice.  {\bf (b)} Extrapolated energy density surface {\bf (c)} On-site energy along selected spin path: $1 \leftrightarrow 4  \leftrightarrow 9  \leftrightarrow 14  \leftrightarrow 18$.  {\bf (d)} Real space representation of the classic-like skyrmion for which the on-site energy has been calculated. }
  \label{fig:12}
\end{figure*}

\begin{figure*}[htbp]
  \centering
  \includegraphics[width=0.8\textwidth]{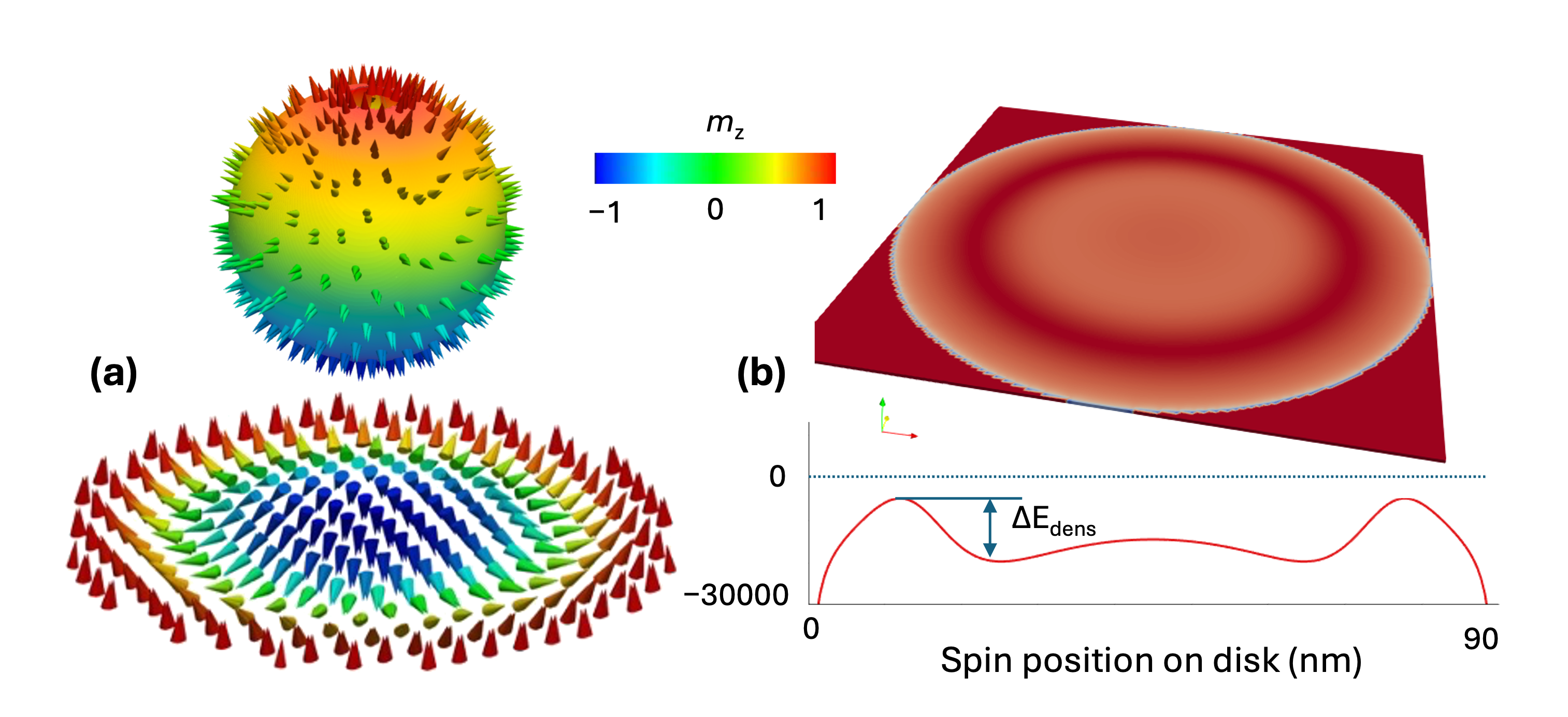}
  \caption{{\bf (a)} Classic N\'eel skyrmion stabilized in a nanodisk calculated micromagnetically.  
  {\bf (b)}  Corresponding energy density surface and energy density line profile illustrating the topological protection barrier $\Delta E_{\rm dens}$ for the classic micromagnetic skyrmion.}
  \label{fig:13}
\end{figure*}

However, contrary to the quantum skyrmions in similar lattice with PBC, the ``classic-like'' skyrmions in lattices with OBC are topologically protected against the reversal, as demonstrated by the on-site energy analysis depicted in Figure~\ref{fig:12}. 
We see that the core spins pointing upwards are located in a quantum well that protects them against reversal, similarly to the classical skyrmions studied in micromagnetism. 
Figure~\ref{fig:13} depicts for comparison a classic skyrmion in a nanodisk calculated with micromagnetic tools~\cite{Nanomaterials2022}.

\begin{figure*}[htbp]
  \centering
  \includegraphics[width=0.8\textwidth]{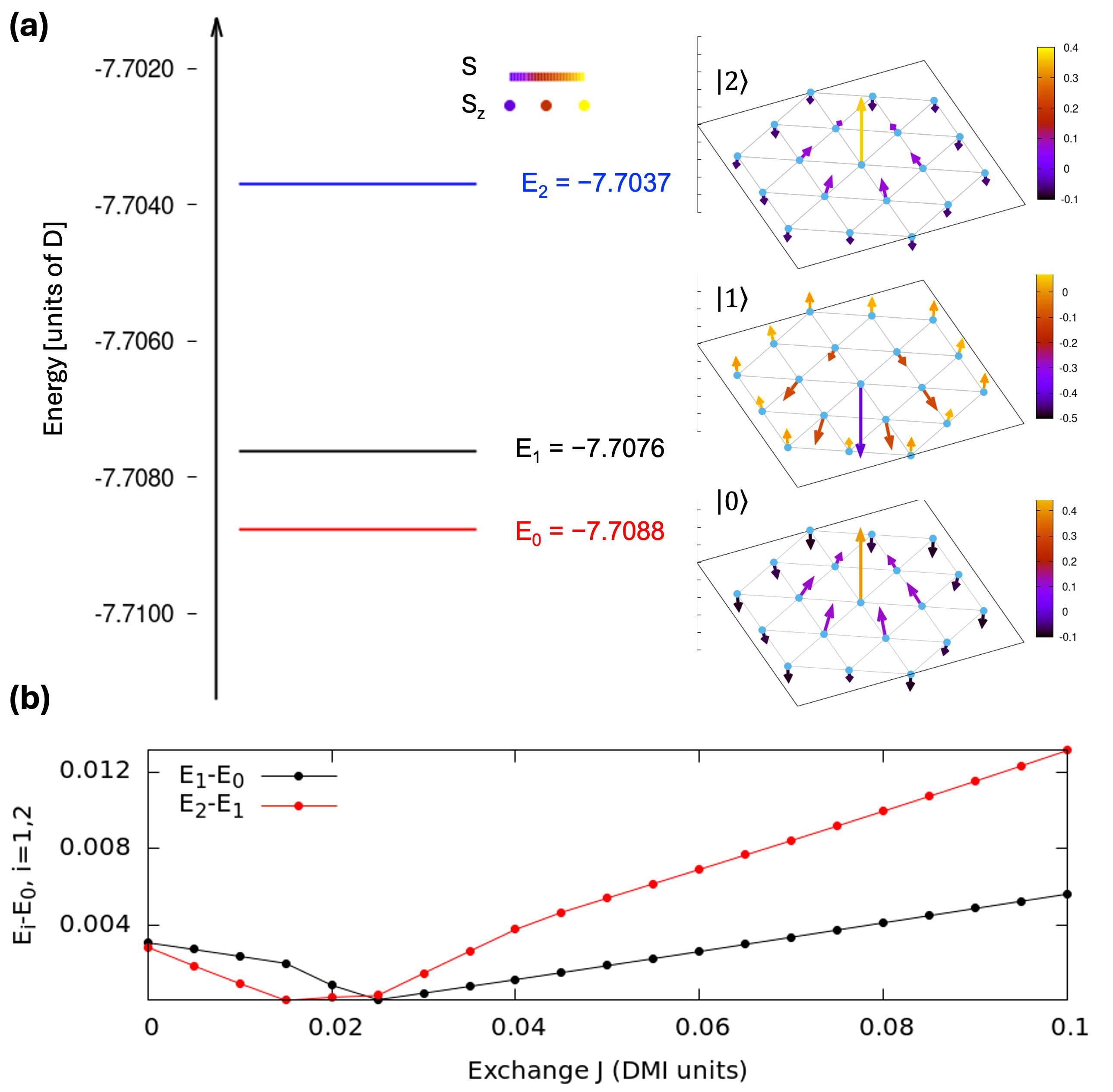}
  \caption{{\bf(a)}  Energy levels and skyrmion structures corresponding to the eigenstates $\ket{0}$-ground state,  $\ket{1}$-first excited state,  $\ket{2}$-second excited state corresponding to a point in the phase diagram [Fig.~\ref{fig:11}(a)] defined by $J/D = 0.04$, $K/D = 0.02$, and $B/D = 0.01$. 
  {\bf (b)} Variation of the spacing between the ground state $E_0$ and first excited state $E_1$, and between first $E_1$ and second excited state $E_2$, tuned at fixed magnetic field $B/D=0.01$ and varying the exchange $J/D$.}
  \label{fig:14}
\end{figure*}

The quantum phase diagram shown in Figure~\ref{fig:11}, corresponds to the ground state of a two-dimensional spin lattice with open boundary conditions. 
Focusing on the skyrmionic region characterized by a topological charge  $Q \sim 1$, we extend our analysis beyond the ground state to include the excited energy levels. 
Within this regime, each eigenstate corresponds to a skyrmion configuration, distinguished primarily by the orientation of the central spin. 
Notably, the energy spectrum is anharmonic, with level spacing that can be tuned by adjusting the parameters $J$ and $B$ within the skyrmionic window of the phase diagram. 
These features enable the construction of a two-level qubit using the ground state and the first excited state, where quantum information is encoded in the orientation of the central spin of the 2D skyrmion.

In Figure~\ref{fig:14}, we illustrate the energy levels and the corresponding eigenstates for a point in the quantum phase diagram defined by  $J=0.04$, $K=0.02$, and $B=0.01$, all in units of $D$. 
Using the eigenvalues corresponding to the $\ket{0}$ $-$ ground state and  $\ket{1}$ $-$ first excited state, we build a qubit that will be dynamically manipulated as Pauli $X$, $Y$, $Z$ and Hadamard gates, as described in the next subsections.

The manipulation of a quantum skyrmion-based qubit through a precessional drive from an external magnetic field, conceptually analogous to standard Nuclear Magnetic Resonance (NMR) or Electron Spin Resonance (ESR) control techniques, does not rely on anharmonicity, unlike qubits controlled via photonic or microwave-cavity driving. In this scheme, control is achieved through direct geometric rotation of the collective spin vector within a genuine two-level Hilbert space, rather than by inducing population transfer across multiple anharmonic oscillator levels. The qubit basis states are simply the Zeeman eigenstates$\ket{0} \equiv \ket{\uparrow \uparrow \cdots \uparrow}$ and $\ket{1} \equiv \ket{\downarrow \downarrow \cdots \downarrow}$, with associated energies $E_{\uparrow} = -\frac{1}{2}\gamma \hbar B_0$ and $E_{\downarrow} = +\frac{1}{2}\gamma \hbar B_0$ , where $\gamma$ is the gyromagnetic ratio and $B_0$ is the applied static magnetic field. The relevant energy scale is therefore the Zeeman splitting, as opposed to an engineered anharmonic level structure found in superconducting circuits or skyrmionic qubits driven optically as we will later illustrate and discuss in the section D.2.  

\subsection{Dynamic quantum analysis of interacting 2D spin lattice with OBC}

\subsubsection{Manipulation by magnetic field drive}
\begin{figure*}[htbp]
\centering
\includegraphics[width=\textwidth]{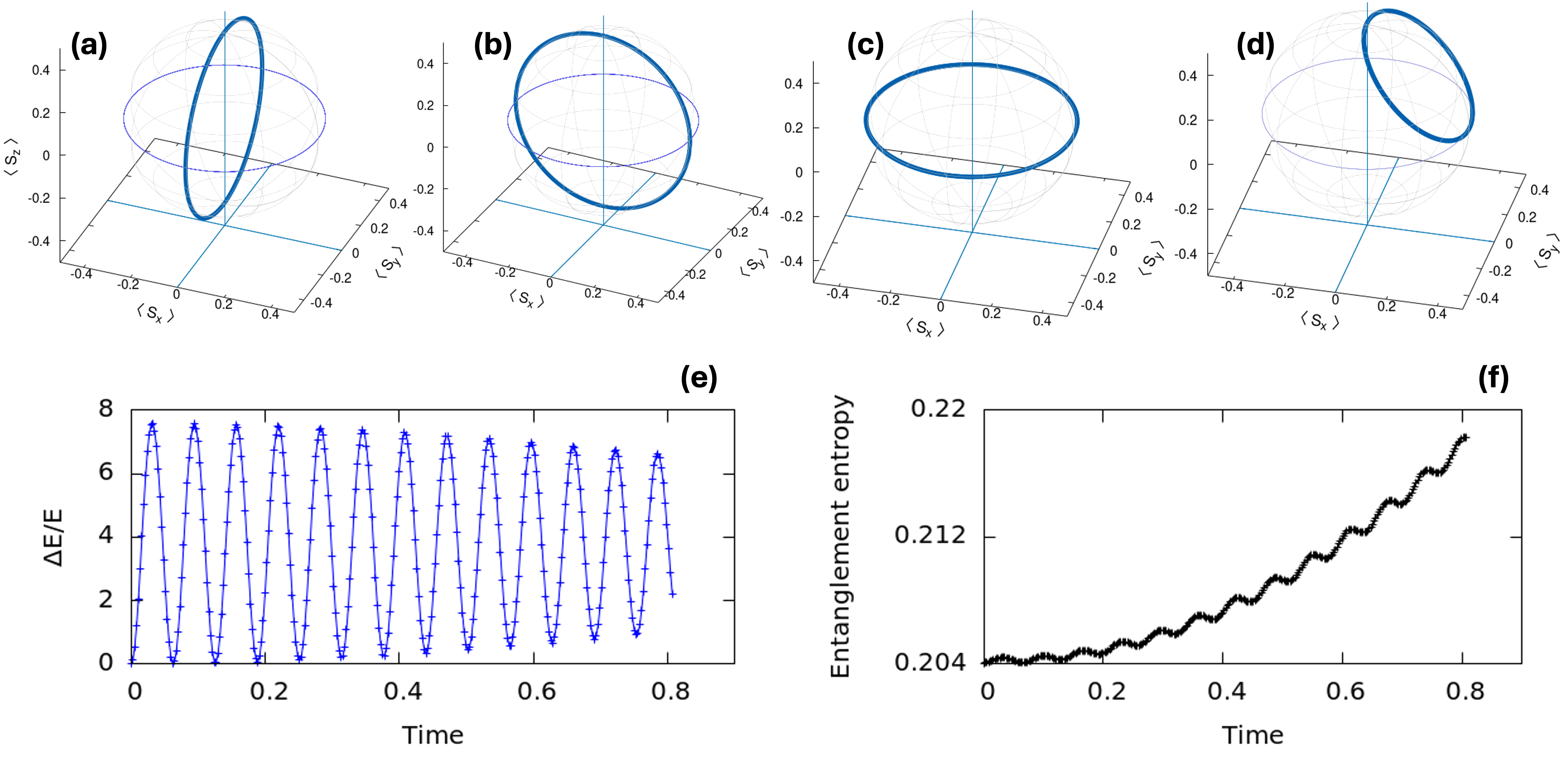}
\caption{Quantum simulation for the Pauli-$X$ {\bf (a)}, $Y$ {\bf (b)}, $Z$ {\bf (c)}, and Hadamard{\bf (d)} gate operations. 
The qubit is build using classic skyrmion, promoted in the 2D spin lattice with OBC, with the energy spectrum and configurations described in Fig.~\ref{fig:14}. 
Rabi oscillations {\bf (e)} and entanglement entropy density variation corresponding to the Pauli-$X$ gate manipulation during about 13 periods.  
Note that exactly similarly behavior has been calculated for all the other type of gates.}
\label{fig:15}
\end{figure*}

Similarly to the quantum skyrmionic qubit, the classic skyrmions promoted in 2D spin lattices with OBC can be manipulated using external applied magnetic-precessional field drive. 
Aligning this external field along the directions $X$, $Y$, $Z$, or simultaneously $X$ and $Z$, one can build, as illustrated for the quantum SK, the corresponding Pauli $X$, $Y$, $Z$ and Hadamard gates, the quantum information being in this case encoded in the orientation of the skyrmion's central spin. 
Like for the quantum SK gates, the applied driving field is chosen to  dominate the DMI interactions, i.e., $B_\alpha/D =100$, $\alpha =x,y,z$.  
As illustrated in Figure~\ref{fig:15}, for a Pauli-$X$ gate corresponding to a drive field  $\bm{B}_{\rm drive}=(B_x,0,0)$, the Rabi oscillations during the gate manipulation take place between the ground state $\ket{0}$ and first excited skyrmionic state $\ket{1}$.
Our calculations show that, during the precessional manipulation of the classical-like skyrmion, similar energy relaxation and reduction in the average value of $\left< S_z \right>$ occur, as observed in the case of the quantum skyrmion. 
Once again, the origin of the decay in both energy and $\left< S_z \right>$, affecting the gate fidelity, can be directly attributed to decoherence mechanisms induced by the DMI. 
This leads to an enhancement of quantum fluctuations, which in turn directly affects spin polarization and the population of the $\ket{0}$ and $\ket{1}$ states. 
This behavior resembles a longitudinal energy relaxation mechanism, typically characterized by a $T_1$ relaxation time.

\begin{figure*}[htbp]
  \centering
  \includegraphics[width=\textwidth]{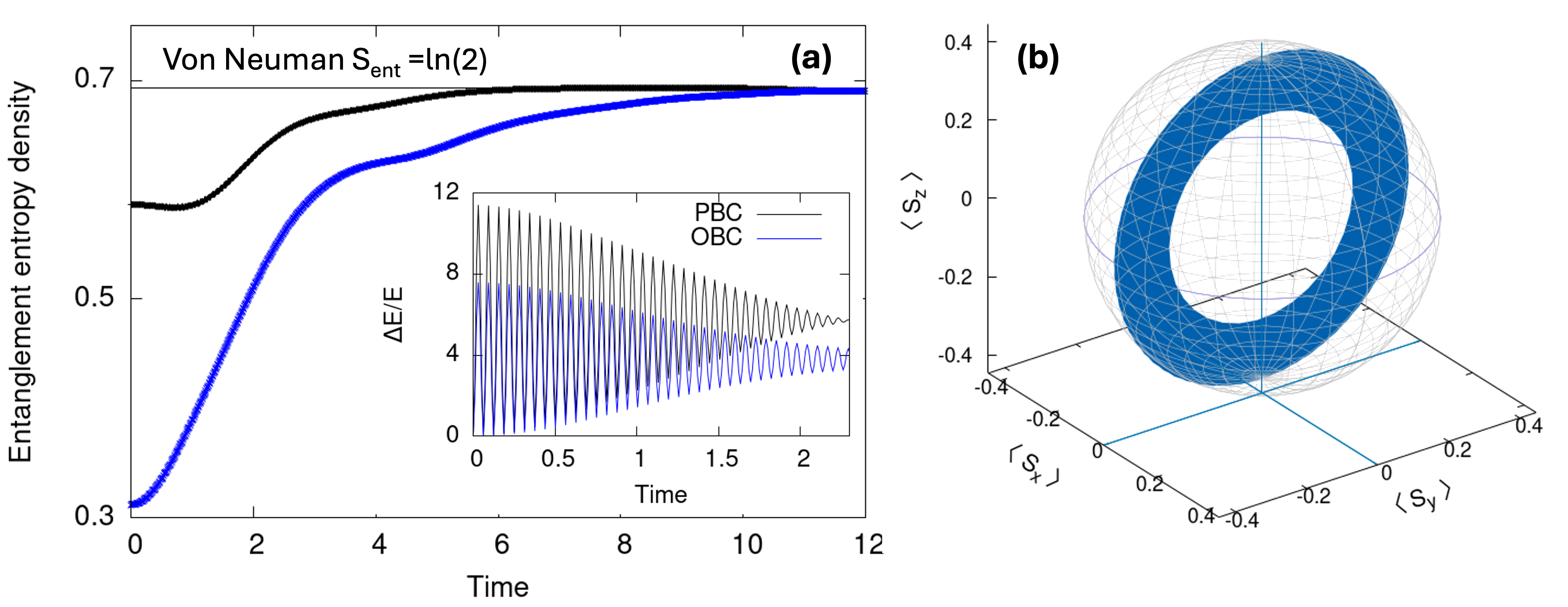}
  \caption{{\bf (a)} Comparison of the entanglement entropy density evolution of the central spin during Pauli-$X$ gate manipulation for a quantum skyrmion in a 2D spin lattice with periodic boundary conditions (PBC) versus a classical skyrmion in a 2D spin lattice with open boundary conditions (OBC). {\bf (b)} Trajectory on the Bloch sphere and $\left< S_z \right>$ decay during the gate manipulation within the timescale of the energy relaxation depicted in the insert of (a)).}
  \label{fig:16}
\end{figure*}

It is important to emphasize that comparing the time evolution of the entanglement entropy density (i.e., the decoherence arrow of time) between the quantum skyrmion qubit without topological protection and the classical-like skyrmion qubit with topological protection reveals the clear benefits of topological protection for qubit coherence, temporal stability, and gate fidelity. 
As shown in Figure~\ref{fig:16}, the topologically protected qubit exhibits a slower growth of entanglement entropy, a smaller initial entropy, and a more gradual energy decay over time. This behavior is consistent with the enhanced quantum fluctuations induced by PBC in the spin lattice, which suppress spin canting. In contrast, spin canting is allowed in classical skyrmionic configurations under OBC. 
For a meaningful comparison, the entanglement entropy density was computed for the central spin in both cases.  
With and without PBC, the entanglement entropy increases asymptotically, approaching the von Neumann entropy density  $S_{\rm ent}=\ln(2)$. 
However, regarding the decay of $\left< S_z \right>$ during the gate manipulation, as shown in Figure~\ref{fig:16}(b), our calculations show a similar decay rate of approximately 1\% per precessional period, for the classical skyrmion with OBC, which is comparable to the quantum analog with PBC (see Figure~\ref{fig:9}(c)). 

Moreover, it is worth noting that during qubit manipulation, the spin energy landscape responsible for topological protection, illustrated in Figure~\ref{fig:12}(b), remains robust, exhibiting only minor fluctuations in the amplitude of the quantum well experienced by the central spins, as compared to the static case at $t=0$. 

\subsubsection{Qubit manipulation by photonic periodic drive}

To advance toward practical experiments, it is essential to explore qubit coupling and its control using either microwave or optical drives. 
We consider again the same two-level qubit, constructed from the lowest two energy levels and their corresponding eigenstates, as illustrated in Figure~\ref{fig:14}. 
The eigenstates shown in Figure~\ref{fig:14}(a) correspond to a specific point in the quantum phase diagram defined by the parameters: $J/D = 0.04$, $K/D = 0.02$, and $B/D = 0.01$. As previously noted, varying the values of $J$, $B$, and $K$ within the skyrmionic region of the quantum phase diagram  only alters the spacing between energy levels, while leaving the qualitative aspects of our analysis unchanged. 
In Figure~\ref{fig:14}(b), we illustrate how the spacing between the ground state $E_0$ and first excited state $E_1$, and the spacing between first $E_1$ and second excited state $E_2$ can be tuned at fixed magnetic field $B/D=0.01$ by varying the exchange $J/D$.

In the next paragraphs, as mentioned in the methodology section~\ref{sec:methods}, we will perform our analysis in several steps and approach complexity.\\

\paragraph{Mapping on the 2-level logical qubit subspace\\ \\}

\begin{figure*}[htbp]
  \centering
  \includegraphics[width=\textwidth]{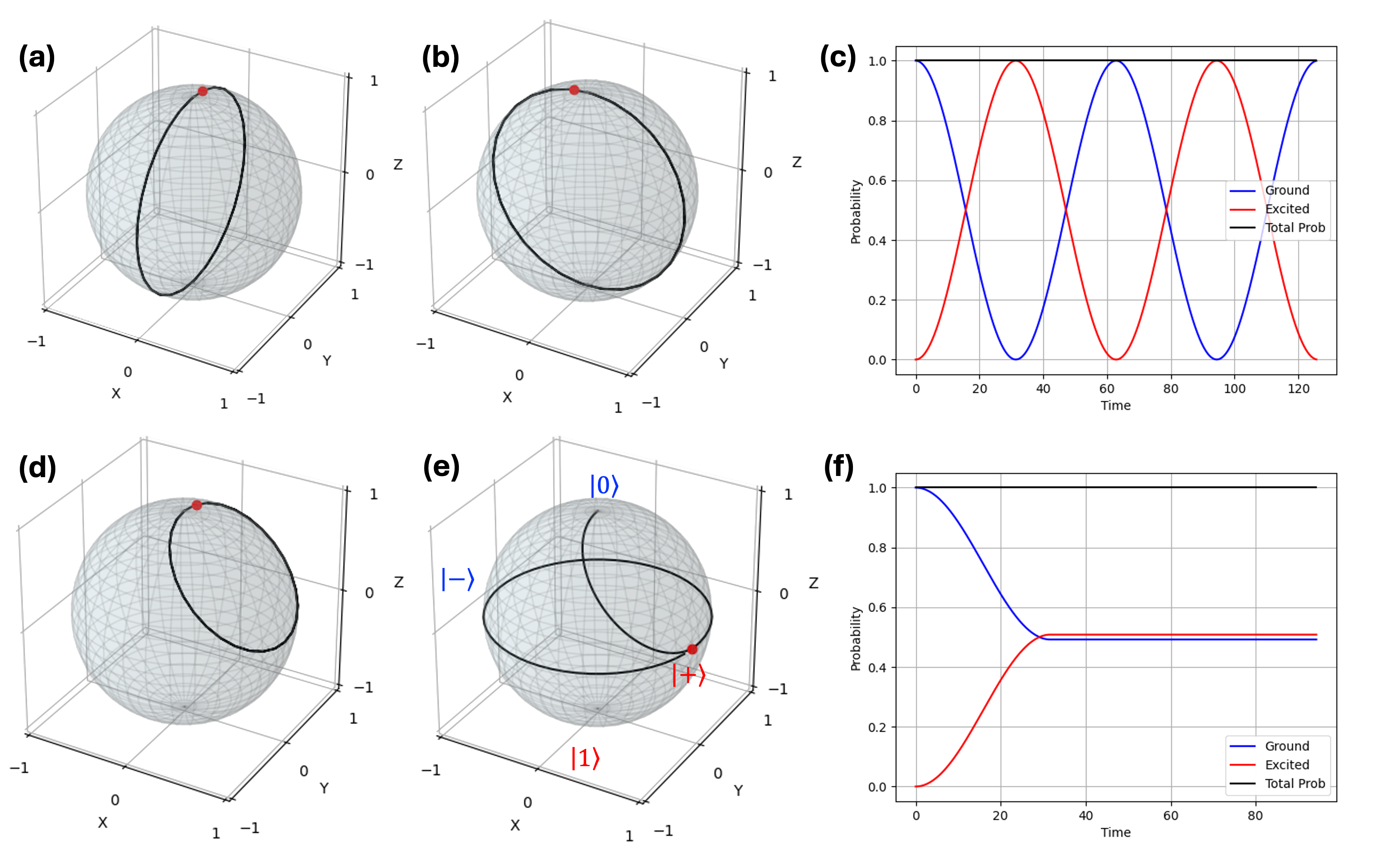}
  \caption{{\bf (a)} Pauli-$X$ and {\bf (b)} $Y$ gates built using a classic-like skyrmion qubit 2-level system coupled to a photonic drive.
  {\bf (c)} Probability of transitions between the ground state $\ket{0}$ and excited state $\ket{1}$ during the Pauli-$X$ and $Y$ gate operation over 2 Rabi oscillation periods. 
  {\bf (d)} Hadamard gate and {\bf (e)} combined Hadamard ($T_{\rm Rabi}/2$)+ Pauli-$Z$ phase gate ($T_{\rm Rabi})$.   
  {\bf (f)} Probability of transitions between the ground state $\ket{0}$ and excited state $\ket{1}$ during the combined Hadamard ($T_{\rm Rabi}/2$) $+$ Pauli-$Z$ phase gate ($T_{\rm Rabi}$).}
  \label{fig:17}
\end{figure*}

If the Hilbert space is large, in our case for 19 spins the dimension being $2^{19}$, and the dynamics is mainly in the $\{ \ket{\Psi_1}$, $\ket{\Psi_2}\}$ subspace, we can project the full Hamiltonian into a $2\times2$ matrix, apply a drive, and simulate the clean Rabi oscillations in the subspace $H_{\rm eff}=\left\{\ket{\Psi_1}, \ket{\Psi_2}\right\}$, as explained in Sec.~\ref{sec:methods}. 
This strategy exploits effective two-level physics in a large Hilbert space, assuming confinement within the subspace, resonant driving with gap $\Delta E$, and well-separated higher energy states. 
The exact diagonalization is first performed in the full Hilbert space to get the two energy levels $E_0$ and $E_1$ defining the qubit. 
Then, based on Eq.~\eqref{eq:drive} and the procedure described in Sec.~\ref{sec:methods}, one can simulate the qubit dynamics for the Pauli $X$, $Y$, $Z$, and Hadamard gates. 
The result is presented in Figure~\ref{fig:17}, considering a drive amplitude $A=0.1$ and $\omega=E_1-E_0$. 

  \begin{figure*}[htbp]
  \centering
  \includegraphics[width=0.8\textwidth]{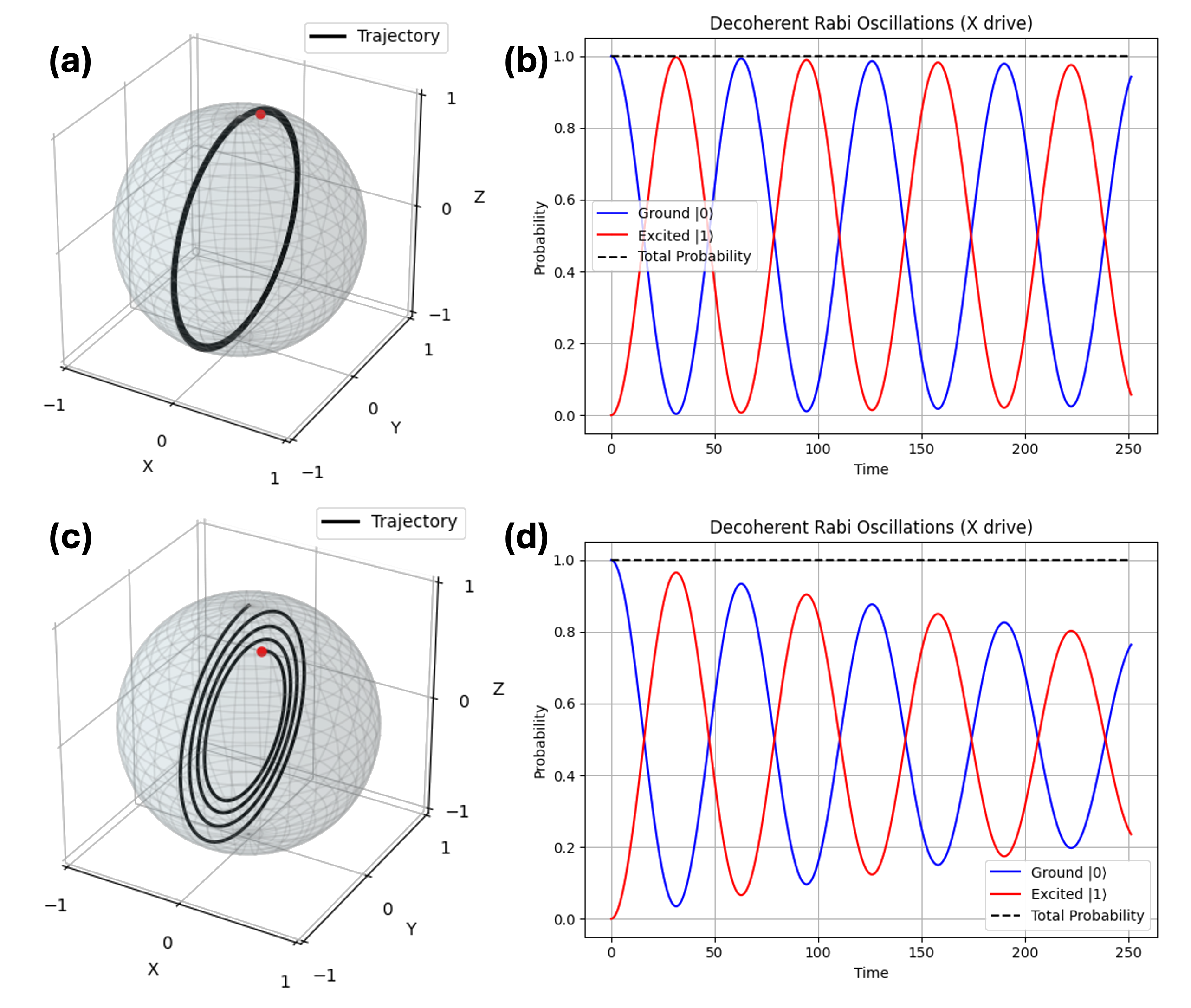}
  \caption{{\bf (a)} Pauli-$X$ gate and {\bf (b)} decoherent Rabi oscillations corresponding to a long longitudinal relaxation time $T_1=10^4$, $T_2=T_1/2$. 
  {\bf (c)} Pauli-$X$ gate and {\bf (d)} decoherent Rabi oscillations corresponding to a 10 times shorter longitudinal relaxation time $T_1=1000$, $T_2=T_1/2$.}
  \label{fig:18}
\end{figure*}

Note that mapping the complex quantum problem 2-level logical qubit subspace is computationally more efficient and gives clean Rabi oscillations, as illustrated in Figure~\ref{fig:17} for simple gates (a), (b), (d) or multiple gate operations (e). 
The transition probabilities $P_{\rm ground} =|C_1|^2$,  $P_{\rm excited} =|C_2|^2$ correspond to the wavefunction at the time $t$ for a 2-level system in a linear combination, $\ket{\Psi (t)}=C_1(t) \ket{\Psi_1}+C_2(t) \ket{\Psi_2}$. 
The trajectories on the Bloch sphere correspond to the spin polarization observables, $\left< S_\alpha(t)\right>=\frac12\bra{\Psi(t)}\sigma_\alpha \ket{\Psi(t)}$, $\alpha=x,y,z$, with $\sigma_\alpha$ being the Pauli matrices.

Note that the drive Hamiltonian $H(t)=A \cos(\omega t) D$ mimics the interaction of the system with a classical (coherent) photon field of frequency $\omega$. 
The dynamic term $A \cos(\omega t) D$ can be a semiclassical electromagnetic drive, i.e., the effect of a coherent photon (like in a microwave cavity) on the system. 
The operator $D$ is often chosen to match the dipole operator (e.g., $\sigma_x$, $\sigma_y$,  etc.) to represent allowed transitions. 
The frequency $\omega$ corresponds to the energy gap that we want to drive, usually $\omega = E_1-E_0$ for resonant transitions. 
This form for the driving Hamiltonian arises from applying a dipole approximation to the interaction Hamiltonian,  $H_{\rm int}=- \bm{d}\cdot \bm{E}(t)$, where $\bm{d}$ is the dipole operator and $\bm{E}(t)$ is an oscillating electric field $\sim \cos(\omega t)$.

Physically, a $\sigma_{x,y,z}$ drive can be implemented in real experiments by shining a microwave (or laser) field onto the qubit. 
The polarization of the field determines which operator is being driven, e.g., linear polarization: $\sigma_x, \sigma_y, \sigma_z$, circular polarization: $\sigma_x \pm \sigma_y$ corresponding to $\sigma^+$ and $\sigma^-$, the frequency $\omega$ being tuned close to the energy gap between the ground state $\ket{0}$ and the excited state $\ket{1}$. \\

\paragraph{2-level logical qubit subspace with decoherence} \leavevmode \\

To simulate a more realistic situation in which the two-level qubit system is not ideally isolated from to the other states of the full Hilbert space or with respect to the outer environment, as explained in the methodology section~\ref{sec:methods}, we added decoherence including longitudinal relaxation, $T_1$ decay, describing the decay from the excited state $\ket{1}$ to the ground state $\ket{0}$, and transverse dephasing, $T_2$ decay,  describing the loss of phase coherence between $\ket{0}$ and $\ket{1}$, without population transfer. 
Instead of solving the Schrödinger equation for a pure state $\ket{\Psi(t)}$, we solved the master Lindblad equation~\eqref{eq:lindblad} for the system density matrix.

Then, based on the eigenvalues, we calculated the spin polarization observables and projected the results on the logical qubit  Bloch sphere, as illustrated in Figure~\ref{fig:18}. 
The calculation illustrates the effect of the $T_1$ and $T_2$ (chosen to be in these simulations $T_1/2$) on the qubit stability, i.e., the evolution in time of the $\ket{0}$ and  $\ket{1}$ populations,  and gate fidelity  illustrated by the decay in time of $\left< S_z \right>$.\\

\paragraph{Full Hamiltonian photon drive} \leavevmode \\

An advanced approach to modeling the classic skyrmion qubit in two-dimensional interacting spin lattices involves going beyond the standard two-level approximation. 
Instead of projecting onto a limited $2 \times 2$ Hilbert subspace, we consider the full Hamiltonian, including a photon drive [as described by Eq.~\eqref{eq:drive}], and solve the corresponding time-dependent Schrödinger equation~\eqref{eq:td_schr} using exact diagonalization within the complete Hilbert space of dimension $2^n$. 
This framework enables the calculation of physical observables, and makes a clear distinction between the dynamics in the full physical spin system and the operations performed in the logical qubit subspace. 

As defined in Eq.~\eqref{eq:Hamiltonian}, the full many-body Hamiltonian incorporates the following interactions: exchange, Dzyaloshinskii-Moriya interactions, Zeeman fields, and magnetic perpendicular anisotropy. 
Logical quantum operations, such as the Pauli $X$, $Y$, $Z$ gates, and the Hadamard gate, are defined within a logical two-dimensional qubit subspace [see Eqs.~\eqref{eq:logical}]. 
These gates are implemented through time-dependent drives that couple the logical states $\ket{\psi_1} \leftrightarrow \ket{\psi_2}$. 
Despite this logical encoding, the actual time evolution of the system takes place in the full physical Hilbert space of dimension $2^n$, where each spin is treated explicitly. 
The physical quantum state $\ket{\Psi(t)}$ evolves according to the exact solution of the Schr\"odinger equation~\eqref{eq:td_schr}, and represents a state vector in the high-dimensional many-body Hilbert space.  
Although the Hamiltonian and its eigenvectors are defined within the full Hilbert space of the physical system consisting of $n$ spins, the dynamics represented on the Bloch sphere is projected onto the logical qubit subspace spanned by $\{ \ket{\psi_1}, \ket{\psi_2} \}$, with a logical Bloch vector defined in equation~\eqref{eq:bloch_logical}. 
In this approximation, the quantum evolution is simulated under the assumption of perfect isolation of the two-level subspace. 
That is, we still neglect any leakage into higher excited states outside the logical qubit subspace, treating the two-level system as closed and coherent.

\begin{figure*}[htbp] 
  \centering
  \includegraphics[width=\textwidth]{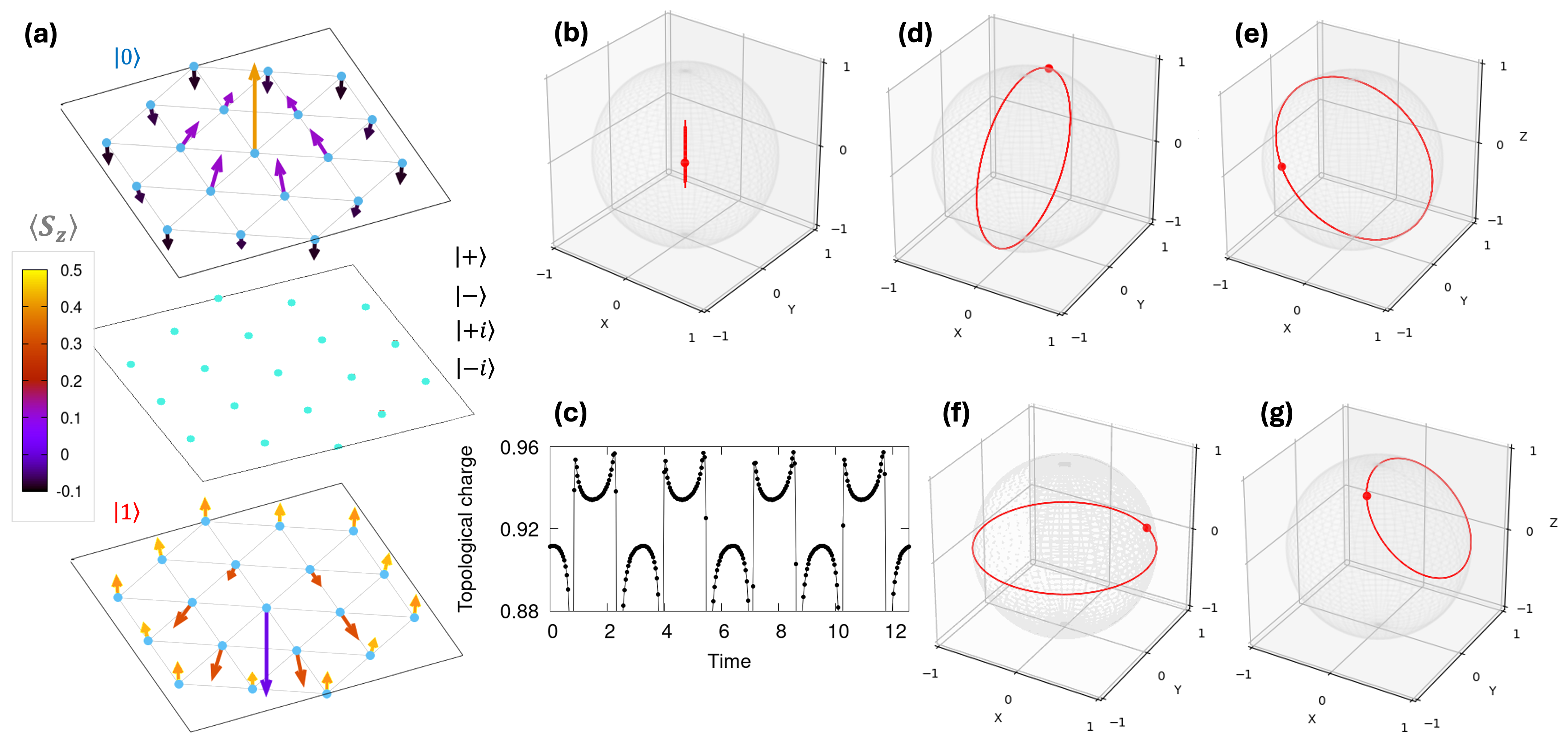}
  \caption{ {\bf (a)} Schematic representation of the on-site spin polarization observables corresponding to following skyrmionic states of the 2-level qubit during the time-dependent dynamics: {\em top} panel, ground state $\ket{0}$, {\em bottom} panel, excited $\ket{1}$ and {\em middle} panel, states in the equatorial plane of the Bloch sphere $\ket{+}$, $\ket{-}$, $\ket{+i}$, $\ket{-i}$. {\bf (b)} Spin trajectory on the Bloch sphere corresponding to the physical spin space. {\bf(c)} Topological charge variation during the gate manipulation. 
  {\bf (d-g)} Central spin trajectories represented in the logical qubit spin subspace. 
  These calculations correspond to a drive amplitude $A=1$.}
  \label{fig:19}
\end{figure*}

\begin{figure}[htbp]
  \centering
  \includegraphics[width=\columnwidth]{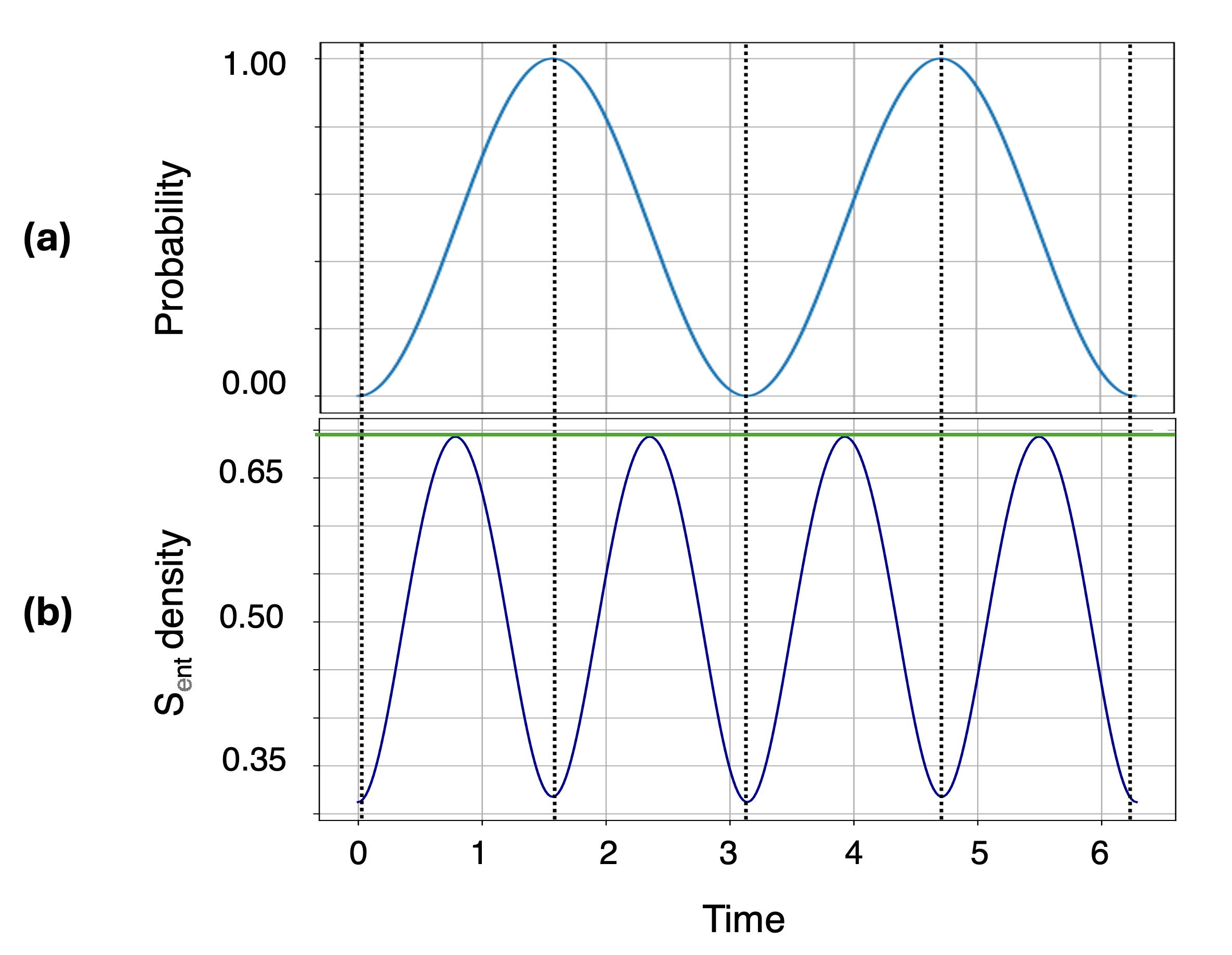}
  \caption{Rabi oscillations: time dependence of the {\bf (a)} transition probability between the 2-level qubit states  $\ket{0}$ and $\ket{1}$ and {\bf (a)}  entanglement entropy, calculated for the correlations between the central spin encoding the quantum information and the other 18 spins of the lattice. 
  The maximum $S_{\rm ent}$ is equal to the von Neumann entropy $\ln(2)$. 
  The dotted vertical  lines  indicate the minimum entanglement entropy values corresponding to the times at which the qubit system is in a pure $\ket{0}$ ot $\ket{1}$ state during the Rabi oscillations, the entropy being maximum in the states belonging to the equatorial plane of the logical Bloch sphere where $\ket{\Psi(t) =1/\sqrt2(\ket{0}+\ket{1})}$.}
  \label{fig:20}
\end{figure}

In Figure~\ref{fig:19}, we illustrate the results for the classical skyrmionic qubit, defined by the same specific point in the quantum phase diagram with parameters: $J/D = 0.04$, $K/D = 0.02$, and $B/D = 0.01$ , as considered previously.  The results have been obtained considering a drive amplitude $A=0.1$ and $\omega=E_1-E_0$. 
Figure~\ref{fig:19}(a) depicts the oscillatory time evolution of the skyrmion between the two states $\{ \ket{\psi_1}, \ket{\psi_2} \}$.
In this panel, the continuous arrows correspond to the ground-state spin configuration, while the dotted arrows indicate the excited state. 
Note that during the time evolution within the full physical spin system (in the laboratory frame) each spin of the 2D lattice composing the skyrmion structure undergoes Rabi oscillations between the upwards and downwards orientations. 
In Figure~\ref{fig:19}(b), we observe the dynamics of the central spin (the spin encoding the quantum information) within the full physical spin system. 
This behavior of the central spin polarization---characterized by an oscillating amplitude that alternates between positive and negative values of $\langle S_z \rangle$, with components solely along the $z$ axis---remains strictly identical when analyzing and representing the Bloch sphere trajectory in the physical spin space for all types of drives implementing the Pauli $X$, $Y$, and $Z$ gates. 
However, if we project the spin trajectory, computed from the spin expectation values, onto the qubit logical subspace, using as definition for the Bloch vectors the expression given by equation~\eqref{eq:bloch_logical}, we recover the well-known gate trajectories (Figures~\ref{fig:19}(d-e)) corresponding to the Pauli $X$, $Y$, $Z$, and Hadamard gates.  
On the other hand, in Figure~\ref{fig:19}(c), we illustrate the variation of the topological charge during the gate manipulation of the classical skyrmion. 
Despite the discontinuities that occur when the on-site spin becomes zero during the \emph{oscillatory}, time evolution of the skyrmionic state between the two levels 
$\{ \ket{\psi_1}, \ket{\psi_2} \}$ [see Fig.~\ref{fig:19}(a)], the topological charge exhibits an \emph{oscillatory} behavior ranging between approximately $0.90$--$0.96$, in correlation with the Rabi oscillations and the entanglement entropy dynamics (see Fig.~\ref{fig:20}).

In Figure~\ref{fig:20}(a), we illustrate the transition probability $P_{\rm exc} = |\langle \psi_2 | \psi(t) \rangle|^2$ as a function of time between the states $\ket{0}$ (ground state) and $\ket{1}=\ket{\psi_2}$ (excited state) of the qubit, together with the corresponding time variation of the entanglement entropy in Figure~\ref{fig:20}(b). 
The two-level system undergoes Rabi oscillations between states, and, when coupled to another quantum system, here the periodic photonic drive, its coherent evolution leads to periodic entanglement dynamics.
The corresponding entropy oscillations reflect the exchange of quantum information between the system and its environment.  
Note that, unlike in precessional manipulation with a magnetic field, the overall entropy does not increase over time, as the two-level qubit system in our current approach is closed and evolves coherently.\\

\paragraph{Dynamics considering the full Hilbert space\\ \\}

The final and most computationally demanding approach addresses spin dynamics induced by coupling with a periodic photonic or microwave drive,  by considering the complete Hilbert space, which for the $19$-spin lattice (Figure~\ref{fig:1}) has a dimension of $2^{19}$.
In this framework, the driven Hamiltonian is constructed using the vector $ \ket{\Psi}$ in the full Hilbert space given by Eq.~\ref{eq:full_drive} with the gate operators given by Eq.~\ref{eq:full_gate}. 

\begin{figure*}[htbp]
  \centering
  \includegraphics[width=\textwidth]{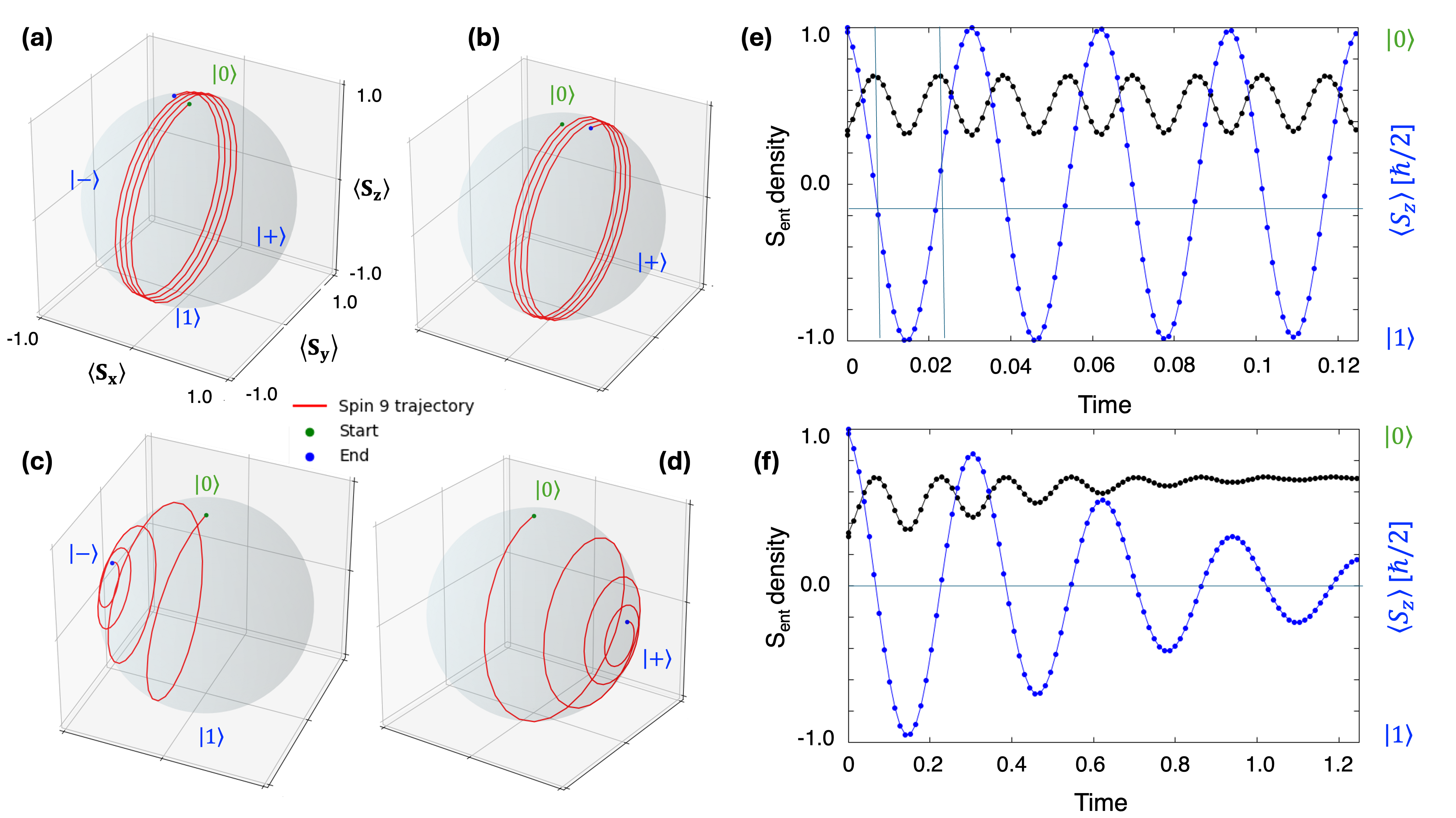}
  \caption{ Central spin trajectory on the logical Bloch sphere describing the dynamics considering the full Hilbert space corresponding to a Pauli-$X$ periodic drive, for different drive amplitudes and sign {\bf (a)} $A =-100$, {\bf (b)} $A =100$, {\bf (c)} $A =-10$, {\bf (d)} $A =10$. For small drive amplitudes the system rapidly decay towards the ground state of the driven Hamiltonian. Time evolution of the entanglement entropy density (black) and average $\langle S_z \rangle$ of the central spin of the skyrmion, calculated for the drive amplitudes {\bf (e)} $A=\pm 100$ and {\bf (f)} $A=\pm 10$. }
  \label{fig:21}
\end{figure*}

Under this driving, the system evolves towards the ground states of the driven Hamiltonian. 
Depending on the sign of the drive $A$ coupled to a periodic Pauli $X$ drive, the system will evolve towards the states $\ket{+}$ and  $\ket{-}$, as illustrated by the Figures~\eqref{fig:21}(1-d), the larger the drive amplitude, the slower the decay. 
Similar to precessional manipulation using an external magnetic field, in this case we found that the drive amplitude is particularly important to preserve the qubit stability and the gate fidelity. 
In Figure~\ref{fig:21}, we illustrate the comparison of gate dynamics in the full Hilbert space for a Pauli-$X$ gate as a function of the drive amplitude and sign. 
Positive drive amplitude $A$ means that the oscillating field drives the qubit in-phase, negative $A$  means the same oscillating field but shifted by $180^\circ$  degrees in phase, i.e., the $180^\circ$ phase shift flips the direction of induced rotations on the qubit. 
Note that similar behavior as the one illustrated in Figure~\ref{fig:21} has been observed for the  Pauli $X$, $Y$, $Z$ and Hadamard gates.

\begin{figure}[htbp]
  \centering
  \includegraphics[width=\columnwidth]{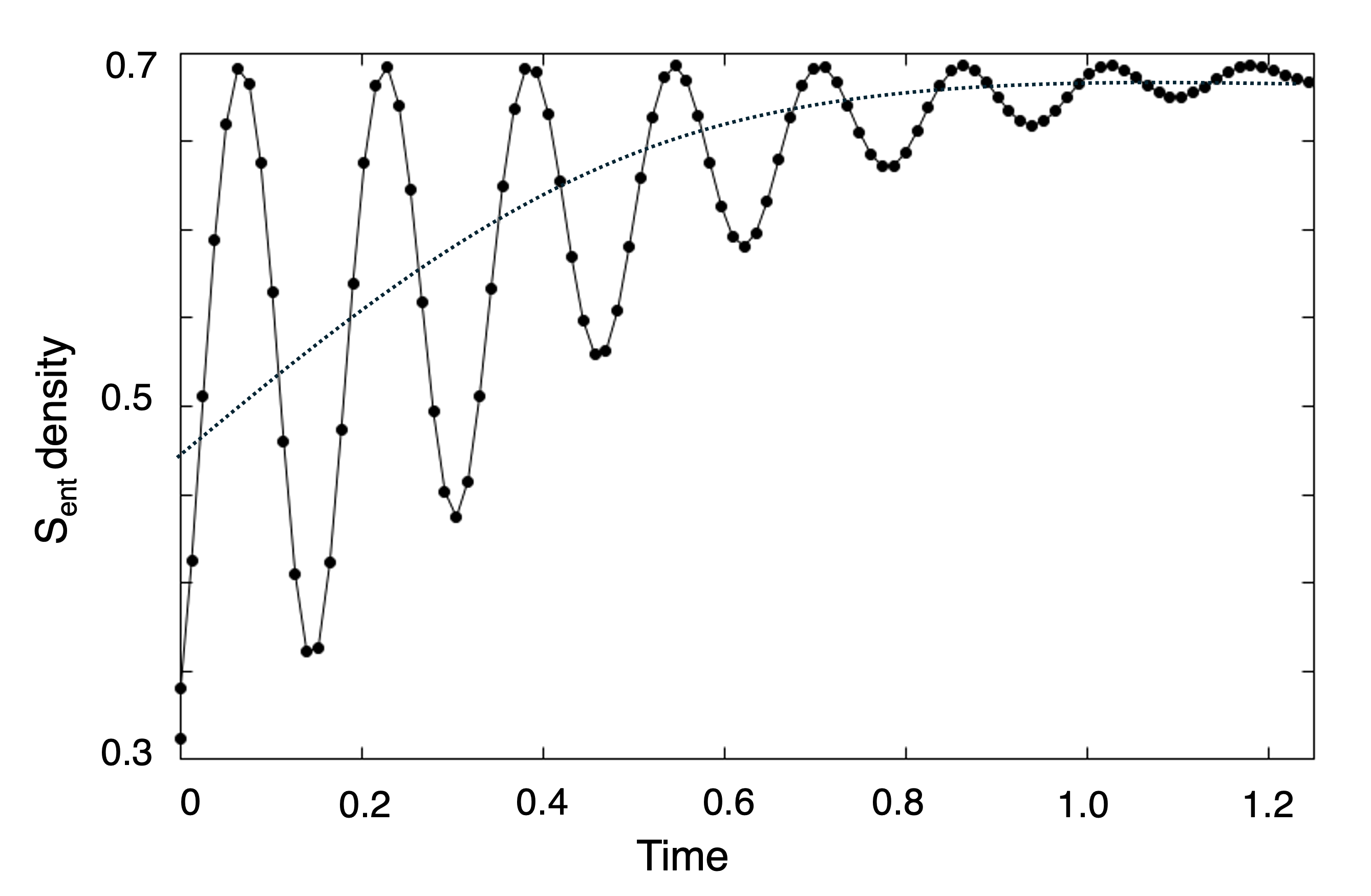}
  \caption{ Time evolution of the entanglement entropy density of the central spin of the skyrmion, calculated for the drive amplitudes $A=\pm 10$, illustrating the gradual oscillatory asymptotic evolution towards the von Neuman maximum entropy density, $\ln(2)$.}
  \label{fig:22}
\end{figure}

In contrast to the previous case, where the two-level subspace remained well isolated, the dynamics in this scenario permits potential leakage into higher excited states. 
As a result, similar to the behavior observed under precessional manipulation, one expects an oscillatory yet gradual increase in the entanglement entropy over time, asymptotically approaching the von Neumann entropy density of $\ln(2)$. 
This trend is more clearly illustrated in Figure~\ref{fig:22}, where the faster relaxation dynamics, due to the smaller drive amplitude, leads the system to converge toward the maximum von Neumann entropy within the timescale of our calculation. 

It is worth noting that simulations involving the full Hilbert space dynamics are extremely time-consuming and require parallel multiprocessing. 
This complexity is further compounded by the large dimensionality of the Hilbert space. 
Additionally, time evolution is performed step-by-step with memory clearing at each step to avoid quickly exhausting the available 64 GB of computational memory.\\

\paragraph{Experimental issues: measuring a logical qubit state physically\\ \\}

\begin{figure*}[htbp]
\centering
\includegraphics[width=0.9\textwidth]{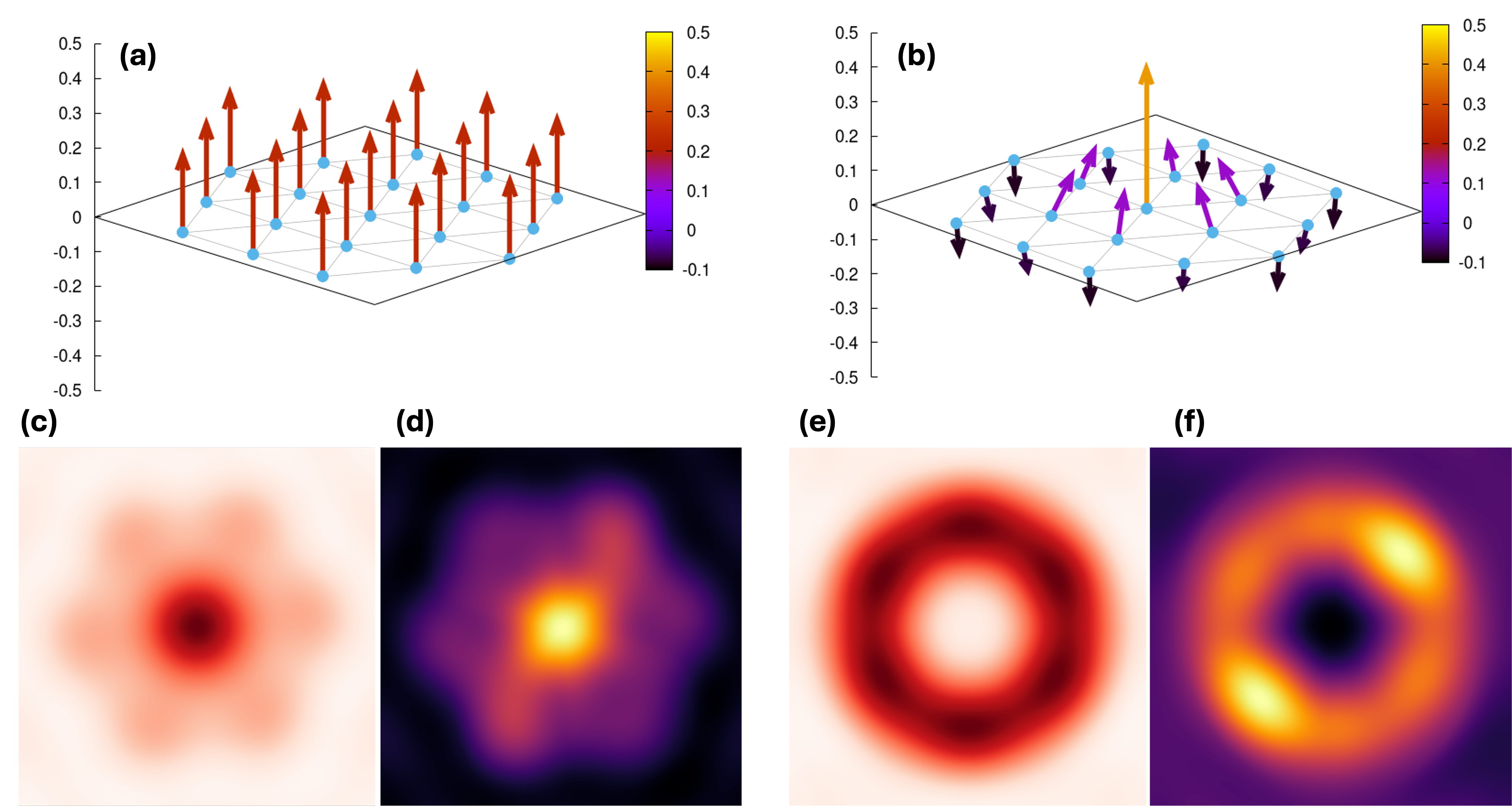}
\caption{Representations in real space of the on-site spin polarizations,  $\langle S_z^i \rangle$,  for a quantum skyrmion  {\bf (a)} and a classic skyrmion {\bf (b)}. 
The FFT of the quantum spin-spin correlation functions for quantum {\bf (c)}  and classic skyrmions {\bf (e)}. 
The elastic magnetic neutron scattering cross section $d\sigma/d\Omega$ at momentum transfer vector $\bm{q}$ corresponding  to quantum {\bf (d)}  and classic skyrmion {\bf (f)}.  
Note that the  images  {\bf (c)}--{\bf (f)} correspond to a 2D  Brillouin zone defined by $q_x/\pi=[-1,+1]; q_y/\pi=[-1,+1]$.}
\label{fig:23}
\end{figure*}

As we illustrated in Figs.~\ref{fig:3} and~\ref{fig:23}, quantum and classic-like skyrmions can be measured via local magnetic polarization measurements, as well as in the spin structure factor obtained through neutron scattering experiments~\cite{Sotnikov2021, Haller2022}. 
However, these type of experiments are not suitable for qubits, especially in quantum computing architectures. 

One of the most important issues for practical implementation of qubits is related to their measurement.  
A logical observable is measured by coupling it to a physical process whose outcome depends on the projection of the quantum state onto the logical subspace. 
Measuring a logical qubit state, i.e., distinguishing {\em logical qubit states} in the energy basis, physically requires the following steps:
\begin{itemize}
\item Manipulate the state using gates via time-dependent drives.
\item Rotate the state into a basis where measurement reveals the desired logical information.
\item Perform a projective measurement in the physical basis (e.g., fluorescence, energy readout, spin alignment). 
For instance, in superconducting qubits, the logical qubit readout is performed by probing a cavity that is sensitive to the qubit state.
\end{itemize}

In a first step, we define the logical qubit subspace using the two lowest energy eigenstates of the full Hamiltonian (see Table~\ref{tab:i}):

\begin{table}[h!]
\centering
\caption{Logical qubit space elements in energy basis. }
\SetTblrInner{colsep=0pt}
\begin{tblr}{
			width=\columnwidth,
      colspec = {|X[c,0.28\columnwidth]|X[c,0.32\columnwidth]|Q[c,0.39\columnwidth]|},
			row{1} = {bg=black!10, font=\bfseries, c}, 
			stretch = 1.5,
			hlines
		}
Energy Levels & Eigenstates & Logical Basis States\\
$E_1$ excited & $\hat{H}_0 \ket{\psi_2} = E_1 \ket{\psi_2}$ & $\ket{1}_L = \ket{\psi_2}$ \\
$E_0$ ground state & $\hat{H}_0 \ket{\psi_1} = E_0 \ket{\psi_1}$ & $\ket{0}_L = \ket{\psi_1}$ \\
\end{tblr}
\label{tab:i}
\end{table}

The $X$ logical qubit states are
\begin{align*}
\ket{+} &= \frac{1}{\sqrt{2}} \left( \ket{0}_L + \ket{1}_L \right), \\
\ket{-} &= \frac{1}{\sqrt{2}} \left( \ket{0}_L - \ket{1}_L \right),
\end{align*}
and the eigenstates of the logical $\sigma_x$ operator:
\begin{align*}
\sigma_x^{(\text{\rm logical})} &= \ket{0}_L \bra{1}_L + \ket{1}_L \bra{0}_L\\
\sigma_x^{(\text{\rm logical})} \ket{+} &= +\ket{+}\\
\sigma_x^{(\text{\rm logical})} \ket{-} &= -\ket{-}
\end{align*}

Similarly, the $Y$ logical qubit states will be:
\begin{align*}
\ket{+i} &= \frac{1}{\sqrt{2}} \left( \ket{0}_L + i \ket{1}_L \right), \\
\ket{-i} &= \frac{1}{\sqrt{2}} \left( \ket{0}_L - i \ket{1}_L \right).
\end{align*}
and the eigenstates of the logical $\sigma_y$ operator:
\begin{align*}
\sigma_y^{(\text{\rm logical})} &= -i \ket{0}_L \bra{1}_L + i \ket{1}_L \bra{0}_L,\\
\sigma_y^{(\text{\rm logical})} \ket{+i} &= +\ket{+i},\\
\sigma_y^{(\text{\rm logical})} \ket{-i} &= -\ket{-i}.
\end{align*}

\noindent {\bf Distinguishing $\ket{+}$ and $\ket{-}$ states in the energy basis} \\

\begin{figure}[htbp]
 \centering
 \includegraphics[width=\columnwidth]{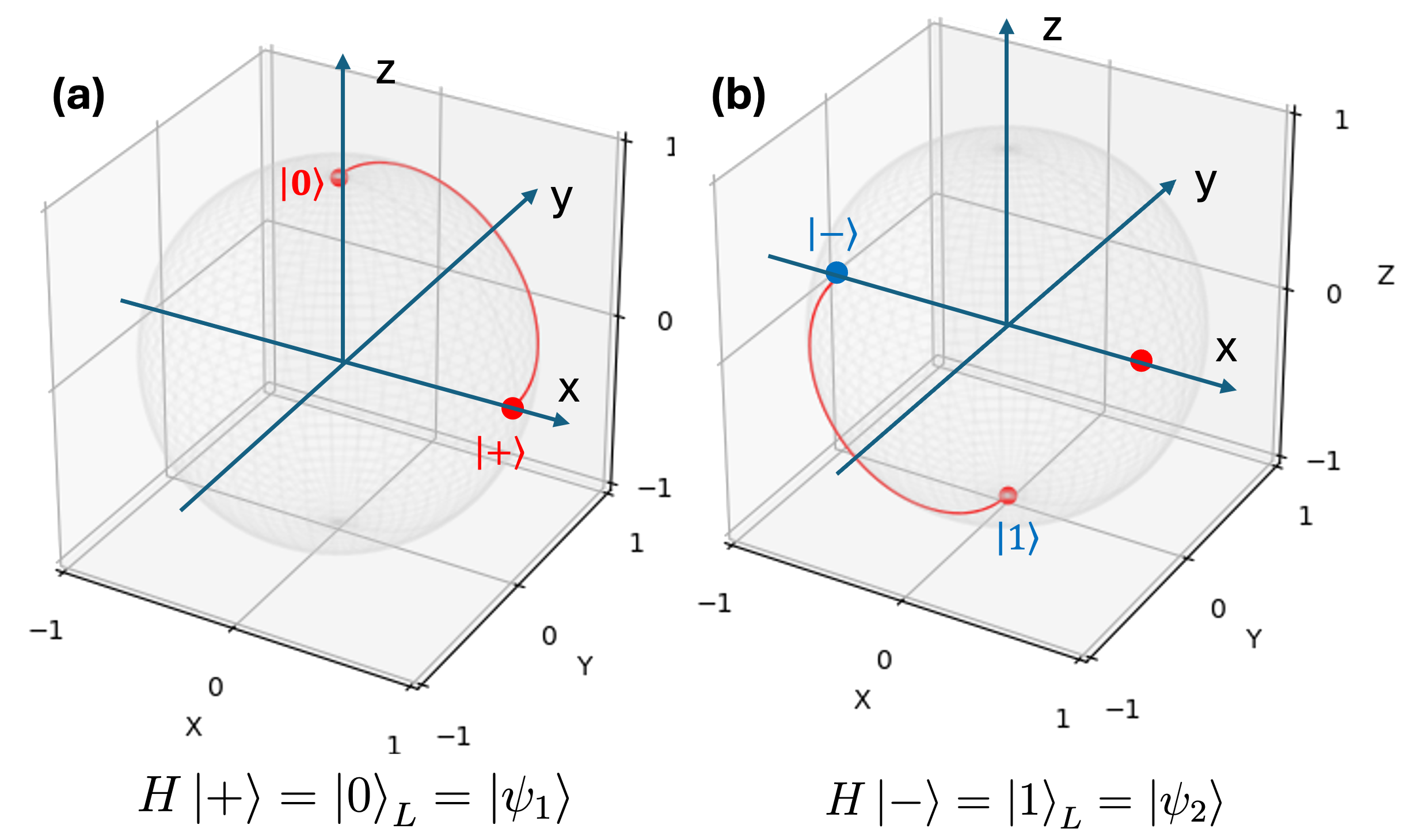}
 \caption{ Simulated evolution of the skyrmionic qubit state after applying a Hadamard rotation gate if initially the system was in {\bf (a)}  $\ket{+}$ and {\bf (b)}  $\ket{-}$ states.}
 \label{fig:24}
\end{figure}

Physically measuring in the energy basis involves applying a rotation that maps the $\ket{+}$ and $\ket{-}$ states onto the energy eigenbasis $\left\{ \ket{\psi_1}, \ket{\psi_2} \right\}$. 
This is achieved via a Hadamard rotation, $H = \frac{1}{\sqrt{2}} (\begin{smallmatrix} 1 & 1 \\ 1 & -1 \end{smallmatrix})$, i.e., coupling the qubit to a circular $X$-$Z$ polarized photon. 
As illustrated in Figure~\ref{fig:24}, if the qubit was in the $\ket{+}$ state, the Hadamard rotation will bring it in the $\ket{0}$ logical state, and, if it was in the $\ket{-}$ state, the Hadamard rotation will bring it to the $\ket{1}$ logical state. 
Then, a projective measurement (e.g., fluorescence, energy readout, spin alignment in the energy basis $\left\{ \ket{\psi_1}, \ket{\psi_2} \right\}$) can distinguish $\ket{+}$ and $\ket{-}$. 
If, after applying Hadamard rotation, a second photon tuned in frequency to be resonant $\omega=E_1-E_0$ is absorbed, then the system was in  $\ket{+}$, if it is not absorbed, the qubit system was initially in $\ket{-}$ state. 
Here we see that it is important that the energy levels are anharmonic, in order to avoid transition with same $\omega$ from $\ket{-}$, the first excited state to the second excited state. \\

\noindent  {\bf Distinguishing $\ket{+i}$ and $\ket{-i}$ states in the Energy Basis} \\

\begin{figure}[htbp]
 \centering
 \includegraphics[width=\columnwidth]{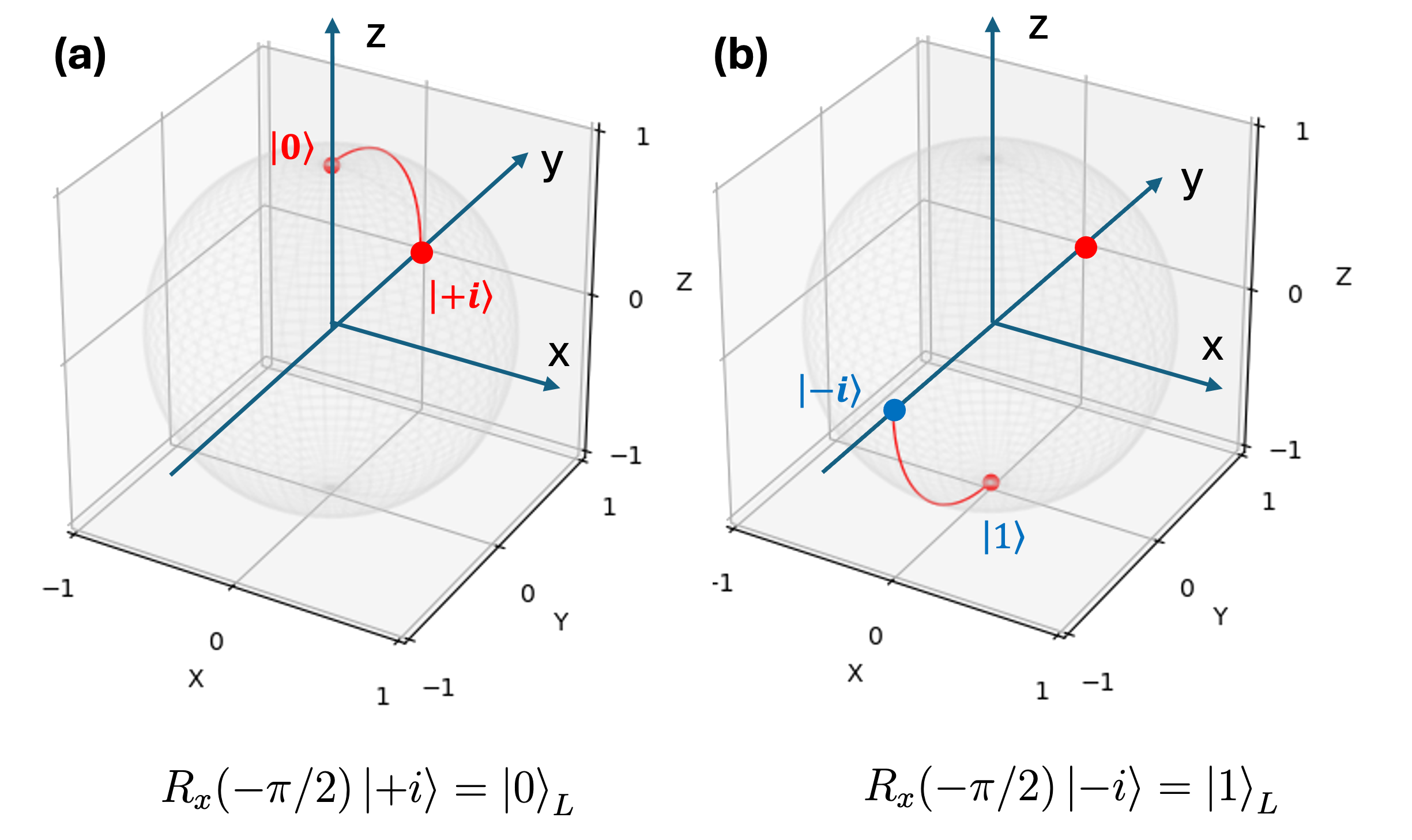}
 \caption{ Simulated evolution of the skyrmionic qubit state after applying a $R_x(-\pi/2)$ rotation gate if initially the system was in {\bf (a)}  $\ket{+i}$ and {\bf (b)}  $\ket{-i}$ states.}
 \label{fig:25}
\end{figure}

To distinguish $|+i\rangle$ and $|-i\rangle$ using a measurement in the energy basis, one has to apply a rotation $R_x(-\pi/2)= \frac{1}{\sqrt{2}} (\begin{smallmatrix} 1 & i \\ i & 1 \end{smallmatrix})$. As illustrated in Figure~\ref{fig:25}, if the qubit was in the $\ket{+i}$ state, a  $R_x(-\pi/2)$ rotation will bring it in the $\ket{0}$ logical state, and, if it was in the $\ket{-i}$ state, a $R_x(-\pi/2)$ rotation will bring it to the $\ket{1}$ logical state. 
Then, the same projective measurement can be performed using a resonant photon to distinguish the logical $\ket{\pm i}$ states.

As the simulations contained in Figs.~\ref{fig:24} and \ref{fig:25} illustrate, these measuring strategies can be applied to the skyrmionic qubits coupled to external periodic photonic or microwave drives. \\

\paragraph{Antiferromagnetic skyrmions as two-qubit quantum gates\\ \\}

For universal computation applications, creating entanglement between individual qubits represents a crucial issue. 
This is not implemented implicitly, and in real quantum hardware the quantum gates are constructed using lower-level operations (like rotations and entangling interactions). 
The specific implementation depends on the quantum computing platform. 

To extend our skyrmionic qubit concept to quantum gate applications, similarly to the proposal of Psaroudaki et al.~\cite{Psaroudaki2021, Psaroudaki2023}, we consider two antiferromagnetically (AF) coupled skyrmions, as found in ferrimagnetic-type materials~\cite{FRadu2023} or in synthetic AF multilayer systems~\cite{Boulle2024}. 
In this case, the entanglement mechanism is the AF exchange coupling, and {\em the key concept for operating the two-qubit skyrmionic quantum gate lies in the ability to manipulate (rotate) individual skyrmionic qubits} for example, through optical-type experiments. 
Recent experimental and modeling studies have demonstrated the all-optical creation and annihilation of skyrmions in chiral magnets using femtosecond pulses, revealing helicity and fluence thresholds, as well as field dependence, thus providing a direct experimental example of helicity-dependent optical control of magnetic textures~\cite{Kalin2024OpticalSkyrmion}. 

\begin{figure}[htbp]
 \centering
 \includegraphics[width=\columnwidth]{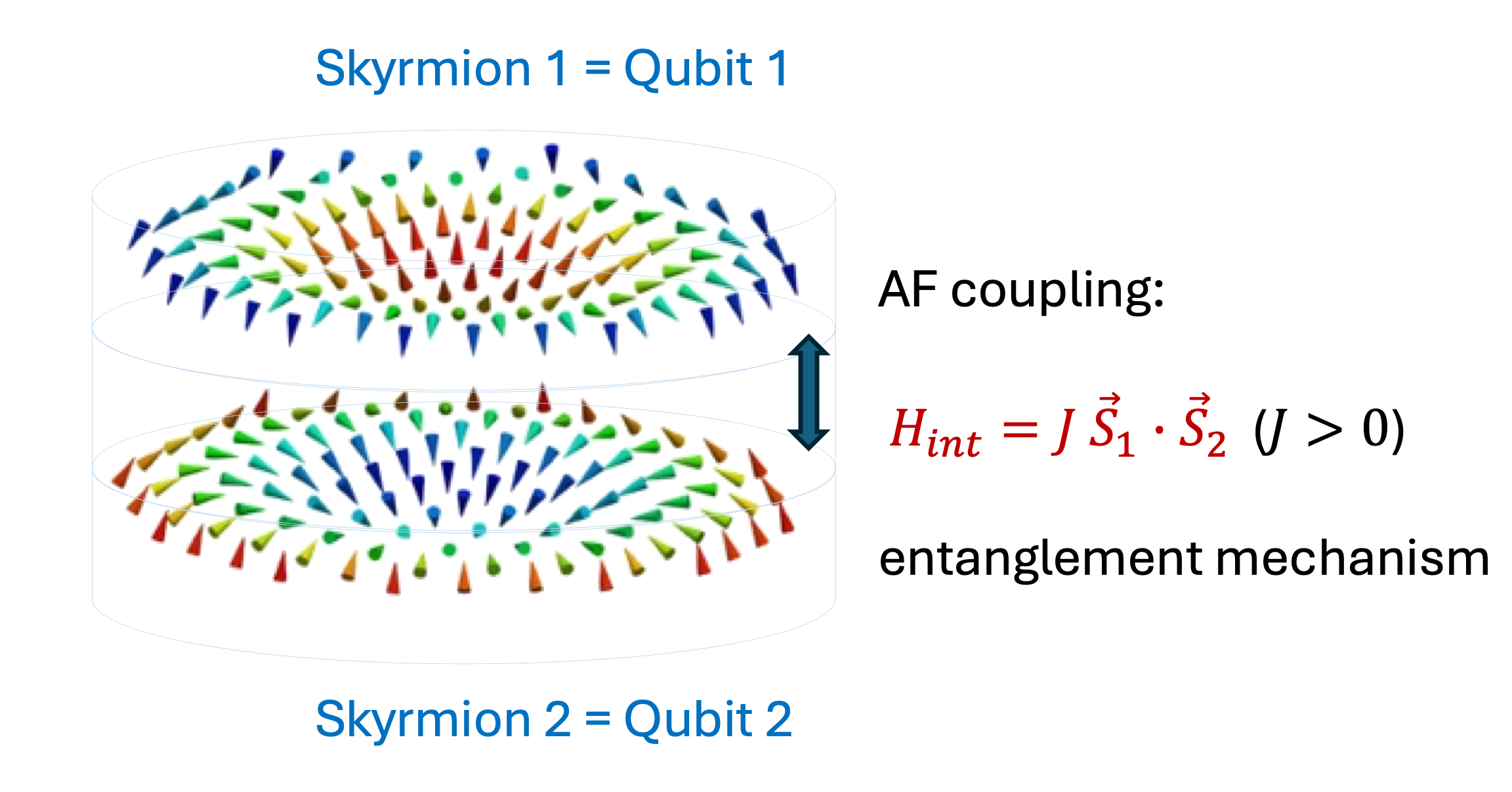}
 \caption{Micromagnetically simulated skyrmionic AF ground state, see for more details~\cite{Nanomaterials2022}.}
 \label{fig:26}
\end{figure}

Figure~\ref{fig:26} illustrates the concept of an antiferromagnetic qubit, with the typical spin configuration obtained from micromagnetic calculations. 
The details of these calculations are provided in one of our previous publications~\cite{Nanomaterials2022}.

In this context, Table~\ref{tab:ii} illustrates examples of two-qubit logical operations that can be implemented. 
While these operations are not specific to the AF skyrmion qubit concept, they can be perfectly adapted to this case. 
A common universal gate set can be constructed based on all single-qubit operations plus a CNOT gate. 
This set can approximate any unitary operation to arbitrary accuracy. 
The required steps to create entanglement using a CNOT gate from a simple superposition are:
\begin{table}[h!]
	\centering
	\caption{Two-bit gate operations and expected result.}
	\begin{tblr}{
			width=\columnwidth,
			colspec = {|X[1,c]|X[2,l]|},
			row{1} = {bg=black!10, font=\bfseries, c}, 
			stretch = 1.5,
			hlines
		}
		2 bit gate & Result \\
		Controlled-NOT (CNOT) & Flips the target qubit if the control qubit is $\ket{1}$. \\
		Controlled-Z (CZ) & Applies a $Z$ gate (phase flip) to the target qubit if the control is $\ket{1}$. \\
		Swap & Exchanges qubit states or creates partial swapping (used in some architectures). \\
	\end{tblr}
  \label{tab:ii}
\end{table}
\begin{enumerate}
\item We prepare the initial state, starting with the two (skyrmionic) qubits initialized as: $\ket{q_1 q_2}=\ket{00}$.
\item We apply a Hadamard gate {\em H} on the control qubit  $\ket{q_1}$. This leads to the following transformation:
$$
|0\rangle \xrightarrow{H} \frac{|0\rangle + |1\rangle}{\sqrt{2}}
$$
so that the combined system becomes: 
$$
|\psi\rangle = \left(\frac{|0\rangle + |1\rangle}{\sqrt{2}}\right) \otimes |0\rangle = \frac{|00\rangle + |10\rangle}{\sqrt{2}}
$$
\item We apply the CNOT gate with $\ket{q_1}$ as control and $\ket{q_2}$ as target.  
This CNOT gate flips the target qubit  $\ket{q_2}$ only if the control qubit  $\ket{q_1}$ = $\ket{1}$.
Therefore, this operation leads to the final state

$$
\ket{\Phi^+}=\frac{\ket{00}+\ket{11}}{\sqrt{2}}
$$

\item {\em Interpretation}. The final state is a maximally entangled Bell state.
We note that neither qubit alone has a definite state and, instead, they are perfectly correlated.
With such Bell state, one can envisage a variety of fundamental tasks such as: quantum teleportation, 
superdense coding, entanglement-Based Quantum Key Distribution (QKD), e.g., Ekert 91 (E91) protocol~\cite{ekert1991quantum},  error correction, resource in stabilizer circuits, in measurement-based quantum computing, where cluster states are built from Bell pairs, and the Bell state $|\Phi^+\rangle$ can be used  to ``teleport'' a qubit through a gate, by performing appropriate measurements and corrections, and, last but not least, for challenging entanglement swapping experiments~\cite{Pan1998EntanglementSwapping, Fiorentino2005SinglePhotonSwap} essential for quantum repeaters. E.g., in the latter case, one takes two $|\Phi^+\rangle$ states, performs a Bell-state measurement on one qubit from each, and ends up with two previously unentangled qubits becoming entangled.
\end{enumerate}

\section{Discussions and conclusions}\label{sec:conclusions}

In our study, we showed that skyrmion-like spin textures can be stabilized in quantum spin lattices incorporating Dzyaloshinskii-Moriya interactions, and these configurations can be coherently manipulated as qubit states. 
A complete quantum mechanical description is achieved within the exact diagonalization framework, considering spin lattices governed by both Heisenberg exchange and DMI terms.
The interplay between Heisenberg coupling  $J$, magnetic field $B$, DMI  strength $D$, and magnetic frustration in our triangular spin lattice gives rise to a rich quantum phase diagram. 
Quantum skyrmionic phases emerge under periodic boundary conditions, forming well-defined regions characterized by uniform local spin polarization and nontrivial skyrmion-like spin-spin correlation patterns---yet lacking topological protection. 
Conversely, under open boundary conditions, classical-like skyrmions are stabilized, exhibiting genuine topological protection. 
In this regime, robust skyrmionic textures enable the realization of quantum logic gates, including Pauli $X$, $Y$, $Z$, and Hadamard operations, alongside the demonstration of entanglement dynamics, entropy generation, and decoherence processes. 
Crucially, while DMI stabilizes skyrmionic configurations akin to classical behavior, it simultaneously introduces quantum fluctuations, which significantly affect spin manipulation protocols and the quantum coherence of the encoded qubit states.

The key effects of the DMI are: (i) Breaks spin-rotational symmetry, favoring chiral configurations. 
These skyrmionic configurations arise either as real-space spin textures formed via on-site spin canting under open boundary conditions, leading to classical-like skyrmions, or as emergent features in spin-spin correlation functions under periodic boundary conditions, where translational invariance enforces a globally aligned spin state along the $z$ axis, yet with reduced spin polarization due to quantum fluctuations. 
(ii) Frustration and entanglement, spreading quantum information across the system.  In the full system, the DMI causes quantum information to spread, resulting in growth of entanglement entropy. 
This mimics interaction-induced chaos or thermalization: the DMI acts like an internal bath, even in a closed quantum system.\\

\noindent {\em DMI impact on the skyrmionic qubit fidelity\\ \\} 
The skyrmionic qubit is built using the ground and first excited states of the 2D interacting spin lattice, in a special parameter window in the quantum phase diagram where the states are skyrmionic-type. 
When modeling the qubit within the truncated Hilbert space, i.e., the two-level system (TLS), we map the full Hilbert space onto a qubit formed from the ground and first excited states. 
This model shows high fidelity, clean Rabi dynamics, and periodic entanglement entropy. 
However, in the full Hilbert space system, the following issues arise: (i) Leakage to higher states due to drive-induced transitions and DMI; (ii) Dephasing from dynamically induced chiral interactions; (iii) Entanglement with internal degrees of freedom, reducing purity and trace fidelity; (iv) DMI tends to close gaps and increase state density near the ground state in frustrated systems. 
This enhances drive-induced transitions out of the TLS. 
Thus, the DMI appears to significantly reduce the coherence time and gate fidelity, unless error suppression techniques are employed.

Despite these DMI induced decoherence effects, coherent manipulation of spins can be achieved by employing large enough drive amplitudes of periodic photonic or microwave drives or external magnetic fields in precessional manipulation. 
Strong drive or Zeeman terms suppress DMI-induced effects, their dominance in the effective field restores coherence, aligning spins and reducing chirality-induced complexity.

Our studies focusing on DMI induced skyrmionic qubits emphasize a very interesting and intriguing result. 
Topological protection in skyrmions is classical (linked to energy barriers) and not  quantum-coherent.
The same DMI interaction that stabilizes skyrmions also introduces internal dynamics that can act as a source of decoherence. 
Additionally, skyrmions are complex many-body objects, and engineering controlled quantum gates between them is nontrivial. 
Environmental couplings that break spin conservation (e.g., magnetic noise or spin flips-often related to spin-orbit mechanisms responsible on DMI) can destroy the topological protection, allowing leakage from an ideal decoherence-free subspace provided by an ideal topological protection. 
As such, while skyrmions provide an intriguing path toward decoherence-resilient encoding, realizing them as fully quantum decoherence-free subspaces suitable for fault-tolerant quantum information processing will require careful Hamiltonian engineering, symmetry protection, and possibly hybrid integration with other quantum technologies. 
The DMI, even in closed systems, acts as a powerful internal decoherence mechanism, leading to energy redistribution and entanglement spreading. Understanding and mitigating these effects would be crucial for maintaining qubit fidelity in real spin-based quantum simulators or devices.\\

\noindent {\em Theoretical and experimental perspectives\\ \\}
Beyond these considerations, although theoretical studies have highlighted the promising potential of skyrmion-based qubits for integration into quantum computing platforms, several open questions and challenges remain. 
A central issue concerns the scale at which quantum effects necessary for qubit functionality can be preserved. 
Classical skyrmions used in racetrack memory technologies are already nanoscale in size---but is this still too large for genuine quantum behavior to persist? 
For instance, can the Zeeman levels of a classical skyrmion be reliably treated as a two-level system suitable for qubit realization?
As with other types of qubits, the influence of finite temperature is a critical factor. 
A functioning two-level qubit system requires an energy gap significantly larger than the thermal broadening of energy levels.
Otherwise, the magnetization and the spin correlation functions follow as thermal averages over multiple states. 
As seen for the lowest two states [e.g., Fig.~\ref{fig:14}(a)], such averaging reduces the on-site magnetization and scalar spin chirality, adversely affecting qubit coherence and control.

Addressing these challenges may necessitate the design of novel skyrmionic quantum materials, where Dzyaloshinskii-Moriya interaction mechanisms are finely tuned to enhance topological protection and suppress decoherence.

Looking forward, once individual skyrmionic qubits are realized, the next objective would be the development of multi-qubit quantum gates. 
The coupling or entanglement mechanism essential for two-qubit gate operations might be proded, as discussed, by antiferromagnetic interactions. 
Notably, antiferromagnetic skyrmions have already been observed at macroscopic scales in ferrimagnetic compounds composed of rare-earth and transition-metal elements~\cite{FRadu2023} and in synthetic AF multilayer systems~\cite{Boulle2024}. 
However, if the skyrmion size proves to be a critical factor for maintaining quantum properties, it may be necessary to explore new two-dimensional materials capable of hosting skyrmions at smaller, quantum-relevant scales~\cite{WdWSkyr2024}. 
In these 2D materials, the relative strength of the DMI interaction and direct exchange can be tuned into the predicted parameter range that favors the formation of skyrmionic states.

From a theoretical standpoint, the exact diagonalization method is limited to solving systems with only a few dozen interacting spins due to computational constraints. 
In contrast, methods like the Density Matrix Renormalization Group (DMRG)~\cite{White1992} can, in principle, handle much larger systems~\cite{Haller2022}. 
However, DMRG is not well-suited for qubit studies that require precise determination of eigenstates and eigenvalues. 
As a variational approach, DMRG can spontaneously break symmetries and, in frustrated systems, may select a single classical-like state from a set of degenerate ones. 
Moreover, it may converge to a local minimum instead of the true ground state.
Even when ED and DMRG yield nearly identical ground state energies, their predictions for observables---such as on-site spin polarization and correlation functions---can differ significantly. 
Extending our study to a larger two-dimensional interacting spin lattice presents a significant theoretical challenge. 
Nevertheless, it is intuitive to expect that as the system size increases, skyrmionic patterns will emerge at higher $J/D$ ratios, with the Dzyaloshinskii-Moriya interaction competing against direct exchange and the magnetic field in stabilizing skyrmions and driving quantum fluctuations. 
Consequently, we anticipate that our results can be reasonably extrapolated to larger systems, up to the point where quantum effects diminish due to dimensional constraints.

\begin{acknowledgments}
C.~T.~acknowledges the funding project UEFISCDI via the project “MODESKY” ID PN-III-P4-ID-PCE-2020-0230-P, grant No.~UEFISCDI: PCE 245/02.11.2021. 
D.~S.~acknowledges the financial support from CNCS/CCCDI-UEFISCDI, under projects number PN-IV-P1-PCE-2023-0159, PN-IV-P1-PCE-2023-0987, PN-IV-P8-8.3-PM-RO-FR-2024-0059, and by the ``Nucleu'' Program within the PNCDI 2022-2027, Romania, carried out with the support of MEC, project no.~27N/03.01.2023, component project code PN 23 24 01 04. 
\end{acknowledgments}

\appendix

\section{Derivation of External Magnetic Fields for Gate Operations}
\label{app:magnetic_fields}

In this appendix, we derive the specific forms of external magnetic fields required to realize the four distinct types of $H_{\rm gate}$ operations in the driven two-level system, along with their physical interpretations.

\subsection{General Framework}

For a two-level quantum system with basis states $\ket{\psi_1}$ and $\ket{\psi_2}$, the interaction with an external time-dependent magnetic field $\bm{B}(t)$ is described by the magnetic dipole Hamiltonian:
\begin{equation}
H_{\text{drive}}(t) = -\boldsymbol{\mu} \cdot \bm{B}(t) = -\gamma \bm{S} \cdot \bm{B}(t) = -\frac{\gamma\hbar}{2} \boldsymbol{\sigma} \cdot \bm{B}(t),
\label{eq:magnetic_interaction}
\end{equation}
where $\boldsymbol{\mu}$ is the magnetic moment operator, $\gamma$ is the gyromagnetic ratio, $\bm{S}$ is the spin angular momentum operator, and $\boldsymbol{\sigma} = (\sigma_x, \sigma_y, \sigma_z)$ represents the vector of Pauli matrices.

To realize a drive of the form given in Eq.~\eqref{eq:drive}, we require:
\begin{equation}
A\cos(\omega t) H_{\rm gate} = -\frac{\gamma\hbar}{2} \boldsymbol{\sigma} \cdot \bm{B}(t).
\label{eq:drive_matching}
\end{equation}

From this matching condition, we can solve for the required magnetic field configuration for each gate operation.

\subsection{Pauli-X Gate: Transverse Driving Field}

\subsubsection{Derivation}

For the Pauli-X gate, $H_{\rm gate} = \sigma_x$. Comparing with Eq.~\eqref{eq:drive_matching}, we identify:
\begin{equation}
A\cos(\omega t) \sigma_x = -\frac{\gamma\hbar}{2} B_x(t) \sigma_x,
\end{equation}
which yields
\begin{equation}
\bm{B}(t) = B_x(t) \hat{\bm{x}} = -\frac{2A}{\gamma\hbar}\cos(\omega t) \hat{\bm{x}}.
\label{eq:Bx_field}
\end{equation}

\subsubsection{Physical Interpretation}

The Pauli-X gate is realized by an oscillating magnetic field along the $x$ direction, transverse to the quantization axis (conventionally taken as the $z$ axis). This configuration corresponds either to the standard setup in magnetic resonance spectroscopy (NMR, ESR) or with a linearly polarized photon along the $x$ direction. 
When the driving frequency $\omega$ matches the transition frequency $\omega_0 = (E_2 - E_1)/\hbar$, the field resonantly drives transitions between states $\ket{\psi_1}$ and $\ket{\psi_2}$, resulting in Rabi oscillations. The population oscillates between the two states with Rabi frequency $\Omega_R = \gamma B_0/2$, where $B_0$ is the amplitude of the oscillating field.

\subsection{Pauli-Y Gate: Phase-Shifted Transverse Field}

\subsubsection{Derivation}

For the Pauli-Y gate, $H_{\rm gate} = \sigma_y$. Following the same procedure:
\begin{equation}
A\cos(\omega t) \sigma_y = -\frac{\gamma\hbar}{2} B_y(t) \sigma_y,
\end{equation}
gives
\begin{equation}
\bm{B}(t) = B_y(t) \hat{\bm{y}} = -\frac{2A}{\gamma\hbar}\cos(\omega t) \hat{\bm{y}}.
\label{eq:By_field}
\end{equation}

\subsubsection{Physical Interpretation}

The Pauli-Y gate requires an oscillating magnetic field along the $y$ direction (e.g., linearly polarized photon along $y$), also transverse to the quantization axis but orthogonal to the $x$-direction. 
This field is phase-shifted by $\pi/2$ relative to the X-gate drive. In experimental implementations, the Y-gate is typically realized by adjusting the phase of the driving field. 
If the X-drive utilizes $B_x(t) = B_0\cos(\omega t)$, the Y-drive employs $B_y(t) = B_0\sin(\omega t) = B_0\cos(\omega t - \pi/2)$. The complex unit $i$ appearing in $\sigma_y$ manifests physically as this $\pi/2$ phase relationship between orthogonal field components.

\subsection{Pauli-Z Gate: Longitudinal Field Modulation}

\subsubsection{Derivation}

For the Pauli-Z gate, $H_{\rm gate} = \sigma_z$. The matching condition gives:
\begin{equation}
A\cos(\omega t) \sigma_z = -\frac{\gamma\hbar}{2} B_z(t) \sigma_z,
\end{equation}
resulting in:
\begin{equation}
\bm{B}(t) = B_z(t) \hat{\bm{z}} = -\frac{2A}{\gamma\hbar}\cos(\omega t) \hat{\bm{z}}.
\label{eq:Bz_field}
\end{equation}

\subsubsection{Physical Interpretation}

The Pauli-Z gate is implemented via an oscillating magnetic field parallel to the quantization axis ($z$ direction, e.g., linearly polarized photon along $z$). Unlike the transverse fields, this longitudinal field does not directly induce transitions between $\ket{\psi_1}$ and $\ket{\psi_2}$ since $\sigma_z$ is diagonal in the computational basis. Instead, it modulates the energy difference between the levels, creating a time-dependent phase accumulation. This mechanism is analogous to the AC Stark shift (or light shift in optical systems). The Z-gate does not produce conventional Rabi oscillations with population transfer; rather, it implements a dynamic phase gate. The relative phase between the two levels evolves according to
\begin{equation}
\phi(t) = \int_0^t \frac{2A}{\hbar}\cos(\omega t') \, dt'.
\end{equation}

\subsection{Hadamard Gate: Tilted Field Configuration}

\subsubsection{Derivation}

The Hadamard gate can be expressed as:
\begin{equation}
H_{\rm gate} = \frac{1}{\sqrt{2}}\begin{pmatrix} 1 & 1 \\ 1 & -1 \end{pmatrix} = \frac{1}{\sqrt{2}}(\sigma_x + \sigma_z).
\label{eq:hadamard_decomposition}
\end{equation}
Substituting into Eq.~\eqref{eq:drive_matching}:
\begin{equation}
\frac{A}{\sqrt{2}}\cos(\omega t) (\sigma_x + \sigma_z) = -\frac{\gamma\hbar}{2} \left[B_x(t) \sigma_x + B_z(t) \sigma_z\right],
\end{equation}
which yields:
\begin{equation}
\bm{B}(t) = -\frac{2A}{\gamma\hbar\sqrt{2}}\cos(\omega t)(\hat{\bm{x}} + \hat{\bm{z}}) = -\frac{\sqrt{2}A}{\gamma\hbar}\cos(\omega t)\left(\frac{\hat{\bm{x}} + \hat{\bm{z}}}{\sqrt{2}}\right).
\label{eq:Bh_field}
\end{equation}

\subsubsection{Physical Interpretation}

The Hadamard gate requires a magnetic field oscillating at $45^{\circ}$ between the $x$ and $z$ axes in the $xz$ plane. 
This configuration, combines equal components of transverse ($\sigma_x$) and longitudinal ($\sigma_z$) driving. The transverse component induces Rabi oscillations and population transfer, while the longitudinal component adds dynamic phase evolution. In the Bloch sphere representation, this corresponds to a rotation about a tilted axis. 

Experimentally, the Hadamard gate is often implemented more practically as a composite pulse sequence:
\begin{equation}
H = R_z(\pi/2) \cdot R_y(\pi/2) \cdot R_z(\pi/2),
\label{eq:hadamard_composite}
\end{equation}
where $R_\alpha(\theta)$ denotes a rotation by angle $\theta$ about axis $\alpha$. This decomposition uses sequential applications of Z and Y rotations, which can be realized through alternating longitudinal and transverse field pulses with appropriate phases and durations.

\subsection{Experimental considerations}

\subsubsection{Rotating Wave Approximation}

In typical experimental scenarios where $\omega \approx \omega_0 = (E_2 - E_1)/\hbar$, the oscillating drive can be decomposed as
\begin{equation}
\cos(\omega t) = \frac{1}{2}\left(e^{i\omega t} + e^{-i\omega t}\right).
\end{equation}
Under the rotating wave approximation (RWA), the counter-rotating term $e^{i\omega t}$ is neglected as it oscillates rapidly and averages to zero over the timescale of system dynamics. The effective time-independent Hamiltonian in the rotating frame becomes:
\begin{equation}
H_{\rm eff} = \frac{\hbar\Delta}{2}\sigma_z + \frac{\hbar\Omega_R}{2}H_{\rm gate},
\label{eq:Heff_RWA}
\end{equation}
where $\Delta = \omega - \omega_0$ is the detuning and $\Omega_R = \gamma B_0/2$ is the Rabi frequency.

\subsubsection{Resonance condition}

Maximum gate efficiency occurs when the driving field is on resonance, $\omega = \omega_0$, eliminating the detuning term in Eq.~\eqref{eq:Heff_RWA}. For a resonant $\pi$-pulse implementing a complete bit flip (X-gate), the pulse duration must satisfy:
\begin{equation}
\Omega_R \tau = \pi \quad \Rightarrow \quad \tau = \frac{2\pi}{\gamma B_0}.
\end{equation}

\subsubsection{Summary of field configurations}

Table~\ref{tab:field_summary} summarizes the magnetic field configurations required for each gate operation.

\begin{table*}[htbp]
\centering
\caption{Summary of magnetic field configurations for quantum gate operations.}
\label{tab:field_summary}
\begin{tabular}{lccc}
\hline\hline
Gate & Field Direction & Physical Effect & Typical Application \\
\hline
Pauli-X & Transverse ($\hat{\bm{x}}$) & Population transfer & $\pi$-pulse, state flip \\
Pauli-Y & Transverse ($\hat{\bm{y}}$) & Phase-shifted transfer & $\pi/2$-pulse (phase) \\
Pauli-Z & Longitudinal ($\hat{\bm{z}}$) & Energy modulation & Phase gate, Stark shift \\
Hadamard & Diagonal ($\hat{\bm{x}}+\hat{\bm{z}}$) & Combined rotation & Superposition creation \\
\hline\hline
\end{tabular}
\end{table*}

The field amplitude required in each case is determined by the desired gate-operation time and the gyromagnetic ratio of the system, with the relationship $B_0 = 2A/(\gamma\hbar)$ for transverse and longitudinal single-axis drives.

\section{DMI-Induced relaxation mechanisms}
\balance 
We mentioned that the same DMI interaction that stabilizes skyrmions also introduces internal dynamics that can act as a source of decoherence, introducing a dissipation mechanism in qubits.  
This is a  complex problem that represents an interesting perspective of the current study,  particularly concerning the  deduction and analysis of the DMI-related disipation term correlated with the relaxation times $T_1$ and $T_2$.  

Several competing mechanisms can be identified and evaluated in terms of their relative impact on the total relaxation time. 
Their contributions depend on the type of manipulation applied: (1) Precessional manipulation using an external magnetic field between the two-level states defined by the Zeeman eigenstates $\ket{0} \equiv \ket{\uparrow \uparrow \cdots \uparrow}$ and $\ket{1} \equiv \ket{\downarrow \downarrow \cdots \downarrow}$ , or (2) manipulation using a periodic photonic drive, in which the system evolves on an anharmonic energy landscape. We will qualitatively discuss the main mechanisms contributing to DMI-induced relaxation within our quantum framework, defined by the Hamiltonian in Eq.~\eqref{eq:Hamiltonian}. 
The key point is that in driven systems, the Dzyaloshinskii--Moriya interaction (DMI) breaks time-reversal symmetry.  Consequently, when a time-dependent field $B(t)$ is applied (for example, a rotating field used to implement an $X$--gate), the DMI gives rise to non--reciprocal dynamics that manifest as effective dissipation even within an otherwise isolated quantum Hamiltonian.

\begin{figure*}[htbp]
	\centering
	\includegraphics[width=\textwidth]{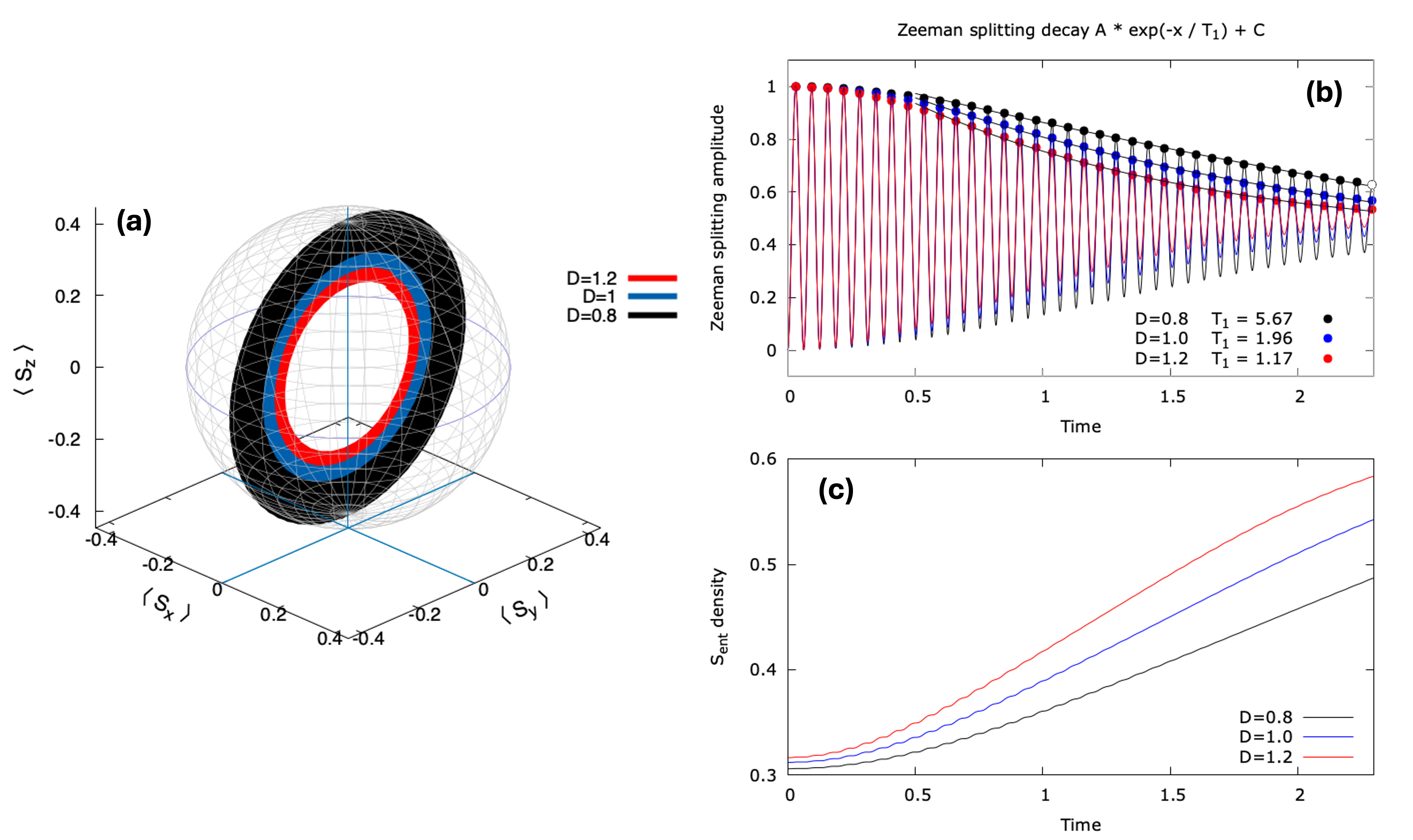}
	\caption{{\bf (a)}  Spin dynamics on the Bloch sphere under a Pauli -X drive for three different values of the DMI. {\bf (b)} Zeeman splitting decay during the qubit manipulation for three values of DMI strength. {\bf (c)} Variation of the entanglement entropy during the qubit manipulation for three values of DMI. }
	\label{fig:27}
\end{figure*}

\begin{figure*}[htbp]
	\centering
	\includegraphics[width=\textwidth]{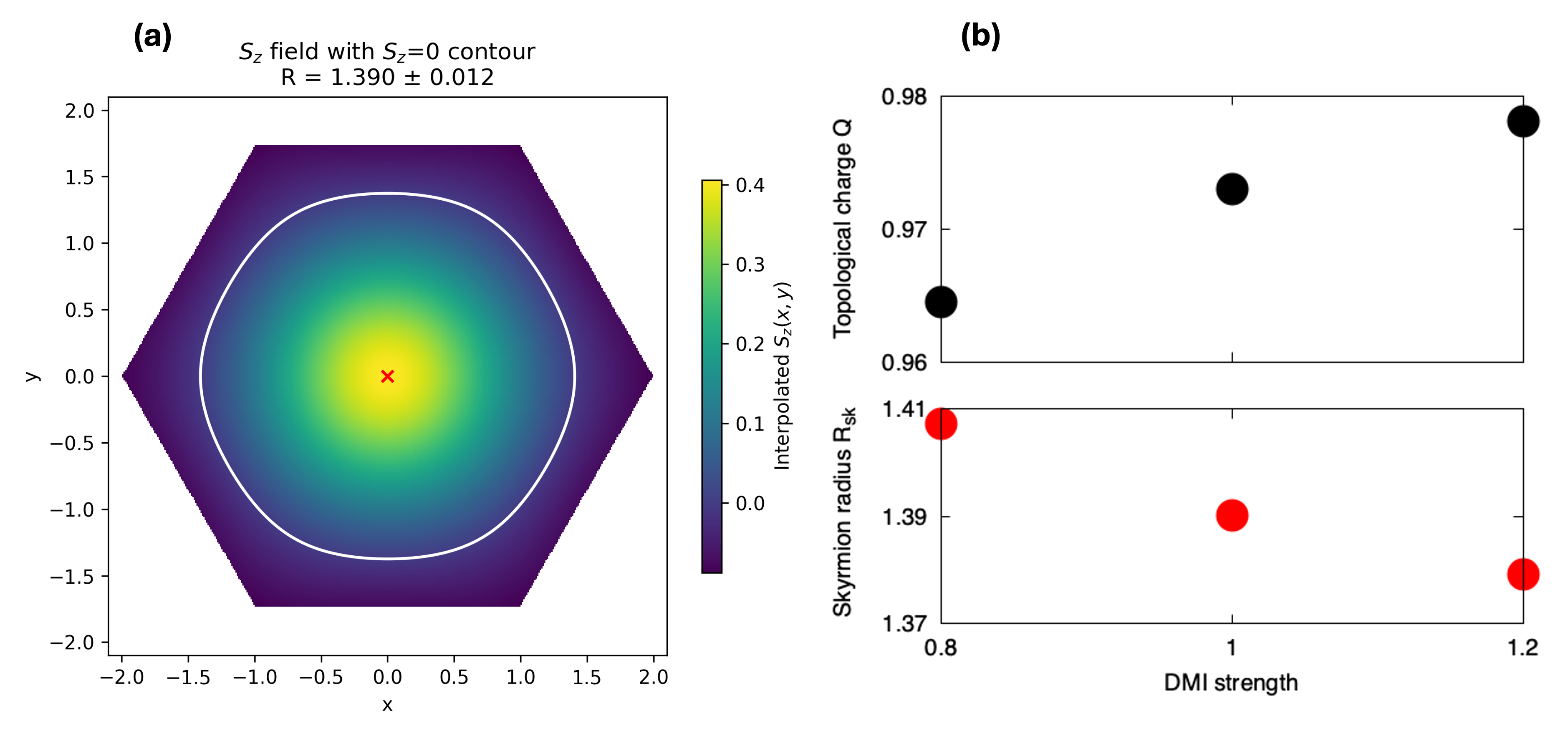}
	\caption{{\bf (a)} Interpolated $S_z$ spin field from the discrete $S_z(x_i, y_i)$  across the grid corresponding to DMI =1.0; the white curve indicates the skyrmion radius. The skyrmion radius on the discrete lattice is obtained by interpolating the discrete $S_z(x_i, y_i)$  across the grid and determining the radial distance from the lattice center at which $S_z(x)=0$ corresponding to the conventional $\pi/2$ angle criterion. {\bf (b)} Topological charge (top panel) and Skyrmion radius $R_{sk}$ as a function of the DMI strength. }
	\label{fig:28}
\end{figure*}

\begin{enumerate}

\item {\it Precessional manipulation under magnetic fields}

The time--dependent field $B(t)$, combined with the DMI term---which does not commute with the Zeeman part of the Hamiltonian---modulates the effective Zeeman splitting of the skyrmion. As a result, this splitting decays over time, and the system gradually relaxes toward a configuration that minimizes the DMI interaction. This behavior corresponds to energy relaxation, as illustrated in Fig.~\ref{fig:16}, and it is accompanied by an increase in the entanglement entropy [Fig.~\ref{fig:16}(a) and inset] and a decrease in $\langle S_z \rangle$ [Fig.~\ref{fig:16}(b)].  

In order to demonstrate the direct influence of the DMI on the decoherence and relaxation dynamics, we have carried out a comparative dynamical analysis of qubit fidelity for several DMI strengths.  
These results, illustrated in Figure \ref{fig:27}, clearly show that stronger DMI leads to stronger fidelity loss (Fig.~\ref{fig:27}(a)),  faster relaxation (Fig.~\ref{fig:27}(b)), and faster increase of the entanglement entropy density, as a measure of decoherence (Fig.~\ref{fig:27}(c)).

\item {\it Manipulation with periodic photonic drive}

In this case, the DMI term induces a time--dependent nonlinear coupling between the qubit states (spanned by $\{\ket{\psi_1}, \ket{\psi_2}\}$) and higher excited modes. 
This coupling progressively leads to leakage into higher--energy states, resulting in relaxation and a decay of qubit fidelity.


\item {\it The DMI influence on the skyrmion size}

Our quantum calculations on a finite discrete lattice confirm that increasing the DMI strength enhances fidelity loss, accelerates relaxation, and drives a more rapid growth of the entanglement-entropy density. Beyond these trends, we uncover another intristing effect: in finite spin lattices the skyrmion radius decreases as the DMI strength increases. At first sight, this trend appears to be opposite to classical micromagnetic expectations \cite{ Rohart2013}, according to which the skyrmion radius in finite-size magnetic disks increases monotonically with the DMI strength, behaviour governed by the balance among exchange, anisotropy, Zeeman energy, dipolar interactions, and boundary confinement. However, the apparent discrepancy, has a physical origin arising from the discrete, finite-lattice character of our model. In such a system, as intuitivelly expected, a larger DMI promotes sharper spin twists, resulting in smaller skyrmion cores. The lattice spacing and finite boundaries introduce additional intrinsic length scales, absent in the continuum description, that critically influence the skyrmion profile.
The dependence of the extracted skyrmion radius on the DMI strength is shown in Fig. 28. As the DMI increases, the twist sharpens, leading to a reduced skyrmion radius, enhanced local curvature, and an increased total scalar chirality (winding number $Q$).


\item{\it Additional DMI--induced effects}

Beyond these two mechanisms, other additional DMI--induced effects can be considered, such as: geometric phase fluctuations (Berry--phase decoherence), where the DMI introduces a topological contribution to the evolving phase; and dynamical hybridization, where under a periodic drive $B(t)$, the effective Floquet Hamiltonian exhibits DMI--induced avoided crossings between the qubit states and dressed excited states.

\end{enumerate}

\bibliography{quantum_references_prb}

@Article{Aldarawsheh2023,
  author               = {Aldarawsheh, Amal and Sallermann, Moritz and Abusaa, Muayad and Lounis, Samir},
  journal              = {Aip. Conf. Proc.},
  title                = {A spin model for intrinsic antiferromagnetic skyrmions on a triangular lattice},
  year                 = {2023},
  issn                 = {2296-424X},
  month                = may,
  pages                = {1175317},
  volume               = {11},
  article-number       = {1175317},
  doi                  = {10.3389/fphy.2023.1175317},
  fjournal             = {Frontiers in Physics},
  orcid-numbers        = {Sallermann, Moritz/0000-0002-2355-9355},
  researcherid-numbers = {Lounis, Samir/G-1875-2010 ABUSAA, Muayad/F-9366-2018},
  unique-id            = {WOS:001000401700001},
}

@Article{Arute2019,
  author     = {F. Arute and others},
  journal    = {Nat},
  title      = {Quantum supremacy using a programmable superconducting processor},
  year       = {2019},
  number     = {7779},
  pages      = {505--510},
  volume     = {574},
  doi        = {10.1038/s41586-019-1666-5},
  fjournal   = {Nature},
}

@Article{Boulle2024,
  author     = {Van Tuong Pham and Naveen Sisodia and Ilaria Di Manici and Joseba Urrestarazu-Larrañaga and Kaushik Bairagi and Johan Pelloux-Prayer and Rodrigo Guedas and Liliana D. Buda-Prejbeanu and Stéphane Auffret and Andrea Locatelli and Tevfik Onur Menteş and Stefania Pizzini and Pawan Kumar and Aurore Finco and Vincent Jacques and Gilles Gaudin and Olivier Boulle},
  journal    = {Science},
  title      = {Fast current-induced skyrmion motion in synthetic antiferromagnets},
  year       = {2024},
  number     = {6693},
  pages      = {307--312},
  volume     = {384},
  doi        = {10.1126/science.add5751},
  fjournal   = {Science},
}

@Article{Camley2023,
  author     = {Camley, Robert E. and Livesey, Karen L.},
  journal    = {Surf. Sci. Rep.},
  title      = {Consequences of the Dzyaloshinskii-Moriya interaction},
  year       = {2023},
  issn       = {0167-5729},
  month      = aug,
  number     = {3},
  pages      = {100605},
  volume     = {78},
  doi        = {10.1016/j.surfrep.2023.100605},
  eissn      = {1879-274X},
  fjournal   = {Surface Science Reports},
}

@Article{Cirac1995,
  author     = {J. I. Cirac and P. Zoller},
  journal    = {Phys. Rev. Lett.},
  title      = {Quantum computations with cold trapped ions},
  year       = {1995},
  number     = {20},
  pages      = {4091--4094},
  volume     = {74},
  doi        = {10.1103/PhysRevLett.74.4091},
  fjournal   = {Physical Review Letters},
}

@Article{Deutsch1985,
  author     = {D. Deutsch},
  journal    = {Proc. R. Soc. Lond. A},
  title      = {Quantum theory, the {Church-Turing} principle and the universal quantum computer},
  year       = {1985},
  number     = {1818},
  pages      = {97--117},
  volume     = {400},
  doi        = {10.1098/rspa.1985.0070},
}

@Article{DeutschJozsa1992,
  author  = {David Deutsch and Richard Jozsa},
  journal = {Proc. R. Soc. Lond. A},
  title   = {Rapid solution of problems by quantum computation},
  year    = {1992},
  number  = {1907},
  pages   = {553--558},
  volume  = {439},
  doi     = {10.1098/rspa.1992.0167},
}

@Article{ekert1991quantum,
  author     = {Ekert, Artur K.},
  journal    = {Phys. Rev. Lett.},
  title      = {Quantum cryptography based on {Bell}’s theorem},
  year       = {1991},
  number     = {6},
  pages      = {661--663},
  volume     = {67},
  doi        = {10.1103/PhysRevLett.67.661},
  fjournal   = {Physical Review Letters},
  publisher  = {American Physical Society},
}

@Article{Ezawa2017,
  author     = {Zhang, Xichao and Xia, Jing and Zhou, Yan and Liu, Xiaoxi and Zhang, Han and Ezawa, Motohiko},
  journal    = {Nat. Commun.},
  title      = {Skyrmion dynamics in a frustrated ferromagnetic film and current-induced helicity locking-unlocking transition},
  year       = {2017},
  month      = nov,
  pages      = {5709},
  volume     = {8},
  doi        = {10.1038/s41467-017-01785-w},
  eissn      = {2041-1723},
  fjournal   = {Nature Communications},
}

@Article{Ezawa2022,
  author     = {Xia, Jing and Zhang, Xichao and Liu, Xiaoxi and Zhou, Yan and Ezawa, Motohiko},
  journal    = {Commun. Mater.},
  title      = {Qubits based on merons in magnetic nanodisks},
  year       = {2022},
  month      = nov,
  number     = {1},
  pages      = {92},
  volume     = {3},
  doi        = {10.1038/s43246-022-00311-w},
  eissn      = {2662-4443},
}

@Article{Fert2017,
  author     = {A. Fert and N. Reyren and V. Cros},
  journal    = {Nat. Rev. Mater.},
  title      = {Magnetic skyrmions: {Advances} in physics and potential applications},
  year       = {2017},
  pages      = {17031},
  volume     = {2},
  doi        = {10.1038/natrevmats.2017.31},
  fjournal   = {Nature Reviews Materials},
}

@Article{Fert2023,
  author     = {Fert, Albert and Chshiev, Mairbek and Thiaville, Andre and Yang, Hongxin},
  journal    = {J. Phys. Soc. Jpn.},
  title      = {From Early Theories of {D}zyaloshinskii-{M}oriya Interactions in Metallic Systems to Today's Novel Roads},
  year       = {2023},
  issn       = {0031-9015},
  month      = aug,
  number     = {8},
  pages      = {081001},
  volume     = {92},
  doi        = {10.7566/JPSJ.92.081001},
  fjournal   = {Journal of The Physical Society of Japan},
}

@Article{Feynman1982,
  author     = {R. P. Feynman},
  journal    = {Int. J. Theor. Phys.},
  title      = {Simulating physics with computers},
  year       = {1982},
  number     = {6/7},
  pages      = {467--488},
  volume     = {21},
  doi        = {10.1007/BF02650179},
  fjournal   = {International Journal of Theoretical Physics},
}

@Article{Fiorentino2005SinglePhotonSwap,
  author   = {Marco Fiorentino and Taehyun Kim and Franco N.~C. Wong},
  journal  = {Phys. Rev. A},
  title    = {Single-photon two-qubit SWAP gate for entanglement manipulation},
  year     = {2005},
  number   = {1},
  pages    = {012318},
  volume   = {72},
  doi      = {10.1103/PhysRevA.72.012318},
  fjournal = {Physical Review A},
}

@Article{FRadu2023,
  author     = {Luo, Chen and Chen, Kai and Ukleev, Victor and Wintz, Sebastian and Weigand, Markus and Abrudan, Radu-Marius and Prokes, Karel and Radu, Florin},
  journal    = {Commun. Phys.},
  title      = {Direct observation of {N\'eel}-type skyrmions and domain walls in a ferrimagnetic {DyCo$_3$} thin film},
  year       = {2023},
  issn       = {2399-3650},
  month      = aug,
  number     = {1},
  pages      = {214},
  volume     = {6},
  doi        = {10.1038/s42005-023-01341-7},
  fjournal   = {Communications Physics},
}

@Article{Gauyacq2019,
  author     = {Gauyacq, J. P. and Lorente, N.},
  journal    = {J. Phys. Condens. Matter},
  title      = {A model for individual quantal nano-skyrmions},
  year       = {2019},
  issn       = {0953-8984},
  month      = aug,
  number     = {32},
  pages      = {325302},
  volume     = {31},
  doi        = {10.1088/1361-648X/ab1f3a},
  eissn      = {1361-648X},
}

@InProceedings{Grover1996,
  author     = {Grover, Lov K.},
  booktitle  = {Proceedings of the 28th annual ACM symposium on Theory of computing},
  title      = {A fast quantum mechanical algorithm for database search},
  year       = {1996},
  pages      = {212--219},
  publisher  = {ACM Press},
  series     = {STOC ’96},
  collection = {STOC ’96},
  doi        = {10.1145/237814.237866},
}

@Article{Haller2022,
  author     = {Haller, Andreas and Groenendijk, Solofo and Habibi, Alireza and Michels, Andreas and Schmidt, Thomas L.},
  journal    = {Phys. Rev. Research},
  title      = {Quantum skyrmion lattices in {Heisenberg} ferromagnets},
  year       = {2022},
  month      = nov,
  number     = {4},
  pages      = {043113},
  volume     = {4},
  doi        = {10.1103/PhysRevResearch.4.043113},
  eissn      = {2643-1564},
  fjournal   = {Physical Review Research},
}

@Article{Josephson1962,
  author     = {B. D. Josephson},
  journal    = {Phys. Lett.},
  title      = {Possible new effects in superconductive tunnelling},
  year       = {1962},
  number     = {7},
  pages      = {251--253},
  volume     = {1},
  doi        = {10.1016/0031-9163(62)91369-0},
  fjournal   = {Physics Letters},
}

@Article{Kalin2024OpticalSkyrmion,
  author   = {Jantje Kalin and Sibylle Sievers and Daniel Kalin and Andreas Bauer and Robin Abram and Heiko Füser and Hans Werner Schumacher and Christian Pfleiderer and Mark Bieler},
  journal  = {Phys. Rev. Applied},
  title    = {Optical creation and annihilation of skyrmion patches in a chiral magnet},
  year     = {2024},
  number   = {3},
  pages    = {034065},
  volume   = {21},
  doi      = {10.1103/PhysRevApplied.21.034065},
  fjournal = {Physical Review Applied},
}

@Article{Kitaev2003,
  author     = {A. Kitaev},
  journal    = {Ann. Phys.},
  title      = {Fault-tolerant quantum computation by anyons},
  year       = {2003},
  number     = {1},
  pages      = {2--30},
  volume     = {303},
  doi        = {10.1016/S0003-4916(02)00018-0},
  fjournal   = {Annals of Physics},
}

@Article{Kjaergaard2020,
  author     = {M. Kjaergaard and M. E. Schwartz and J. Braum{\"u}ller and P. Krantz and J. I.-J. Wang and S. Gustavsson and W. D. Oliver},
  journal    = {Annu. Rev. Condens. Matter Phys.},
  title      = {Superconducting qubits: Current state of play},
  year       = {2020},
  pages      = {369--395},
  volume     = {11},
  doi        = {10.1146/annurev-conmatphys-031119-050605},
  fjournal   = {Annual Review of Condensed Matter Physics},
}

@Article{Knill2001,
  author     = {E. Knill and R. Laflamme and G. J. Milburn},
  journal    = {Nature},
  title      = {A scheme for efficient quantum computation with linear optics},
  year       = {2001},
  number     = {6816},
  pages      = {46--52},
  volume     = {409},
  doi        = {10.1038/350510091},
  fjournal   = {Nature},
}

@Article{Lohani2019,
  author     = {Lohani, Vivek and Hickey, Ciaran and Masell, Jan and Rosch, Achim},
  journal    = {Phys. Rev. X},
  title      = {Quantum Skyrmions in Frustrated Ferromagnets},
  year       = {2019},
  issn       = {2160-3308},
  month      = dec,
  number     = {4},
  pages      = {041063},
  volume     = {9},
  doi        = {10.1103/PhysRevX.9.041063},
  fjournal   = {Physical Review X},
}

@Article{Loss1998,
  author     = {D. Loss and D. P. DiVincenzo},
  journal    = {Phys. Rev. A},
  title      = {Quantum computation with quantum dots},
  year       = {1998},
  number     = {1},
  pages      = {120--126},
  volume     = {57},
  doi        = {10.1103/PhysRevA.57.120},
  fjournal   = {Physical Review A},
}

@Article{Ma2025,
  author   = {Ma, Jiantao and Yang, Jiawei and Liu, Shunfa and Chen, Bo and Li, Xueshi and Song, Changkun and Qiu, Guixin and Zou, Kai and Hu, Xiaolong and Li, Feng and Yu, Ying and Liu, Jin},
  journal  = {Nat. Phys.},
  title    = {Nanophotonic quantum skyrmions enabled by semiconductor cavity quantum electrodynamics},
  year     = {2025},
  issn     = {1745-2473},
  month    = {2025 JUL 9},
  doi      = {10.1038/s41567-025-02973-y},
  eissn    = {1745-2481},
  fjournal = {Nature Physics},
}

@Article{Mazurenko2023,
  author     = {Mazurenko, Vladimir V. and Iakovlev, Ilia A. and Sotnikov, Oleg M. and Katsnelson, Mikhail I.},
  journal    = {J. Phys. Soc. Jpn.},
  title      = {Estimating Patterns of Classical and Quantum Skyrmion States},
  year       = {2023},
  issn       = {0031-9015},
  month      = aug,
  number     = {8},
  pages      = {081004},
  volume     = {92},
  doi        = {10.7566/JPSJ.92.081004},
  fjournal   = {Journal of the Physical Society of Japan},
}

@Article{Microsoft2025,
  author     = {Microsoft Azure Quantum and M. Aghaee and A. Alcaraz Ramirez and others},
  journal    = {Nat},
  title      = {Interferometric single-shot parity measurement in InAs--Al hybrid devices},
  year       = {2025},
  pages      = {651--655},
  volume     = {638},
  doi        = {10.1038/s41586-024-08445-2},
  fjournal   = {Nature},
}

@Article{Moessner2006,
  author     = {Moessner, R and Ramirez, AP},
  journal    = {Phys. Today},
  title      = {Geometrical frustration},
  year       = {2006},
  issn       = {0031-9228},
  month      = feb,
  number     = {2},
  pages      = {24--29},
  volume     = {59},
  doi        = {10.1063/1.2186278},
  eissn      = {1945-0699},
  fjournal   = {Physics Today},
}

@Article{Mohylna2021,
  author               = {Mohylna, M. and Busa, Jr., J. and Zukovic, M.},
  journal              = {J. Magn. Magn. Mater.},
  title                = {Formation and growth of skyrmion crystal phase in a frustrated {Heisenberg} antiferromagnet with {D}zyaloshinskii-{M}oriya interaction},
  year                 = {2021},
  issn                 = {0304-8853},
  month                = jun,
  pages                = {167755},
  volume               = {527},
  doi                  = {10.1016/j.jmmm.2021.167755},
  eissn                = {1873-4766},
  fjournal             = {Journal of Magnetism and Magnetic Materials},
  researcherid-numbers = {{\v Z}ukovi{\v c}, Milan/H-1600-2016 Zukovic, Milan/H-1600-2016 Bu{\v s}a, J{\'a}n/AAB-6463-2019},
}

@article{Nanomaterials2022,
	author = {R.-A. One and S. Mican and A.-G. Cimpoeșu and M. Joldos and R. Tetean and Tiusan},
	doi = {10.3390/nano12244411},
	journal = {Nanomaterials},
	number = {24},
	pages = {4411},
	title = {Micromagnetic Design of Skyrmionic Materials and Chiral Magnetic Configurations in Patterned Nanostructures for Neuromorphic and Qubit Applications},
	volume = {12},
	year = {2022}}

@InBook{neutron,
  author    = {Jensen, Jens and Mackintosh, Allan R},
  chapter   = {4.2},
  publisher = {Oxford University Press},
  title     = {Rare Earth Magnetism: {Structures} and Excitations},
  year      = {1991},
  address   = {Oxford},
  month     = jun,
  doi       = {10.1093/oso/9780198520276.001.0001},
  url       = {https://doi.org/10.1093/oso/9780198520276.001.0001},
}

@Book{Nielsen2010,
  author     = {M. A. Nielsen and I. L. Chuang},
  publisher  = {Cambridge University Press},
  title      = {Quantum Computation and Quantum Information},
  year       = {2010},
  address    = {Cambridge},
  doi        = {10.1017/CBO9780511976667},
}

@Article{Ochoa2019,
  author     = {Ochoa, Hector and Tserkovnyak, Yaroslav},
  journal    = {Int. J. Mod. Phys. B},
  title      = {Quantum skyrmionics},
  year       = {2019},
  issn       = {0217-9792},
  month      = aug,
  number     = {21},
  pages      = {1930005},
  volume     = {33},
  doi        = {10.1142/S0217979219300056},
  eissn      = {1793-6578},
  fjournal   = {International Journal of Modern Physics B},
}

@Article{Ornelas2024,
  author     = {Ornelas, Pedro and Nape, Isaac and de Mello Koch, Robert and Forbes, Andrew},
  journal    = {Nat. Photonics},
  title      = {Non-local skyrmions as topologically resilient quantum entangled states of light},
  year       = {2024},
  issn       = {1749-4885},
  month      = mar,
  number     = {3},
  pages      = {258--266},
  volume     = {18},
  doi        = {10.1038/s41566-023-01360-4},
  eissn      = {1749-4893},
  fjournal   = {Nature Photonics},
}

@Article{Ornelas2025,
  author     = {Ornelas, Pedro and Nape, Isaac and Koch, Robert de Mello and Forbes, Andrew},
  journal    = {Nat. Commun.},
  title      = {Topological rejection of noise by quantum skyrmions},
  year       = {2025},
  month      = mar,
  number     = {1},
  pages      = {1349},
  volume     = {16},
  doi        = {10.1038/s41467-025-58232-4},
  eissn      = {2041-1723},
  fjournal   = {Nature Communications},
}

@Article{Pan1998EntanglementSwapping,
  author   = {Jian-Wei Pan and Dik Bouwmeester and Harald Weinfurter and Anton Zeilinger},
  journal  = {Phys. Rev. Lett.},
  title    = {Experimental Entanglement Swapping: Entangling Photons That Never Interacted},
  year     = {1998},
  month    = may,
  number   = {18},
  pages    = {3891--3894},
  volume   = {80},
  doi      = {10.1103/PhysRevLett.80.3891},
  fjournal = {Physical Review Letters},
}

@Article{Preskill2018,
  author     = {J. Preskill},
  journal    = {Quantum},
  title      = {Quantum computing in the {NISQ} era and beyond},
  year       = {2018},
  pages      = {79},
  volume     = {2},
  doi        = {10.22331/q-2018-08-06-79},
}

@Article{Psaroudaki2021,
  author     = {Psaroudaki, Christina and Panagopoulos, Christos},
  journal    = {Phys. Rev. Lett.},
  title      = {Skyrmion qubits: {A} new class of quantum logic elements based on nanoscale magnetization},
  year       = {2021},
  pages      = {067201},
  volume     = {127},
  doi        = {10.1103/PhysRevLett.127.067201},
  fjournal   = {Physical Review Letters},
}

@Article{Psaroudaki2023,
  author     = {C. Psaroudaki, E. Peraticos, C. Panagopoulos},
  journal    = {Appl. Phys. Lett.},
  title      = {Skyrmion qubits: Challenges for future quantum computing applications},
  year       = {2023},
  number     = {26},
  pages      = {260501},
  volume     = {123},
  doi        = {10.1063/5.0177864},
  fjournal   = {Applied Physics Letters},
}

@Article{Quspin2017,
  author    = {Phillip Weinberg and Marin Bukov},
  journal   = {SciPost Phys.},
  title     = {{QuSpin: a {Python} package for dynamics and exact diagonalisation of quantum many body systems part I: spin chains}},
  year      = {2017},
  pages     = {003},
  volume    = {2},
  doi       = {10.21468/SciPostPhys.2.1.003},
  publisher = {SciPost},
}

@Article{Quspin2019,
  author     = {P. Weinberg and M. Bukov},
  journal    = {SciPost Phys.},
  title      = {QuSpin: a {Python} package for dynamics and exact diagonalisation of quantum many body systems. {Part II}: bosons, fermions and higher spins},
  year       = {2019},
  pages      = {020},
  volume     = {7},
  doi        = {10.21468/SciPostPhys.7.2.020},
  fjournal   = {SciPost Physics},
}

@article{Rohart2013,
  title = {Skyrmion confinement in ultrathin film nanostructures in the presence of Dzyaloshinskii-Moriya interaction},
  author = {Rohart, S. and Thiaville, A.},
  journal = {Phys. Rev. B},
  volume = {88},
  issue = {18},
  pages = {184422},
  numpages = {8},
  year = {2013},
  month = {Nov},
  publisher = {American Physical Society},
  doi = {10.1103/PhysRevB.88.184422},
  url = {https://link.aps.org/doi/10.1103/PhysRevB.88.184422}
}

@Article{Salvati2024,
  author     = {Salvati, Fabio and Katsnelson, Mikhail I. and Bagrov, Andrey A. and Westerhout, Tom},
  journal    = {Phys. Rev. B},
  title      = {Stability of a quantum skyrmion: {Projective} measurements and the quantum {Zeno} effect},
  year       = {2024},
  issn       = {2469-9950},
  month      = feb,
  number     = {6},
  pages      = {064409},
  volume     = {109},
  doi        = {10.1103/PhysRevB.109.064409},
  eissn      = {2469-9969},
  fjournal   = {Physical Review B},
}

@InProceedings{Shor1994,
  author     = {Shor, P.W.},
  booktitle  = {Proceedings 35th Annual Symposium on Foundations of Computer Science},
  title      = {Algorithms for quantum computation: {Discrete} logarithms and factoring},
  pages      = {124--134},
  publisher  = {IEEE Comput. Soc. Press},
  series     = {SFCS-94},
  collection = {SFCS-94},
  doi        = {10.1109/sfcs.1994.365700},
}

@Article{Siegl2022,
  author     = {Siegl, Pia and Vedmedenko, Elena Y. and Stier, Martin and Thorwart, Michael and Posske, Thore},
  journal    = {Phys. Rev. Research},
  title      = {Controlled creation of quantum skyrmions},
  year       = {2022},
  month      = may,
  number     = {2},
  pages      = {023111},
  volume     = {4},
  doi        = {10.1103/PhysRevResearch.4.023111},
  eissn      = {2643-1564},
  fjournal   = {Physical Review Research},
}

@Article{Sotnikov2021,
  author     = {O. M. Sotnikov and V. V. Mazurenko and J. Colbois and F. Mila and M. I. Katsnelson and E. A. Stepanov},
  journal    = {Phys. Rev. B},
  title      = {Probing the topology of the quantum analog of a classical skyrmion},
  year       = {2021},
  number     = {6},
  pages      = {L060404},
  volume     = {103},
  doi        = {10.1103/PhysRevB.103.L060404},
  fjournal   = {Physical Review B},
}

@Article{Stepanov2019,
  author     = {Stepanov, E. A. and Nikolaev, S. A. and Dutreix, C. and Katsnelson, I, M. and Mazurenko, V. V.},
  journal    = {J. Phys. Condens. Matter},
  title      = {Heisenberg-exchange-free nanoskyrmion mosaic},
  year       = {2019},
  issn       = {0953-8984},
  month      = may,
  number     = {17},
  pages      = {175802},
  volume     = {31},
  doi        = {10.1088/1361-648X/ab02b9},
  eissn      = {1361-648X},
}

@Article{Vijayan2023,
  author     = {Vijayan, Vipin and Chotorlishvili, L. and Ernst, A. and Parkin, S. S. P. and Katsnelson, M. I. and Mishra, S. K.},
  journal    = {Phys. Rev. B},
  title      = {Topological dynamical quantum phase transition in a quantum skyrmion phase},
  year       = {2023},
  issn       = {2469-9950},
  month      = mar,
  number     = {10},
  pages      = {L100419},
  volume     = {107},
  doi        = {10.1103/PhysRevB.107.L100419},
  eissn      = {2469-9969},
  fjournal   = {Physical Review B},
}

@Article{WdWSkyr2024,
  author     = {Liu, Chen and Zhang, Senfu and Hao, Hongyuan and Algaidi, Hanin and Ma, Yinchang and Zhang, Xi-Xiang},
  journal    = {Adv. Mater.},
  title      = {Magnetic Skyrmions above Room Temperature in a van der {Waals} Ferromagnet {Fe$_3$GaTe$_2$}},
  year       = {2024},
  issn       = {0935-9648},
  month      = may,
  number     = {18},
  pages      = {2311022},
  volume     = {36},
  doi        = {10.1002/adma.202311022},
  eissn      = {1521-4095},
  fjournal   = {Advanced Materials},
}

@Article{Wineland2013,
  author     = {D. J. Wineland},
  journal    = {Rev. Mod. Phys.},
  title      = {Nobel lecture: Superposition, entanglement, and raising {Schr{\"o}dinger}'s cat},
  year       = {2013},
  number     = {3},
  pages      = {1103--1114},
  volume     = {85},
  doi        = {10.1103/RevModPhys.85.1103},
}

@Article{XFPan2024,
  author     = {X-F Pan and X-L Hei and X. Yao and J. Chen and others},
  journal    = {Phys. Rev. Research},
  title      = {Skyrmion-mechanical hybrid quantum systems: Manipulation of skyrmion qubits via phonons},
  year       = {2024},
  pages      = {023067},
  volume     = {6},
  doi        = {10.1103/PhysRevResearch.6.023067},
  fjournal   = {Physical Review Research},
}

@Article{Yu2023,
  author     = {Yu, Dongxing and Yang, Hongxin and Chshiev, Mairbek and Fert, Albert},
  journal    = {Natl. Sci. Rev.},
  title      = {Skyrmions-based logic gates in one single nanotrack completely reconstructed via chirality barrier},
  year       = {2022},
  issn       = {2095-5138},
  month      = dec,
  number     = {12},
  pages      = {nwac021},
  volume     = {9},
  doi        = {10.1093/nsr/nwac021},
  eissn      = {2053-714X},
  fjournal   = {National Science Review},
}

@Article{Zhang2023,
  author     = {Zhang, Huai and Zhang, Yajiu and Hou, Zhipeng and Qin, Minghui and Gao, Xingsen and Liu, Junming},
  journal    = {Mater. Futures},
  title      = {Magnetic skyrmions: {Materials}, manipulation, detection, and applications in spintronic devices},
  year       = {2023},
  month      = sep,
  number     = {3},
  pages      = {032201},
  volume     = {2},
  doi        = {10.1088/2752-5724/ace1df},
  eissn      = {2752-5724},
}

@Article{Paul1990,
  author     = {W. Paul},
  journal    = {Rev. Mod. Phys.},
  title      = {Electromagnetic traps for charged and neutral particles},
  year       = {1990},
  number     = {3},
  pages      = {531--540},
  volume     = {62},
  doi        = {10.1103/RevModPhys.62.531},
}

@Article{Bruzewicz2019,
  author  = {Bruzewicz, Colin D. and Chiaverini, John and McConnell, Robert and Sage, Jeremy M.},
  journal = {Appl. Phys. Rev.},
  title   = {Trapped-ion quantum computing: Progress and challenges},
  year    = {2019},
  issn    = {1931-9401},
  month   = may,
  number  = {2},
  pages   = {021314},
  volume  = {6},
  doi     = {10.1063/1.5088164},
}

@Article{White1992,
  author    = {White, Steven R.},
  journal   = {Phys. Rev. Lett.},
  title     = {Density matrix formulation for quantum renormalization groups},
  year      = {1992},
  month     = nov,
  pages     = {2863--2866},
  volume    = {69},
  doi       = {10.1103/PhysRevLett.69.2863},
  issue     = {19},
  numpages  = {0},
  publisher = {American Physical Society},
}

@Article{Lindblad1976,
  author    = {Lindblad, G.},
  journal   = {Commun. Math. Phys.},
  title     = {On the generators of quantum dynamical semigroups},
  year      = {1976},
  issn      = {1432-0916},
  month     = jun,
  number    = {2},
  pages     = {119--130},
  volume    = {48},
  doi       = {10.1007/bf01608499},
  fjournal  = {Communications in Mathematical Physics},
  publisher = {Springer Science and Business Media LLC},
}

@Article{Gorini1976,
  author    = {Gorini, Vittorio and Kossakowski, Andrzej and Sudarshan, E. C. G.},
  journal   = {J. Math. Phys.},
  title     = {Completely positive dynamical semigroups of {N}-level systems},
  year      = {1976},
  issn      = {1089-7658},
  month     = may,
  number    = {5},
  pages     = {821--825},
  volume    = {17},
  doi       = {10.1063/1.522979},
  fjournal  = {Journal of Mathematical Physics},
  publisher = {AIP Publishing},
}

@Article{Daley2022,
  author   = {Daley, Andrew J. and Bloch, Immanuel and Kokail, Christian and Flannigan, Stuart and Pearson, Natalie and Troyer, Matthias and Zoller, Peter},
  journal  = {Nature},
  title    = {Practical quantum advantage in quantum simulation},
  year     = {2022},
  issn     = {1476-4687},
  number   = {7920},
  pages    = {667--676},
  volume   = {607},
  doi      = {10.1038/s41586-022-04940-6},
  fjournal = {Nature},
  refid    = {Daley2022},
}

@Article{Steane1998,
  author    = {Steane, Andrew},
  journal   = {Rep. Prog. Phys.},
  title     = {Quantum computing},
  year      = {1998},
  issn      = {1361-6633},
  month     = feb,
  number    = {2},
  pages     = {117--173},
  volume    = {61},
  doi       = {10.1088/0034-4885/61/2/002},
  publisher = {IOP Publishing},
}

@Article{Krantz2019,
  author    = {Krantz, P. and Kjaergaard, M. and Yan, F. and Orlando, T. P. and Gustavsson, S. and Oliver, W. D.},
  journal   = {Appl. Phys. Rev.},
  title     = {A quantum engineer’s guide to superconducting qubits},
  year      = {2019},
  issn      = {1931-9401},
  month     = jun,
  number    = {2},
  pages     = {021318},
  volume    = {6},
  doi       = {10.1063/1.5089550},
  publisher = {AIP Publishing},
}

@Article{Blais2021,
  author    = {Blais, Alexandre and Grimsmo, Arne L. and Girvin, S. M. and Wallraff, Andreas},
  journal   = {Rev. Mod. Phys.},
  title     = {Circuit quantum electrodynamics},
  year      = {2021},
  month     = may,
  pages     = {025005},
  volume    = {93},
  doi       = {10.1103/RevModPhys.93.025005},
  issue     = {2},
  numpages  = {72},
  publisher = {American Physical Society},
}

@Article{Burkard2023,
  author    = {Burkard, Guido and Ladd, Thaddeus D. and Pan, Andrew and Nichol, John M. and Petta, Jason R.},
  journal   = {Rev. Mod. Phys.},
  title     = {Semiconductor spin qubits},
  year      = {2023},
  month     = jun,
  pages     = {025003},
  volume    = {95},
  doi       = {10.1103/RevModPhys.95.025003},
  issue     = {2},
  numpages  = {58},
  publisher = {American Physical Society},
}

@Article{Kok2007,
  author    = {Kok, Pieter and Munro, W. J. and Nemoto, Kae and Ralph, T. C. and Dowling, Jonathan P. and Milburn, G. J.},
  journal   = {Rev. Mod. Phys.},
  title     = {Linear optical quantum computing with photonic qubits},
  year      = {2007},
  issn      = {1539-0756},
  month     = jan,
  number    = {1},
  pages     = {135--174},
  volume    = {79},
  doi       = {10.1103/revmodphys.79.135},
  publisher = {American Physical Society (APS)},
}

@Article{Nayak2008,
  author    = {Nayak, Chetan and Simon, Steven H. and Stern, Ady and Freedman, Michael and Das Sarma, Sankar},
  journal   = {Rev. Mod. Phys.},
  title     = {Non-Abelian anyons and topological quantum computation},
  year      = {2008},
  issn      = {1539-0756},
  month     = sep,
  number    = {3},
  pages     = {1083--1159},
  volume    = {80},
  doi       = {10.1103/revmodphys.80.1083},
  publisher = {American Physical Society (APS)},
}

@Article{Kitaev2001,
  author    = {Kitaev, A Yu},
  journal   = {Phys.-Uspekhi},
  title     = {Unpaired {Majorana} fermions in quantum wires},
  year      = {2001},
  issn      = {1468-4780},
  month     = oct,
  number    = {10S},
  pages     = {131--136},
  volume    = {44},
  doi       = {10.1070/1063-7869/44/10s/s29},
  fjournal  = {Physics-Uspekhi},
  publisher = {Uspekhi Fizicheskikh Nauk (UFN) Journal},
}

@Article{Beenakker2013,
  author   = {Beenakker, C.W.J.},
  journal  = {Annu. Rev. Condens. Matter Phys.},
  title    = {Search for {Majorana} {Fermions} in {Superconductors}},
  year     = {2013},
  issn     = {1947-5462},
  number   = {Volume 4, 2013},
  pages    = {113--136},
  volume   = {4},
  doi      = {10.1146/annurev-conmatphys-030212-184337},
}

@Article{Aguado2017,
  author  = {Aguado, Ramón},
  journal = {Riv. Nuovo Cim.},
  title   = {Majorana quasiparticles in condensed matter},
  year    = {2017},
  issn    = {1826-9850},
  month   = nov,
  number  = {11},
  pages   = {523--593},
  volume  = {40},
  doi     = {10.1393/ncr/i2017-10141-9},
}

@Article{Sotnikov2023,
  author    = {Sotnikov, O. M. and Stepanov, E. A. and Katsnelson, M. I. and Mila, F. and Mazurenko, V. V.},
  journal   = {Phys. Rev. X},
  title     = {Emergence of Classical Magnetic Order from {Anderson} Towers: {Quantum} {Darwinism} in Action},
  year      = {2023},
  month     = nov,
  pages     = {041027},
  volume    = {13},
  doi       = {10.1103/PhysRevX.13.041027},
  issue     = {4},
  numpages  = {17},
  publisher = {American Physical Society},
}

@Article{Joshi2024,
  author    = {Joshi, Ashish and Peters, Robert and Posske, Thore},
  journal   = {Phys. Rev. B},
  title     = {Quantum skyrmion dynamics studied by neural network quantum states},
  year      = {2024},
  month     = sep,
  pages     = {104411},
  volume    = {110},
  doi       = {10.1103/PhysRevB.110.104411},
  issue     = {10},
  numpages  = {8},
  publisher = {American Physical Society},
}

@Article{Xia2023,
  author     = {Xia, Jing and Zhang, Xichao and Liu, Xiaoxi and Zhou, Yan and Ezawa, Motohiko},
  journal    = {Phys. Rev. Lett.},
  title      = {Universal Quantum Computation Based on Nanoscale Skyrmion Helicity Qubits in Frustrated Magnets},
  year       = {2023},
  issn       = {0031-9007},
  month      = mar,
  number     = {10},
  pages      = {106701},
  volume     = {130},
  doi        = {10.1103/PhysRevLett.130.106701},
  eissn      = {1079-7114},
  fjournal   = {Physical Review Letters},
  issue      = {10},
  numpages   = {6},
  publisher  = {American Physical Society},
}

@Article{Petrovic2025,
  author     = {Petrovi{\'c}, Alexander P. and Psaroudaki, Christina and Fischer, Peter and Garst, Markus and Panagopoulos, Christos},
  journal    = {Rev. Mod. Phys.},
  title      = {\textit{{Colloquium}} : {Quantum} properties and functionalities of magnetic skyrmions},
  year       = {2025},
  issn       = {0034-6861, 1539-0756},
  month      = jul,
  number     = {3},
  pages      = {031001},
  volume     = {97},
  doi        = {10.1103/RevModPhys.97.031001},
  fjournal   = {Reviews of Modern Physics},
  shorttitle = {\textit{{Colloquium}}},
}

@Article{Cai2023,
  author    = {Cai, Zhenyu and Babbush, Ryan and Benjamin, Simon C. and Endo, Suguru and Huggins, William J. and Li, Ying and McClean, Jarrod R. and O'Brien, Thomas E.},
  journal   = {Rev. Mod. Phys.},
  title     = {Quantum error mitigation},
  year      = {2023},
  month     = {Dec},
  pages     = {045005},
  volume    = {95},
  doi       = {10.1103/RevModPhys.95.045005},
  issue     = {4},
  numpages  = {37},
  publisher = {American Physical Society},
  url       = {https://link.aps.org/doi/10.1103/RevModPhys.95.045005},
}

@Book{Lidar2013,
  author    = {Lidar, Daniel A and Brun, Todd A},
  publisher = {Cambridge University Press},
  title     = {Quantum error correction},
  year      = {2013},
  address   = {Cambridge},
  doi       = {https://doi.org/10.1017/CBO9781139034807},
}

\end{document}